\documentclass[]{aa} % for a referee version
\usepackage{graphicx}
\usepackage{txfonts}
\usepackage{subfigure}
\usepackage{longtable,lscape}
\usepackage{rotating}
\usepackage{natbib}

\begin{document}

\newcommand{\be}{\begin{equation}}
\newcommand{\ee}{\end{equation}}
\newcommand{\cmc}{cm$^{-3}$}
\newcommand{\kms}{km/s}
\newcommand{\um}{$\mu m$}
\newcommand{\sori}{$\sigma$~Ori}
\newcommand{\Msun}{M$_\odot$}
\newcommand{\Lsun}{L$_\odot$}
\newcommand{\Ha}{H$\alpha$}
\newcommand{\Pab}{Pa$\beta$}
\newcommand{\Pag}{Pa$\gamma$}
\newcommand{\Brg}{Br$\gamma$}
\newcommand{\Teff}{T$_{eff}$}
\newcommand{\Lstar}{L$_{star}$}
\newcommand{\Rstar}{R$_{star}$}
\newcommand{\Mstar}{M$_{star}$}
\newcommand{\Lacc}{L$_{acc}$}
\newcommand{\Lx}{L$_{\rm X}$}
\newcommand{\Macc}{$\dot M_{acc}$}
\newcommand{\Mloss}{$\dot M_{loss}$}
\newcommand{\Mwind}{$\dot M_{wind}$}
\newcommand{\Myr}{M$_\odot$/yr}
\newcommand {\Lline}{L$_{line}$}
\newcommand {\OIA}{[\ion{O}{i}]\,630.03}
\newcommand {\OIB}{[\ion{O}{i}]\,557.79}
\newcommand {\SIIA}{[\ion{S}{ii}]\,406.86}
\newcommand {\SIIB}{[\ion{S}{ii}]\,673.08}
\newcommand {\NII}{[\ion{N}{ii}]\,658.34}
\newcommand {\OII}{[\ion{O}{ii}]\,372.60}
\newcommand {\OIIB}{[\ion{O}{ii}]\,731.89}
\newcommand {\sig}{$\sigma$~Ori} 
 
\newcommand{\simless}{\mathbin{\lower 3pt\hbox
      {$\rlap{\raise 5pt\hbox{$\char'074$}}\mathchar"7218$}}}
\newcommand{\simgreat}{\mathbin{\lower 3pt\hbox
     {$\rlap{\raise 5pt\hbox{$\char'076$}}\mathchar"7218$}}}

   \title{X-Shooter spectroscopy of young stellar objects: 
         \\
	  V -- Slow winds in T Tauri stars  
   \thanks{Based on 
observations collected at the European Souther Observatory at
Paranal, under programs 084.C-0269(A), 085.C-0238(A), 086.C-0173(A), 
087.C-0244(A) and 089.C-0143(A).}}

\titlerunning{Slow winds in  TTS}

   \author{
          A. Natta\inst{1,2},  L. Testi\inst{1,3}, J.M. Alcal\'a\inst{4}, E. Rigliaco\inst{5}, E. Covino\inst{4}, B. Stelzer\inst{6} and V. D'Elia\inst{7}.
          }

   \institute {INAF/Osservatorio Astrofisico of Arcetri, Largo E. Fermi, 5, 50125 Firenze, Italy
              \email{natta@arcetri.astro.it} 
	\and
   {School of Cosmic Physics, Dublin Institute for Advanced Studies, 31 Fitzwilliams Place, Dublin 2, Ireland}
\and 
   {ESO/European Southern Observatory, Karl-Schwarzschild-Strasse 2
D-85748 Garching bei M\"unchen, Germany}
\and 
 {INAF/Osservatorio Astronomico di Capodimonte, Salita Moiariello, 16  80131, Napoli, Italy}
         \and
   {Department of Planetary Science, Lunar and Planetary Lab, University of Arizona, 1629, E. University Blvd, 85719, Tucson, AZ, USA}
	\and 
   {INAF/Osservatorio Astronomico di Palermo, Piazza del Parlamento 1, 90134 Palermo, Italy }
        \and
 {ASI-Science Data Center, Via del Politecnico snc, I-00133 Rome, Italy }\\
}
	\offprints{natta@arcetri.astro.it}
   \date{Received ...; accepted ...}

% \abstract{}{}{}{}{} 
% 5 {} token are mandatory
 
  \abstract
   { Disks around T Tauri stars are known to lose mass, as best
shown by the profiles of forbidden emission lines
of low ionization species. At least two separate
kinematic components have been identified, one characterised by velocity shifts of tens to hundreds km/s (HVC) and one with much lower velocity  of few km/s (LVC).
The HVC are convincingly associated to the emission of jets, but
the origin of the LVC is still unknown. In this paper we analyze the
forbidden line spectrum of a sample of   44 mostly low mass young stars in Lupus and \sori\ observed with the X-Shooter ESO spectrometer.
We detect forbidden line emission of \ion{O}{i},
\ion{O}{ii}, \ion{S}{ii}, \ion{N}{i}, and \ion{N}{ii}, and characterize the line profiles as LVC, blue-shifted HVC and red-shifted HVC.
We focus our study on the LVC.
We show that there is a good correlation between line luminosity and both
\Lstar\ and the accretion luminosity (or the mass-accretion rate) over a large
interval of values (\Lstar $\sim 10^{-2} - 1$ \Lsun; \Lacc $\sim 10^{-5} - 10^{-1}$ \Lsun;
\Macc $\sim 10^{-11} - 10^{-7}$ \Myr).
The lines show the presence of a slow wind ($V_{peak}<20$ km/s), dense ($n_H>10^8$ \cmc), warm (T$\sim 5000-10000$ K), mostly neutral.
We estimate the mass of the emitting gas and provide a value for
the maximum volume it occupies. Both quantities increase steeply
with the stellar mass, from $\sim 10^{-12}$ \Msun\ and $\sim 0.01$ AU$^3$ for \Mstar$\sim 0.1$ \Msun, to
$\sim 3 \times 10^{-10}$ \Msun\ and $\sim 1$ AU$^3$ for \Mstar$\sim 1$ \Msun, respectively.
These results  provide quite stringent constraints to wind models in 
low mass young stars, that need to be explored further.}
%   {b.}
%   {c.}
%   {d.}

   \maketitle
%
%________________________________________________________________

\section{Introduction}\label{introduction}

Circumstellar disks of gas and dust surround young stars from their birth for a period of few million years; 
during this time planetary systems may form and evolve. The disk structure and its evolution affect the 
conditions for the formation of planets and their properties 
\citep[][]{dutrey14, alexander14}. 
%{\bf (Dutrey et al. 2014; Alexander et al. 2014)}. 
Disks are dynamical structures, with gas and dust
accreting  onto the central stars, but  also being expelled from the systems. In particular, 
the way mass-loss occurs and evolves with time has important consequences
on the disk survival and on its properties.

Mass-loss from magnetized accretion disks is expected, due to the combined effect of rotation and magnetic fields; 
it is probably a very complex phenomenon, with different components: a so-called disk wind, due to the relic 
magnetic field that threads the disk at all distances from the star, an X-wind, which is launched
 in the region close to the star where the stellar magnetic field provides the pressure required to overcome 
 the stellar gravity, and, possibly, a stellar wind, where gas from the central star is expelled from the system 
\citep[see, e.g.][and references therein]{ferreira13}.  Under the action of the magnetic field, the gas is collimated 
and accelerated to terminal velocities of few hundred km/s, forming the bright jets observed in several 
young objects \citep[e.g.][]{frank14}.  As magnetocentrifugal winds extract angular momentum from the disk, they can control  
accretion \citep[e.g.][]{turner14}.

Disks can lose mass also when their
upper layers are heated to temperatures such that the gas  thermal energy exceeds its binding energy, 
and the gas escapes from the system (photoevaporation). This mechanism was firstly recognized to lead 
to dispersal of disks in the vicinity of a hot star \citep[e.g.,][]{odell94}. 
More recently, it has been realized that the combined effect of the high energy photons (UV, FUV, X) 
emitted by low-mass stars can heat the disk surface to sufficiently high temperatures to produce 
a centrifugally launched ouflow, driven by thermal pressure. 
Photoevaporation winds are possibly the cause of the quick disk dissipation, which ends their much longer 
phase of viscous evolution  
\citep[see, e.g.] [and references therein] {alexander14}. 
%Simon \& Prato 1995; Wolk and Walter, 1996; Andrews and Williams, 2005; Luhman et al., 2010; 
%Koepferl et al., 2013) {\bf these are all the references included in Alexander 2014. Should we quote only 
%this latter paper again? If we keep them they have to be included in the ref list}).

Observational evidence of mass-loss from young stars (T Tauri stars; TTS) with  disks (Class II) is provided by the intensity 
and profiles of forbidden lines of atomic and low-ionization species, which present 
at least two different components \citep[][]{hartigan95}. One is emitted by gas moving at high velocity (HVC), 
which  is clearly identified with  the jets that have been imaged and carefully 
studied in several objects. 
The other component is originated by a much slower moving gas 
(low-velocity component-LVC). The LVC is detected in most Class II objects, and its origin is still unknown. 
It  could be emitted at the base of a magnetically driven disk wind, as suggested by \citet[][]{hartigan95} for 
the \ion{O}{i} lines, but it can also be a tracer of a photoevaporative disk wind, as shown by, e.g.,  
\citet[][]{pascucciandsterzik09} for the LVC of the \ion{Ne}{ii} mid-IR emission lines.  
Lately, \citet[][]{rigliaco13} have shown that the LVC \ion{O}{i} lines  could also have multiple components, 
with one component tracing gas in keplerian rotation, and another component tracing a photoevaporative wind.   
\citet[][]{acke05} find that in Herbig Ae/Be stars, the \OIA nm emission could come from the disk surface layers. 
In this paper we will focus on the properties of the LVC components of the winds from low-mass T Tauri stars.
%The other component comes from a much slower moving gas (LVC), of lower 
%excitation, closer to the disk plane than the HVC (Hirth et al.??). The LVC emission is detected in 
%almost all Class II objects,  and its origin is still not known.
%It has been suggested that it could be emitted at the base of a  magnetically-driven disk wind (??),
%but it can also be the tracer of a photoevaporative wind, and provide the best
%direct diagnostics of its presence and properties (ref).
%In some more luminous objects, there is evidence that the \OIA\ line traces the keplerian rotation of 
%the circumstellar disk (Acke et al.??; see also Rigliaco et al. 2013), and an origin in the heated disk 
%surface has been proposed (ref.??).

As for other disk and stellar diagnostics,  
%we expect that the mass-loss properties will vary over a large range for  any given will also
%have a large spread of properties. 
to find meaningful trends and to
make full use of the diagnostic potential of the LVC forbidden line emission, it is necessary to analyze 
simultaneously as many lines as possible and to have access to  a large sample of objects, covering a 
wide range of stellar and accretion properties. The best sample available so far is still that of 
\citet[][]{hartigan95}, who observed   42 TTS  in Taurus and detected forbidden line emission in 
about 30 of them. The number of stars with reliable measurements of the mass-accretion
properties is, however, smaller, of the order of 12--15 objects 
\citep[][]{gullbring98, rigliaco13}, 
mostly of relatively high mass and accretion rate.
In this paper, we analyze a sample of 44 low-mass stars  in two star-forming regions, Lupus and \sori, 
of age ~1--3 My. The two regions differ mostly due to the presence of a massive star (\sori) in 
the \sori\ region. The spectra have been obtained with X-Shooter@VLT \citep[][]{vernet11}, which gives medium-resolution
spectra simultaneously over the spectral range from 310 to 2500 nm. They have been collected as part 
of the Italian GTO \citep[][]{alcala11} and have already been used to derive the stellar parameters and the accretion 
properties of the stars \citep[][]{rigliaco12, alcala14}. This is, therefore, the largest 
sample of TTS with well-known accretion properties studied so far with the aim of characterizing the 
mass-loss phenomenon and its origin.

In general, forbidden lines in the optical and near-IR are  emitted by warm gas (temperature of few 
thousand degrees), where collisions with electrons or neutral hydrogen excite the upper levels 
of the transitions, and we will assume that this is the case throught this paper. However, an alternative, 
non-thermal mechanism has been proposed for the formation of the  \ion{O}{i} lines, namely that they have 
origin in a cooler disk region where OH is photodissociated by the stellar FUV photons 
\citep[][]{acke05, gorti11}. We will come back to this point at the end of the paper.

The paper is organized as follows. \S 2 and 3 briefly summarize the observations and data reduction, and 
the sample properties, already discussed in \citep[][]{alcala14,rigliaco12}; \S 4 presents the forbidden line 
spectra; \S 5 the line profiles and the separation of the HVC and LVC. \S 6 is dedicated to the LVC. 
Discussion and summary follow in \S 7 and \S 8, respectively.

\section {Observations}

In this work we use the spectra published in \citet{alcala14}
and \citet[][]{rigliaco12}. The instrument setup, including 
the spectral resolution in each wavelength range and the corresponding signal-to-noise ratio, as well as the
data reduction and calibration procedures and the derivation of extinction, spectral type and accretion luminosity are described in those papers. 
All the spectra used in this paper have been corrected for extinction.
For most targets, the slit width was 1 arcsec, yielding a resolution
of $\sim$60 km/s in the UVB arm and $\sim$35 km/s in the VIS arm.

In addition to the wavelength calibrations described in \citet{alcala14}
and \citet[][]{rigliaco12} we have cross checked and adjusted
the wavelength scale of the UVB and VIS arms by aligning the \OIB\ line,
which is covered by both arms. This line is detected in all objects for
which we have detections of other lines in the UVB arm, with the exception
of SO587 and SO646. 
%For these two objects the wavelength scale of the UVB
%arm has been adjusted assuming that the \SIIA\ line is centered at the 
%photosphere velocity. 
 We detected  and applied an offset to the UVB wavelength 
calibration only in 8 cases out of 31 objects, in those cases the offset
between the two scales was less than 24~km/s. We assumed zero offset for
SO587 and SO646. The emission line velocities 
reported in this paper are all relative to the photospheric rest velocity
as measured by the \ion{Li}{i} line at 670.78~nm.

All the measurements of the \OIB\ line used for the scientific analysis in this paper come from the
UVB arm. While the spectral resolution in this arm is lower than in the VIS,
the signal to noise at the wavelength of the \OIB\ line is better.

\section {The sample}
The sample studied in this paper comprises 36 low-mass Class II (i.e., with
evidence of disk from the infrared) stars in Lupus 
(2 of these are brown dwarfs)
and 8 in \sori.
 The spectra are publicly  available in Vizier (J/A+A/561/A2 and  J/A+A/548/A56).
The Lupus sample contains about 50\% of the total Class II population
and is well representative of the properties of the region, while
the number of objects in \sori\ is small and the targets were selected among
low-mass Class II objects in different locations with respect to the
bright star \sori.
 The objects are listed in Table~\ref{table_stars}, which gives the stellar 
parameters and the  accretion luminosity and mass accretion rates from 
\citet[][]{alcala14} for objects in Lupus and  \citet[][]{rigliaco12} for
the \sori\ stars.  As detailed in those papers, the X-Shooter
spectrum of each object is described as the sum of
the photospheric emission, taken to be that of  non-accreting diskless young objects (Class III; see \citet[]{manara13}), and of a slab of hydrogen that represents the emission of the accretion shock.
The spectral type, extinction and slab luminosity (i.e., the accretion luminosity) are therefore self-consistently determined.
In the following, we will refer to stars in Table~\ref{table_stars}   as the Lupus and \sori\ GTO (GTO for simplicity) sample.

\begin{figure}
	\begin{center}
		\includegraphics[width=9cm]{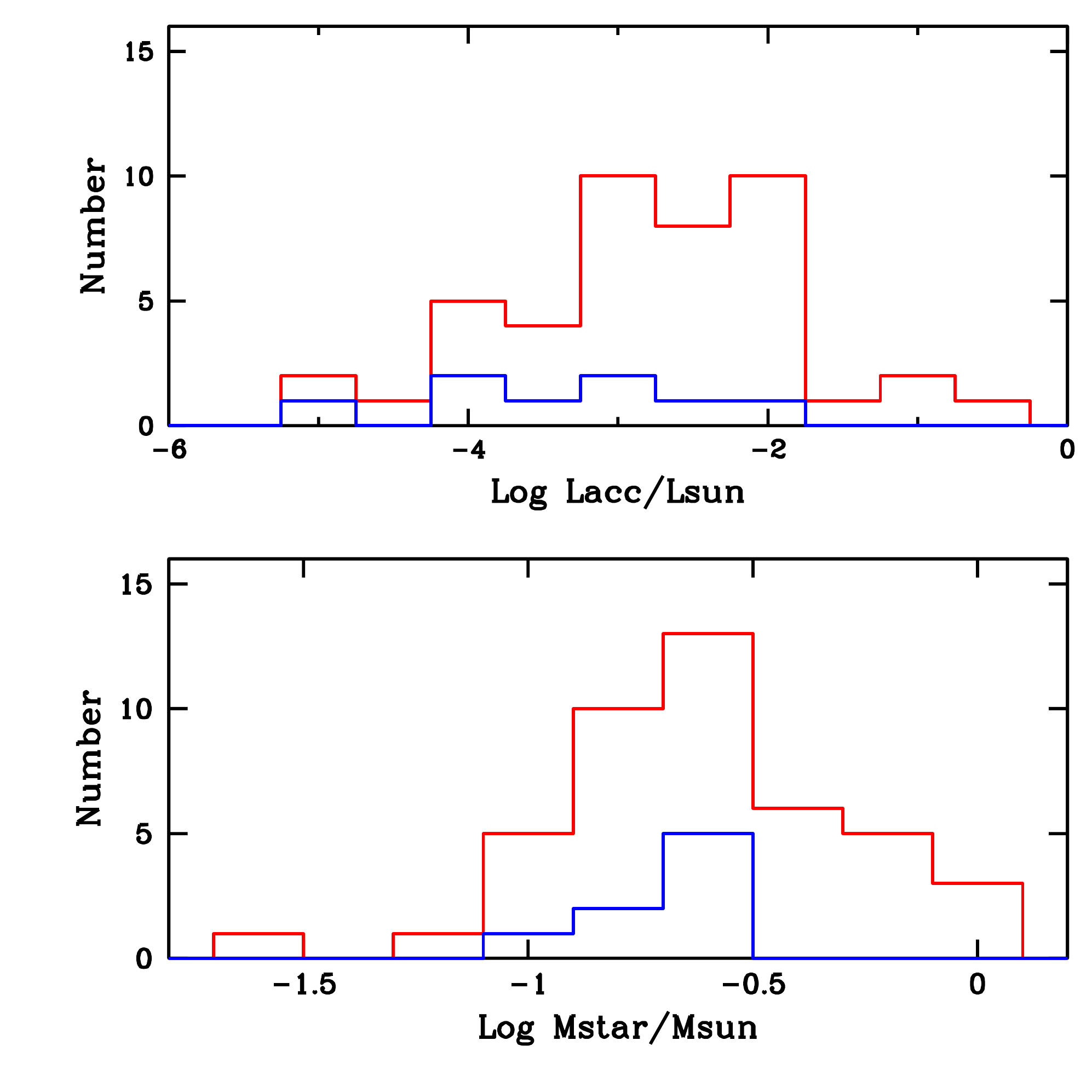}
	\end{center}
	\caption{Accretion luminosity (top) and stellar mass (bottom)
distribution. Red histogram for stars in Lupus, blue for stars in \sori. }
\label{fig_sample}
\end{figure}

\begin{figure}
	\begin{center}
		\includegraphics[width=9cm]{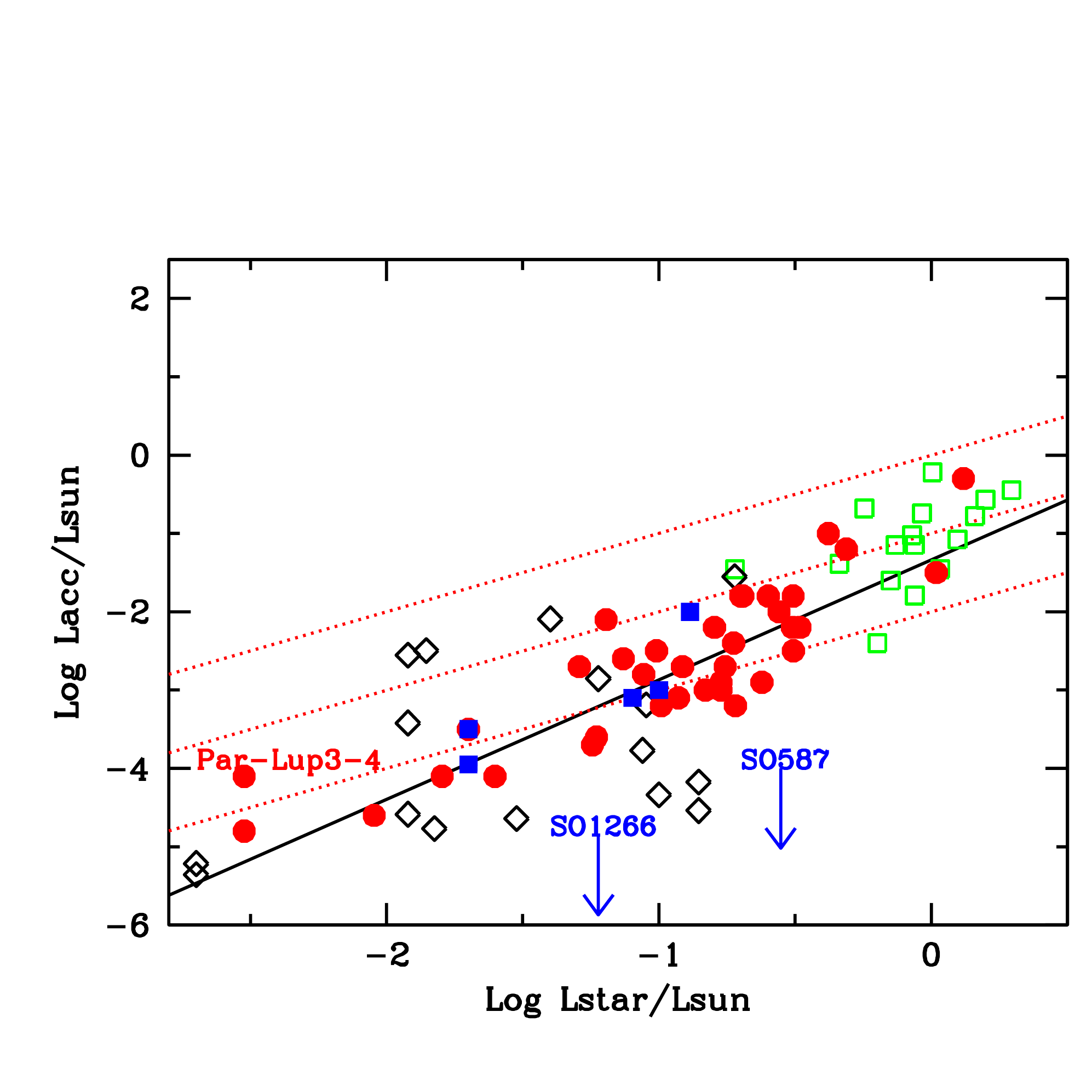}
	\end{center}
	\caption{ \Lacc\ as function of \Lstar\ for objects in Lupus (red dots) 
	         and \sori\ (filled blue squares). Open green squares plot the Taurus TTS of 
		 \citet[][]{gullbring98}; black diamonds the results of 
		 \citet[][]{HH08}. 
%The black cross shows the location of DR Tau. 
The black line shows the best-fitting relation  (GTO objects only)
		 with slope 1.53; dashed lines are the locus \Lacc/\Lstar=1, 0.1, 0.01 
		 from top to bottom, respectively. }
             \label{fig_lacc_lstar}
\end{figure}

The distribution of the stars in mass and accretion luminosity is shown in 
Figure~\ref{fig_sample}. About 70\% of the objects have masses $< 0.3$\,\Msun; the
accretion luminosity ranges from $\sim$$10^{-5}$ to $\sim$$1$\,\Lsun\ and
the corresponding mass accretion rates between $\sim$$10^{-12}$ and
$\sim$$10^{-7}$ \Myr, with most of the objects having \Macc$< 2\times 10^{-9}$ \Myr. 
%In the GTO sample, there are correlations between stellar properties;
%namely, since  stars have similar ages, more massive stars are  more luminous.
In the GTO sample, the accretion luminosity correlates with the stellar luminosity
and the mass accretion rate with the mass of the star, with dispersions of
about one order of magnitude, which is much smaller than in other samples,
as discussed in \citet[][]{alcala14}. 
The correlation between \Lacc\ and \Lstar\ for the GTO sample
%is shown in Figure \ref{fig_lacc_lstar}; in the GTO sample, it 
is significantly steeper than linear, with slope $1.53\pm 0.18$, computed using 
ASURV \citep[parametric EM algorithm;][]{feigelson85} (Figure~\ref{fig_lacc_lstar}).
We have not included Par-Lup3-4, for reasons discussed in \S~6.2.
This relation extends  also to the more massive  TTS analyzed by 
\citet[][]{gullbring98}, \citet[][]{ingleby13} and to the low-mass ($\simless 0.15$ \Lsun) objects studied 
by \citet[][]{HH08, HH14}, which, however, have a large dispersion 
at low \Lstar. 

%The relation between \Macc\ and \Lacc\ is very tight (see Appendix),
%with a very small dispersion introduced by the factor \Mstar/\Rstar.
Given the tight correlation between  \Lacc\ and \Macc, in the following
we will use \Lacc, which is derived directly from the observations, as a proxy for  \Macc.

%\begin{figure*}
%	\begin{center}
%		\label{table_stars}
%		\includegraphics[width=20cm]{table_stars.pdf}
%	\end{center}
%	\caption{Sample Properties }
%\end{figure*}

\section {Results}

%\begin{figure*}
%	\begin{center}
%		\label{table_lines}
%		\includegraphics[width=20cm]{table_lines.pdf}
%	\end{center}
%	\caption{Observed forbidden lines }
%\end{figure*}

%table 2
%
   \begin{table*}
   \centering
      \caption[]{Observed Forbidden Lines}
         \label{table_lines}
    \begin{tabular}{c c c c c c c c c c c c c }
        \hline

 Element&  $\chi_{ion}$ & $\lambda_{vacuum}$& $\lambda_{air}$& Lower& Upper & $g_1$& $g_2$& T$_{ex}$ & A$_{21}$&n$_{cr}$& \multicolumn {2}{c} {Number of detections}\\
        &  (eV)&   (nm)& (nm)&     level& level&  &     & (K)&  (s$^{-1}$)& (cm$^{-3}$)& Lupus& $\sigma$~Ori\\
                \hline
\ion{O}{i}&     0&  557.89& 557.7339& $^1D_2$& $^1S_0$& 5& 1& 48619&  1.26 &1.0e+8 &    28&    4\\
\ion{O}{i}&     "&  630.20& 630.0304& $^3P_2$& $^1D_2$& 5& 5& 22830&  5.6e-3 &1.8e+6 &   31&    7\\
\smallskip
\ion{O}{i}&     "&  636.55& 636.3776& $^3P_1$& $^1D_2$& 3& 5& 22830&  1.8e-3 &1.8e+6 &  25&    5\\
\ion{O}{ii}&  13.618&  372.98& 372.8815& $^4S^0_{3/2}$& $^2D^0_{5/2}$& 4 & 4& 38574& 2.0e-5& 3.3e+3& 3& 4 \\
\ion{O}{ii}&  "&  372.71& 372.6032& $^4S^0_{3/2}$& $^2D^0_{3/2}$& 4 & 6& 38603& 1.8e-4& 3.8e+3&  9& 3 \\
%\ion{O}{ii}&  "&  372.71+372.98& 372.73+372.92& -&-&- & -& -& -& -&  9& 5 \\
\ion{O}{ii}&  "&  732.22& 731.89& $2D^0_{5/2}$& $^2P^0_{3/2}$& 4& 6& 58224&  9.9e-2& 1.3e+7& 2&    1\\
\medskip
\ion{O}{ii}&  "&  733.17& 732.97& $2D^0_{3/2}$& $^2P^0_{1/2}$& 6& 2& 58121&  8.7e-2& 2.2e+6& 2&    1\\
%\smallskip
%\ion{O}{ii}&  "&  733.28& 733.07&$2D^0_{3/2}$& $^2P^0_{3/2}$& 6& 4& 58227&  3.8e-2& 1.3e+7& 2&    1\\
%\ion{O}{III}&  35.117& 496.03& 495.89& $^3P_1$& $^1D_2$& 3& 5& 29168& 6.2e-3& 1.4e+6&    0&    2\\
%\ion{O}{III}&  "& 500.82& 500.68& $^3P_2$& $^1D_2$& 5& 5& 29168& 1.8e-2& 1.4e+6&    0&    1\\
%\ion{O}{III}&  "& 436.44& 436.32&$^1D_2$& $^1S_0$& 1& 5& 62135& 1.7& 3.2e+7&    0&    0\\
\ion{S}{ii}&   10.360& 407.75& 407.6349&$^4S^0_{3/2}$& $^2P^0_{1/2}$& 4 & 2 & 35430& 7.7e-2 & 1.9e+6 &   8& 2\\
\ion{S}{ii}&   "& 406.98& 406.8600&$^4S^0_{3/2}$& $^2P^0_{3/2}$& 4 & 2 & 35354& 1.9e-1 & 2.6e+6 &   21& 5\\
\ion{S}{ii}&   "& 671.83& 671.644& $^4S^0_{3/2}$& $^2D^0_{5/2}$& 4& 6& 21416&  2.0e-4 & 1.7e+3 &  5&  3\\
\medskip
\ion{S}{ii}&   "& 673.27& 673.081& $^4S^0_{3/2}$& $^2D^0_{3/2}$& 4 & 4 & 21370& 6.8e-4 & 1.6e+4 &   11&  3\\
%\ion{S}{ii}&   "& 1028.95& 1028.67& $^2D^0_{3/2}$&$^2P^0_{3/2}$& 4 & 4 & 35353&  1.1e-1 & 2.6e+6 &   1& 0\\
%\ion{S}{ii}&   "& 1032.33& 1032.05& $^2D^0_{5/2}$&$^2P^0_{3/2}$& 6 & 4 & 35353&  1.6e-1 &1.9e+6 &   1& 0\\
%\ion{S}{ii}&   "& 1033.92& 1033.64& $^2D^0_{3/2}$&$^2P^0_{1/2}$& 4 & 2 & 35430& 1.4e-1 & 1.9e+6 &   1& 0\\
%\medskip
%\ion{S}{ii}&   "& 1037.33& 1037.05& $^2D^0_{5/2}$&$^2P^0_{1/2}$& 6 & 2 & 35430& 6.8e-2 & 2.6e+6 &   1& 0\\
%\ion{N}{i}&   0&   346.749& 346.6497& $^4S^0_{3/2}$& $^2P^0_{3/2}$& 4& 4& 41495& 6.5e-3 & 3.1e+6 &- &  -\\
%\ion{N}{i}&   "&   346.754& 346.6543& $^4S^0_{3/2}$& $^2P^0_{1/2}$& 4& 2& 41494& 2.6e-3 & 7.1e+6 &- & -\\
\ion{N}{i}& 0  &   519.93& 519.7902& $^4S^0_{3/2}$& $^2D^0_{3/2}$& 4& 4& 27673& 2.0e-5 & 2.2e+3 &11 &  1\\
\smallskip
\ion{N}{i}&   "&   520.17& 520.0257& $^4S^0_{3/2}$& $^2D^0_{5/2}$& 4& 6& 27660& 7.6e-6 &1.2e+3 & 1 &  1\\
%\ion{N}{i}&   "& -& 1039.77& $^2D^0_{5/2}$ & $^2P^0_{3/2}$& 6& 4& 41495& 6.1e-2& 3.1e+6& ??& ??\\ 
%\ion{N}{i}&   "& -& 1039.81& $^2D^0_{5/2}$ & $^2P^0_{1/2}$& 6& 4& 41494& 3.45e-2& 7.1e+6& ??& ??\\ 
%\ion{N}{i}&   "& -& 1040.72& $^2D^0_{3/2}$ & $^2P^0_{3/2}$& 4& 4& 41495& 2.7e-2& 3.1e+6& ??& ??\\ 
%\ion{N}{i}&   "& -& 1040.76& $^2D^0_{3/2}$ & $^2P^0_{1/2}$& 6& 4& 41494& 5.3e-2& 7.1e+6& ??& ??\\ 
%\ion{N}{ii}& 14.534& 575.62& 575.46& $^1D_2$& $^1S_0$& 5& 1& 47030& 1.1& 1.1e+7& 1& 1\\ 
\ion{N}{ii}& 14.534& 654.99& 654.805& $^3P_1$& $^1D_2$& 3& 5& 22037& 9.8e-4& 8.5e+4& 1& 2\\ 
\ion{N}{ii}& "& 658.53& 658.345& $^3P_2$& $^1D_2$& 5& 5& 22037& 2.9e-3& 8.5e+4& 13& 2\\ 
%\ion{C}{i} & 0 &985.30& 985.03& $^3P_2$& $^1D_2$& 5& 5& 14664& 2.2e-4& ?&  1&   0\\
\hline
\medskip
%\ion{C}{i} & " &982.68& 982.41& $^3P_1$& $^1D_2$& 3& 5& 14664& 7.3e-5& ?&  1&   1\\
%%\ion{Ca}{ii}& 6.113& --& 729.15& $2S_{1/2}$& $2D_{5/2}$& 2& 6& 19726& 1,3 & ?? & 1& 0\\
%%\ion{Ca}{ii}& " & --& 732.39& $2S_{1/2}$& $2D_{3/2}$& 2& 4& 19639& 1.3 & ?? & 1& 0\\

        \end{tabular}

   \end{table*}

%\begin{figure*}
%	\begin{center}
%		\label{table_components}
%		\includegraphics[width=20cm]{table_components.pdf}
%	\end{center}
%	\caption{Detection Statistics of LVC (c), blushifted HVC (b) and redshifted HVC (r) for the lines analysed in this paper.}
%\end{figure*}

%table 3
%
   \begin{table*}
   \centering
      \caption[]{Kinematical Components: "L": LVC; "Hb": blue-shifted HVC; "Hr": red-shifted HVC.}
         \label{table_components}
    \begin{tabular}{l c  c   c c c c}
        \hline
\medskip
Name&  [\ion{O}{i}]& [\ion{O}{i}]& [\ion{S}{ii}]& [\ion{S}{ii}]& [\ion{O}{ii}] &  [\ion{N}{ii}] \\
& 557.73& 630.03& 406.86& 673.08& 372.60 & 658.34 \\
\hline
       Sz66& L& L& L& L& L&  L\\
       AKC2006-1& -& -& -& -& -&  -\\
        Sz69& L& L,Hb,Hr& L,Hb& Hb,Hr& Hr&  Hb,Hr\\
        Sz71& L& L& -& -& -&  -\\
        Sz72& L& L,Hb& Hb&  Hb& Hb&   -\\
        Sz73& L& L,Hb& Hb& Hb& Hb& Hb\\
        Sz74& L&L&  -& -& -&  -\\
        Sz84& L& L& -& -& L&  -\\
       Sz130& L& L,Hb& Hb& -& -&  Hb\\
       Sz88A& L& L& L& -& -&  -\\
       Sz88B& L& L& L& -& -&  -\\
        Sz91& L& L& -& -& -&  -\\
      Lup713& -& -& -& -& -&  -\\
     Lup604s& -& L& -& -& -&  -\\
        Sz97& -& L& -& -& -&  -\\
        Sz99& L& L,Hr& Hr& Hr&  -\\
       Sz100 & L& L,Hb,Hr& L,Hb& L,Hb& Hb&  Hb\\
       Sz103 &L&  L& L& L& -&  Hb\\
       Sz104 & L& L& -& -& -&  -\\
      Lup706 & -& -& -& -& -&  -\\
       Sz106 & L& L& L& L& L&  L\\
  Par-Lup3-3 & -& L& -& -& -&  -\\
  Par-Lup3-4 & L& L& L& L& L&  L\\
       Sz110 & L& L& L& -& -&  -\\
       Sz111&  L& L& -& -& -&  -\\
       Sz112 & L& L& L& -& -&  -\\
       Sz113 & L& L,Hb& Hb& Hb& - & Hb\\
     J160859 & L& L& L& -& -&  -\\
    c2dJ1609 & L& L& L& -& -&  L\\
       Sz114 & L& L,Hb& L& -& -&  -\\
       Sz115 & -& - & -& -& -&  -\\
     Lup818s & L&L& -& -& -&  -\\
      Sz123A& L& L,Hb& L,Hb& -& -&  Hb\\
      Sz123B&L&  L,Hb& L,Hb& -& -&  Hb\\
\smallskip
  SST-Lup3-1 & -& -& -& -& -&  -\\
       SO397 & -& -& -& -& -&  -\\
       SO490 & L& L& -& -& -&  -\\
       SO500 & -& L,Hb& -& -& -&  -\\
       SO587 & -& L& L& L& L&  L\\
       SO646 & -& L& L& -& -&  -\\
       SO848 & L& L,Hb& Hb& L,Hb& L&  L,Hb\\
      SO1260 & L& L& L& -& -&  -\\
      SO1266 & L& L& Hb& L& Hb&  -\\
\end{tabular}
\end{table*}

We have searched the spectra of our targets for evidence of emission in the
forbidden lines of \ion{O}{i}, \ion{O}{ii}, \ion{O}{iii}, \ion{S}{ii}, \ion{S}{iii}, 
\ion{N}{i}, \ion{N}{ii}, \ion{C}{i}, \ion{Ca}{ii} in the blue, visual and near-IR 
wavelengths. The lines detected are \OIB, the \ion{O}{i} doublet at 630.03,~636.37nm, 
the \ion{O}{ii} doublets at 372.88,~372.60nm and 731.89,~732.97nm, the \ion{S}{ii} doublets
at 407.63,~406.86nm and 671.64,~673.08nm, the \ion{N}{ii} doublet at 654.80,~658.34nm.
The [\ion{N}{i}]\,519.79nm is detected in many objects, but it is contaminated
by Fe lines that cannot be reliably deconvolved at the resolution of our spectra, 
so that its intensity is very uncertain. In few cases, we detect also the 
[\ion{N}{i}]\,346.649,~346.654 doublet, but in this spectral region the
signal-to-noise ratio achieved is  generally very low; 
there are no detections of the \ion{N}{i} quadruplet around 1 micron.
In Par-Lup3-4, which has the richest emission line spectrum,
we detect also  the [\ion{Ca}{ii}]\,729.15nm, [\ion{C}{i}]\,982.40,~985.30nm lines
and the  \ion{S}{ii} quadruplet at $\sim$1030nm.
In addition, there are some lines of Fe in different ionization states
\citep[][]{giannini13}. 
%that will be the subject of a forthcoming paper (Giannini et al. in prep).

Table~\ref{table_lines} lists the spectroscopic parameters of the  lines. 
For each line, it gives the ionization potential of the
ion, the line wavelength in the vacuum and in air, the quantum number and multiplicity of the 
lower and upper state, the excitation temperature of the upper state of
the transition, the value of 
the A$_{21}$ coefficient, the critical density for collisions with electrons and
the number of stars in which the line is detected. 
The emissivity of the lines used in this paper have been kindly provided
to us by Bruce Draine and computed using a
5-level atomic model with collisional rates for electrons or atomic hydrogen
\citep[][]{draine11}.

Some lines are detected in a large fraction of stars, for example the [\ion{O}{i}]\,630.03nm
(detected in 38/44 stars) or the [\ion{S}{ii}]\,406.98nm which is detected in 26/44 stars;
in   6 stars (AKC2006-1, Lup713, Lup706, Sz115, SST-Lup3-1, SO397)
no forbidden line is detected. 
Some lines of ions with high ionization potential, such as the \NII\, are also seen 
in several stars.

There is a large variation in the number of forbidden lines detected in different stars.
%Fig.~\ref{fig_detections}  shows the number of detected lines as function of the stellar 
%luminosity (bottom panel) and of the accretion luminosity (top panel).
None of the low luminosity and low accretion luminosity objects  
show a rich forbidden line spectrum, with the exception of four objects 
(Par-Lup3-4, SO587,  SO848 and SO1266), that will be discussed in more detail in
the following.

Of the stars with no forbidden line detections, 4 (AKC2006-1, Lup713, Lup706, SST-Lup3-1) are among the low luminosity, low \Lacc\ objects, while 2 (Sz115 and SO397) are typical TTS and the lack of lines is somewhat surprising. Note, however, that in all cases the  upper limits to the line luminosities are of the order of the detections, not significantly lower.

%\begin{figure}
%	\begin{center}
%	\includegraphics[width=9cm]{detections}
%	\end{center}
%	\caption{Number of forbidden lines detected as function of the stellar 
%	        luminosity (bottom panel) and the  accretion luminosity  (top panel) 
%		of each star. Red dots refer to stars in Lupus, blue dots to stars 
%		in \sori.}
%\label{fig_detections}
%\end{figure}

\section {Line Profiles}

Forbidden lines from TTS are known to show multiple components. Typically,
one can identify a low-velocity component, broad and roughly symmetric,
with a slightly blue-shifted peak velocity, and high-velocity components,
with peaks shifted to the blue and/or to the red (much less frequently)
by tens of km/s. 

We have examined  the lines \OIA, \OIB, \SIIA, \SIIB, \OII, \OIIB\ and \NII\
%that are detected in a significant number of targets 
and classified them as high-velocity blue-shifted (HVC-blue), 
high-velocity red-shifted (HVC-red) and low-velocity (LVC) components. 

Because the spectral resolution of X-Shooter is relatively low, we have
adopted a conservative criterion to identify the presence of a HVC
component in each line:
i) when a line has a single component shifted by more than $\pm 40$~km/s
this is identified as an HVC; ii) when more than one component is present
the HVC is identified if it is well resolved from the LVC; iii) when a
line shows a broad wing, this wing is identified as an HVC if another line
has an HVC at the same velocity defined by one of the previous criteria.
In Table~\ref{table_components} we report the components identified in each
line for all stars in our sample. In this Table ``L'' stands for LVC, ``Hb''
for HVC-blue, and ``Hr'' for HVC-red.
%Even with our rather drastic criterium,
LVC and HVC components are clearly detected in several objects.
The star Sz83 (RU~Lup) has very complex line profiles, not only for the 
forbidden lines but also for the permitted lines that trace directly the 
accreting matter \citep[][]{alcala14}, and will not be discussed in 
this paper.

Our criteria in  classifying the different components  differ from those adopted by
\citet[][]{hartigan95}. These authors 
define the LVC as the portion of the spectrum between -60 and +60 km/s
and measured the HVC as 
the total flux (integrated over all the profile) 
minus the LVC flux. Note also that the spectral resolution they used is about 12 km/s, 
which allowed them a better definition of the line wings.

\begin{figure}
	\begin{center}
		\includegraphics[width=9cm]{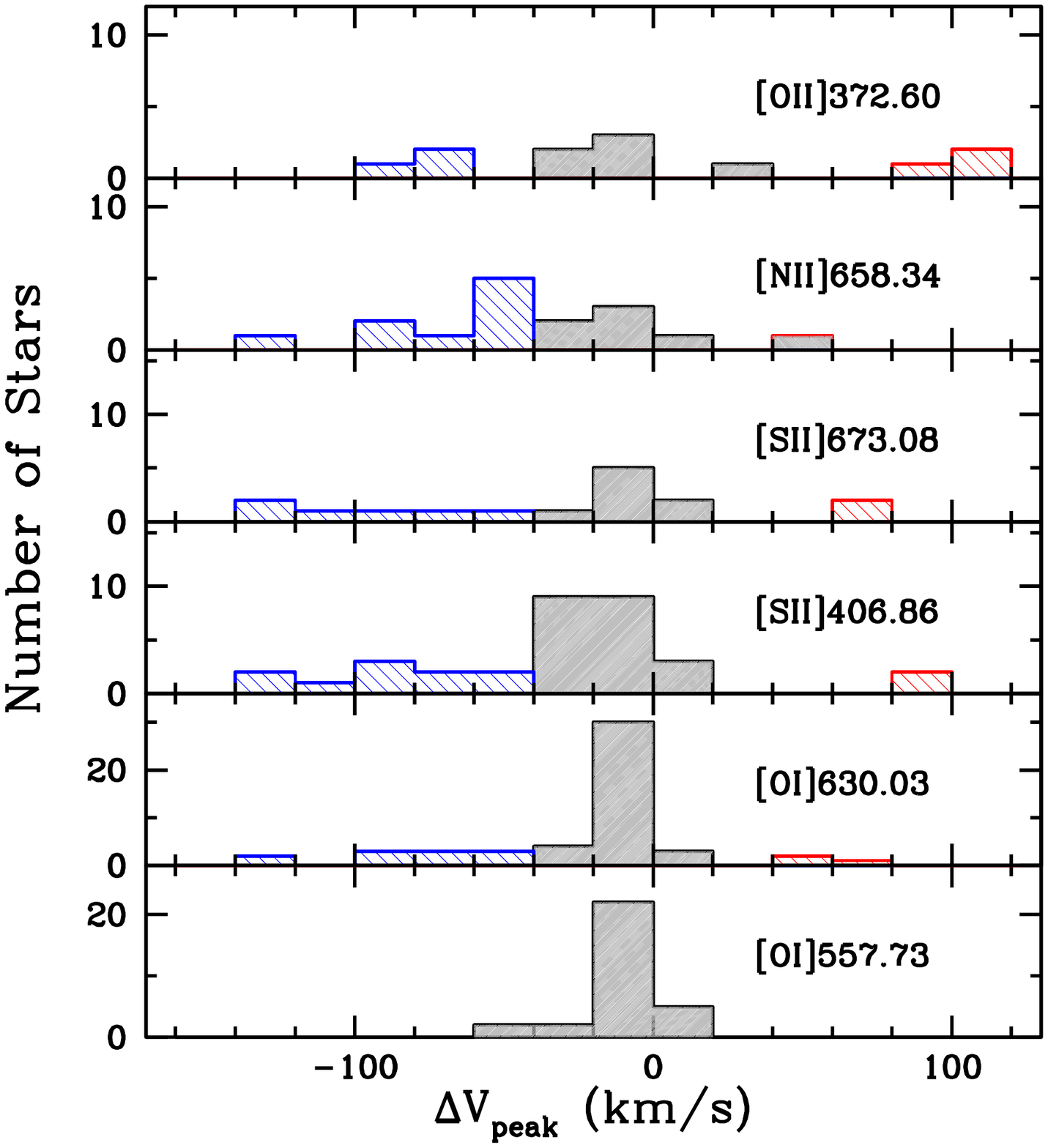}
	\end{center}
	\caption{Distribution of the peak velocity shift with respect to the star (solid for LVC,
hatched  blue 
	        for HVC-blue, red for HVC-red) for different lines, as labelled. }
\label{fig_vpeak}
\end{figure}

We compute
the intensity and the peak velocity of each component by
fitting a gaussian profile to the line. 
%The intrinsic FWHM is derived from the 
%fit by deconvolving a gaussian corresponding to the spectral resolution at the line wavelength. 
Table~\ref{tab_tabellone_1}, \ref{tab_tabellone_2},
\ref{tab_tabellone_3} and \ref{tab_tabellone_4} give for each star the properties
(flux, uncertainty, peak velocity and FWHM) of each component for the set of lines 
defined above.
Upper limits ($3\sigma$) have been computed from the rms of the continuum, assuming 
a line FWHM width of 50 km/s ($3{\rm rms}\times FWHM/\sqrt{\rm(nres)}$, where 
${\rm nrs}$ is the number 
of resolution elements within 50 km/s). The  uncertainty on $V_{peak}$ is $\sim$$10$ km/s.  
Figure~\ref{fig_vpeak} shows the distribution of the peak velocity; almost all LVC have 
a blue-shifted peak, with $V_{peak}$ smaller than $\sim 20$ km/s. Note that the gaussian fit tends to 
overestimate $V_{peak}$ if there is a significant unresolved asymmetric wing.
In general, our values of $V_{peak}$ are consistent, given the resolution of our spectra, 
with the typical value $\sim 5$ km/s of \citet[][]{hartigan95}. The large majority of 
the lines, both LVC and HVC, are broad and well resolved, with FWHM
ranging from the resolution limit to $\simgreat 100$ km/s.

 Examples of the observed line profiles are shown in 
Appendix~\ref{app_profiles}. Fig~\ref{fig_OI_all} shows the observed profiles of the \OIA\ line for all the GTO objects. 
Fig.~\ref{fig_Sz66} -- \ref{fig_SO848} show line profiles for a selection of objects where several lines were detected, together with the results of the gaussian fitting and deconvolution.

\begin{figure}
	\begin{center}
		\includegraphics[width=9cm]{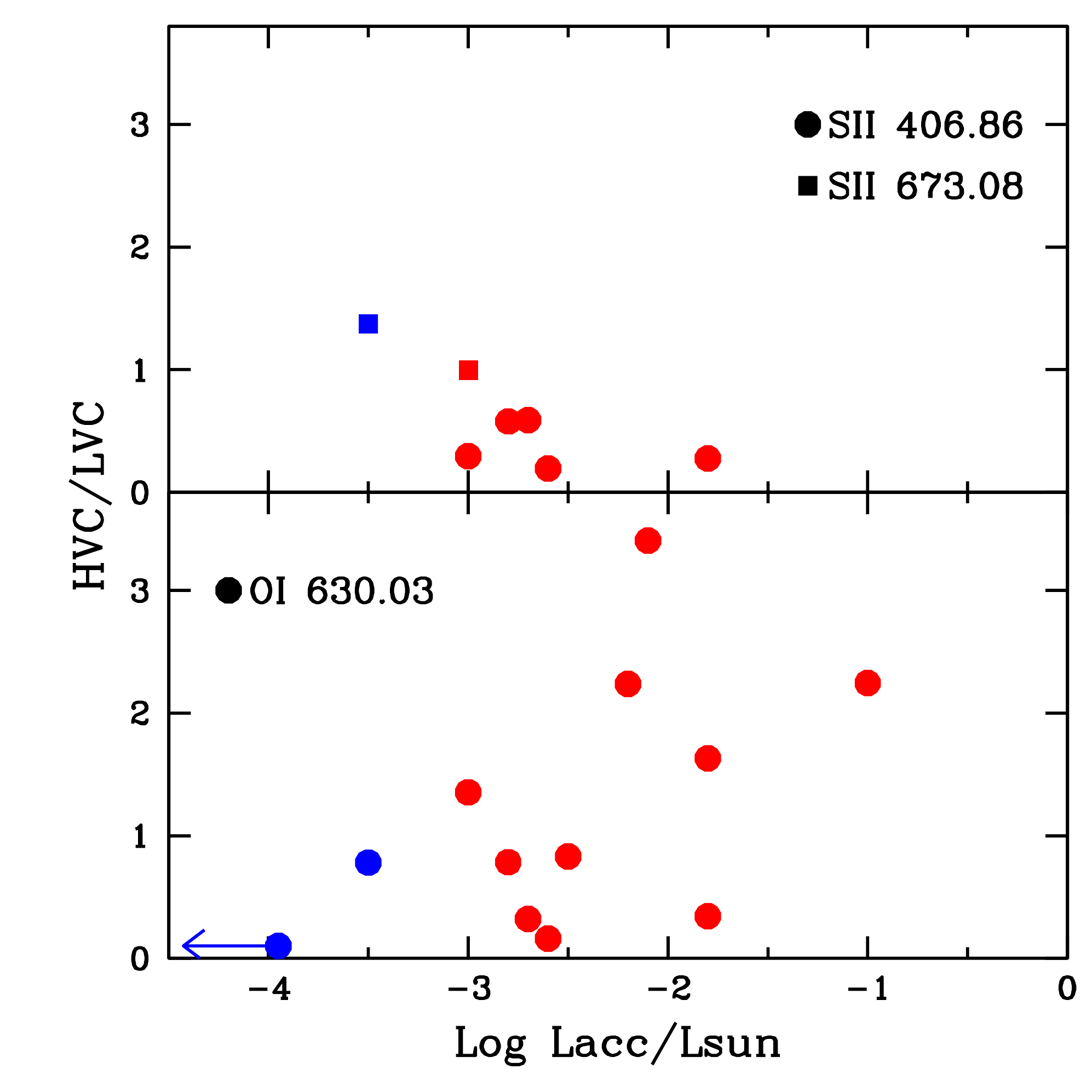}
	\end{center}
	\caption{Ratio of the HVC to the LVC intensity in objects where both 
	        components are simultaneously detected. The bottom panel refers
to the \OIA\ lines; the top panel to the \SIIA| and \SIIB\ lines, as labelled.
Red and blue for Lupus and \sori, as before.}
\label{fig_LVC_HVC_ratio}
\end{figure}

\subsection {Statistics}

For each star, Table~\ref{table_components} characterizes the profile of 
the lines discussed in the following. 
LVC, HVC-blue and/or HVC-red may be present in all the lines we
observe; as also shown by previous studies, 
the HVC-red is very rare.
The   exception  is the \OIB\ that does not show any HVCs
clearly identifiable in our spectra, although in some cases there is a
hint of emission in the line wings that could be resolved as a separate
component in higher sensitivity and resolution spectra.
This is in agreement with the results
of other authors \citep[i.e.][]{hartigan95, rigliaco13}.

When detected, the \OIA\ always has a LVC,  while a HVC is detected in 12/37 of 
cases. In general, HVCs are detected proportionally more often in high
excitation/high ionization lines such as \NII\ and \OII, or in lines with
low critical density, such as the \SIIB\ doublet. 

%The \SIIA\ line has a LVC in 22 stars, 9 objects have a HVC blue-shifted, 2 a
%red-shifted one; of these, 1 star has both blue and redshifted components (Sz66?). 

Of particular interest are stars where both LVC and HVC are detected, since 
in these cases the potential confusion between the LVC emission and that of 
a jet projected on the plane of the sky does not exist. Our sample includes 
12 stars with both LVC and HVCs in the \OIA\ line; of these, 5 show both
components also in the \SIIA\ line and 2 in the \SIIB\ one.
%Note that no  star  with \OIA\ LVC only shows multiple  components in  other lines. 
The intensity ratio  between the two components is shown in Figure~\ref{fig_LVC_HVC_ratio} 
for the \OIA, \SIIA\ and \SIIB\ lines. The ratios range from very small to $\sim 3.5$;  
however, only few objects have a very strong HVC (high value of the ratio), and
in  most cases the HVC has intensity lower than that of the LVC. 

%The \SIIB\ has a LVC in 8 stars, a blue-shifted HVC in 6, a red-shifted HVC in 2, 1 star (Sz69) has both. In 2 stars (SO848, ?) there is both a LVC and a HVC, their ratio is shown in Figure~\ref{fig_LVC_HVC_ratio}, these stars have similar components also in \OIA.
%The \NII\ line has has a LVC in  6 stars, a blue-shifted HVC in 9, a red-shifted HVC in 1. Only SO848 has both a LVC and a HVC, with ratio ??.
%The \OII\ line has a LVC in 6 stars, a blue-shifted HVC in 7; no star has a red-shifted HVC, nor both a LVC and a HVC component.

%SII67: HVCb in 6 (1 in sigma, SO848), HVCr in 2, 1(Sz69) has both. LVC in 8.
%Ratio HVC/LVC in 2 (1 is SO848), both 2-comp in OI63.
%\ion{N}{ii}: HVCb in 9 (1 is SO848), HVCr in 1, LVC in 6. 1 star has HVC/LVC (SO848).
%OII3727: HVCb in 7 (2 in sigma), HVCr in 0, LVC in 6. Ratio HVC/LVC in 0.
%

\subsection {Contamination of the LVC sample by HVC in the plane of the sky}

When only one component with small peak velocity is detected in any given object,
it is possible that, rather than a bona-fide LVC, it is a HVC
misclassified because of projection effects.

This is certainly the case of  Par-Lup3-4, which has a well resolved jet
and an edge-on disk obscuring the central star so that both its stellar
and its accretion luminosity are largely underestimated \citep[][]{bacciotti11, alcala14}.
At the  spatial resolution of our spectra ($\sim 1$ arcsec, due to seeing),  forbidden lines from Par-Lup3-4 have very small $V_{peak}$ 
and we classify them as LVC. Contrary to many other cases, however, 
the three lines \SIIB, \OII\ and \NII\  are  quite
strong in this object, with ratios to the \OIA\ line of 0.3, 0.1 and 0.05, 
respectively.

Two additional objects in Lupus also have a rich LVC spectrum, with all the lines
clearly detected. They are Sz66 and Sz106. The high-excitation lines are
particularly strong, with ratios \OII/\OIA\ and \NII/\OIA\ of 0.02 and 0.05 for
Sz66 and 0.25 and 0.3 for Sz106. For the latter object there is evidence
in the literature for an edge-on geometry \citep[][]{comeron03, alcala14}. 
It is probable that, as in Par-Lup3-4, we are detecting the emission of a jet 
with axis close to the plane of the sky.
% Note that these ratios are likely underestimated, as some of the \OIA\ emission
% may be due to a proper LVC.

The presence of 3 "spurious" LVC objects in Lupus (about 10\%) is consistent
with the expectations for a randomly oriented sample.
Assuming a  velocity of 200 km/s for the HVC \citep[e.g.][]{appenzellerandbertout13},
we expect that about 12\% have projected velocities less than $\sim 40$ km/s 
% (and $\sim 20$\% less than 60 km/s) 
and would therefore be classified as LVC 
in our analysis. 

Given the small number of potentially spurious LVC, we include all the objects 
in the GTO sample in the following analysis (with the already noted exception 
of Sz83).

%In the case of the \OIA\ line, since we detect 12 object with HVC emission, this would give about 1-2 "false LVC" cases; similar number for the \SIIA. In all
%cases, the contamination of the LVC sample would be small. The situation is
%somewhat different for the other lines, where the "false LVC" can account
%for 1/8, 1-2/10, and 1/6 for the \SIIB, \NII\ and \OII\ lines, respectively.
%The possibility that the LVC sample contains some "odd" objects is
%confirmed when we consider that Par-Lup3-4 (a strong jet in the plane of the sky)
%%and SO587 (a likely photoevaporative wind heated by the nearby O9.5 star \sig)
%has strong emission in all the forbidden lines  centered at zero velocity, with no evidence of blue-shifted or red-shifted emission.
%This, however, should not affect our conclusions.

\subsection {Doublets}
All the lines in our sample are doublets, with the exception of the \OIB. 
In some cases (i.e., the \OIA\ and the \NII\ doublets), the ratio of the intensity 
of the two  components depends on atomic parameters only;  in others, it depends also 
on the physical conditions that determine the level population. If, for example, 
electronic collisions dominate, the intensity ratio depends on the electron density and, 
to a  lower degree, on the temperature, as for the \SIIA\ and \SIIB\ case.
The range, however, is not large. The \SIIB\ ratio ranges from $\sim 1.5$ for density much 
lower than the critical density, to $\sim 0.5$ in the high density limits. In the case of 
the \SIIA\ doublet, the [\ion{S}{ii}]\,407.63 is always weaker than the \SIIA, with a ratio of $\sim 0.3$ 
for $n_e \ll n_{cr}$ to $\sim 0.2$ in the high-density limit. Our observed ratios and upper 
limits are always consistent with the predictions. Unfortunately, the large errors, especially 
on the weaker components, prevent us from using these line ratios to constrain the  
density of the emitting regions.

\section {The Low-Velocity Component }

In the following, we will focus our discussion on the region emitting the LVC and on its properties.

\subsection {Line Intensities}

\begin{figure*}
	\begin{center}
		\includegraphics[width=18cm]{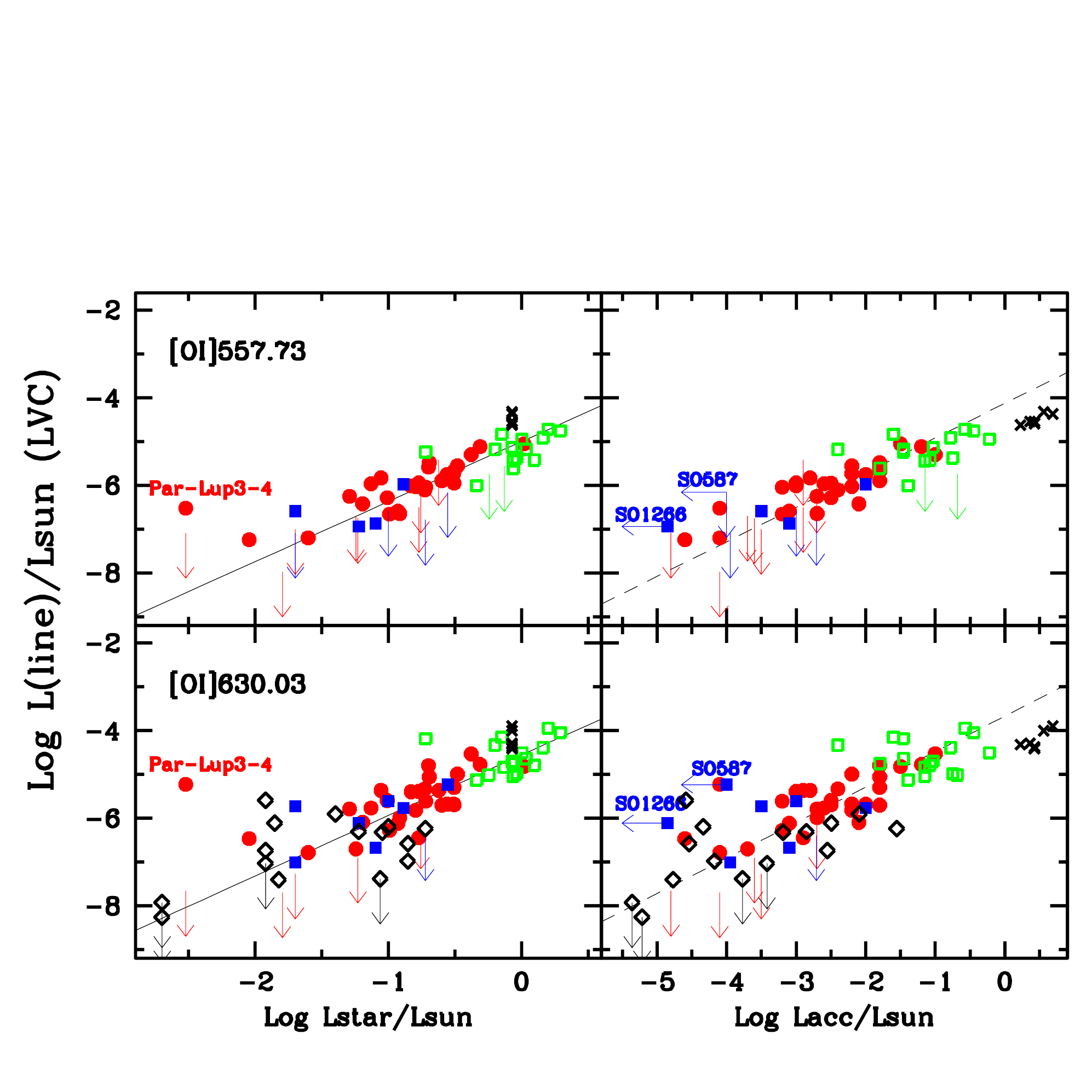}
	\end{center}
	\caption{ Line luminosities for the \OIA\ and \OIB\ are plotted as function of the stellar 
	         luminosity (left Panels) and of the accretion luminosity (right Panels). Red filled dots are 
		 the Lupus objects, blue filled squares are objects in \sig; arrows show 3$\sigma$ upper limits. 
		 The black crosses show the same quantities for the strongly accreting TTS DR Tau at 
		 different epochs \citep[][]{banzatti14}. Black diamonds are values for Taurus 
		 very low mass objects from \citet[][see text]{HH08, HH14}; green open 
		 squares denote Taurus TTS from \citet[][]{gullbring98} and 
\citet [][]{hartigan95}, 
		 as described in the text. 
Few objects discussed in \S~6.2 are labelled in some panels only, to avoid confusion.}
\label{fig_llines_OI}
\end{figure*}

Figure~\ref{fig_llines_OI} shows the luminosity of the LVC of the  lines \OIA, \OIB\ 
as function of \Lstar\ (left panels) and \Lacc\ (right panels).

For the two \ion{O}{i} lines, which are detected in a large fraction of objects,  there is an 
excellent correlation of \Lline\ with both \Lstar\ and \Lacc, with similar slopes for 
the two lines; using ASURV, EM method \citep[][]{feigelson85}, we find:
\begin {equation}
 Lg L(\ion{O}{i}\,630.03)= (1.37\pm 0.18) Lg L_{star} + (-4.56 \pm 0.18)
\end{equation}
\begin {equation}
 Lg L(\ion{O}{i}\,630.03)= (0.81\pm 0.09) Lg L_{acc} + (-3.66 \pm 0.27)
\end{equation}
\begin {equation}
 Lg L(\ion{O}{i}\,557.79)= (1.37\pm 0.16) Lg L_{star} + (-5.00 \pm 0.16)
\end{equation}
\begin {equation}
 Lg L(\ion{O}{i}\,557.79)= (0.79\pm 0.07) Lg L_{acc} + (-4.13 \pm 0.20)
\end{equation}
%The correlation coefficients are all in the range $0.32 - 0.40$.
Upper limits to the line luminosity are taken into account.
We have not included in the correlations the object Par-Lup-3-4, because of its very uncertain
values of \Lstar\ and \Lacc, 
and, in the case of the correlations with \Lacc, the two objects in \sig\ which have only \Lacc\ 
upper limits. If they are included, the correlations will be slightly
flatter.

Figure~\ref{fig_llines_OI} plots also the observed \OIA\ and \OIB\ luminosities for other stars in the literature. 
The black crosses, in particular, are values for DR Tau at different epochs; stellar, accretion and line 
luminosities have been obtained from X-Shooter spectra in the same manner as the objects analyzed in 
this paper \citep[][]{banzatti14}. The green points refer to the sample of TTS in Taurus firstly 
studied by \citet[][]{hartigan95}, but with extinction, \Lstar\ and \Lacc\ as re-measured by 
\citet[][]{gullbring98}. The black diamonds are the low-mass objects studied by \citet[][]{HH08}, 
corrected according to the new values of extinction, \Lstar\ and \Lacc\ given by 
 \citet[][]{HH14}, when available.

The  correlations of eq.(1)--(4) describe quite accurately also the higher luminosity TTS from the literature. 
The slope of the \Lline-\Lacc\ correlation for the \OIA\ line is 
similar within the uncertainties  to that
 of \citet[][]{HH08}. On the other hand, 
it is marginally steeper than  the slope ($0.52 \pm 0.07$) found by \citet[][]{rigliaco13}  using the \OIA\ LVC luminosities of 
\citet[][]{hartigan95} and new determinations of \Lacc\ from the \Ha\ luminosity. 
Inspection of Figure~\ref{fig_llines_OI} shows that  in the TTS sample alone
(green squares)
L(\OIA)  has a  flat dependence on \Lacc, and it is possible that the 
discrepancy comes from the different range of \Lacc\ covered by the
two samples. 
The possibility of a change of the slope at high \Lacc\ is also suggested by the
location of DR Tau (black crosses); however, there
are differences in the quality of the data and in the
methods used to determine stellar and accretion properties for the various
samples that prevent any conclusion at this stage.

\begin{figure*}
	\begin{center}
		\includegraphics[width=18cm]{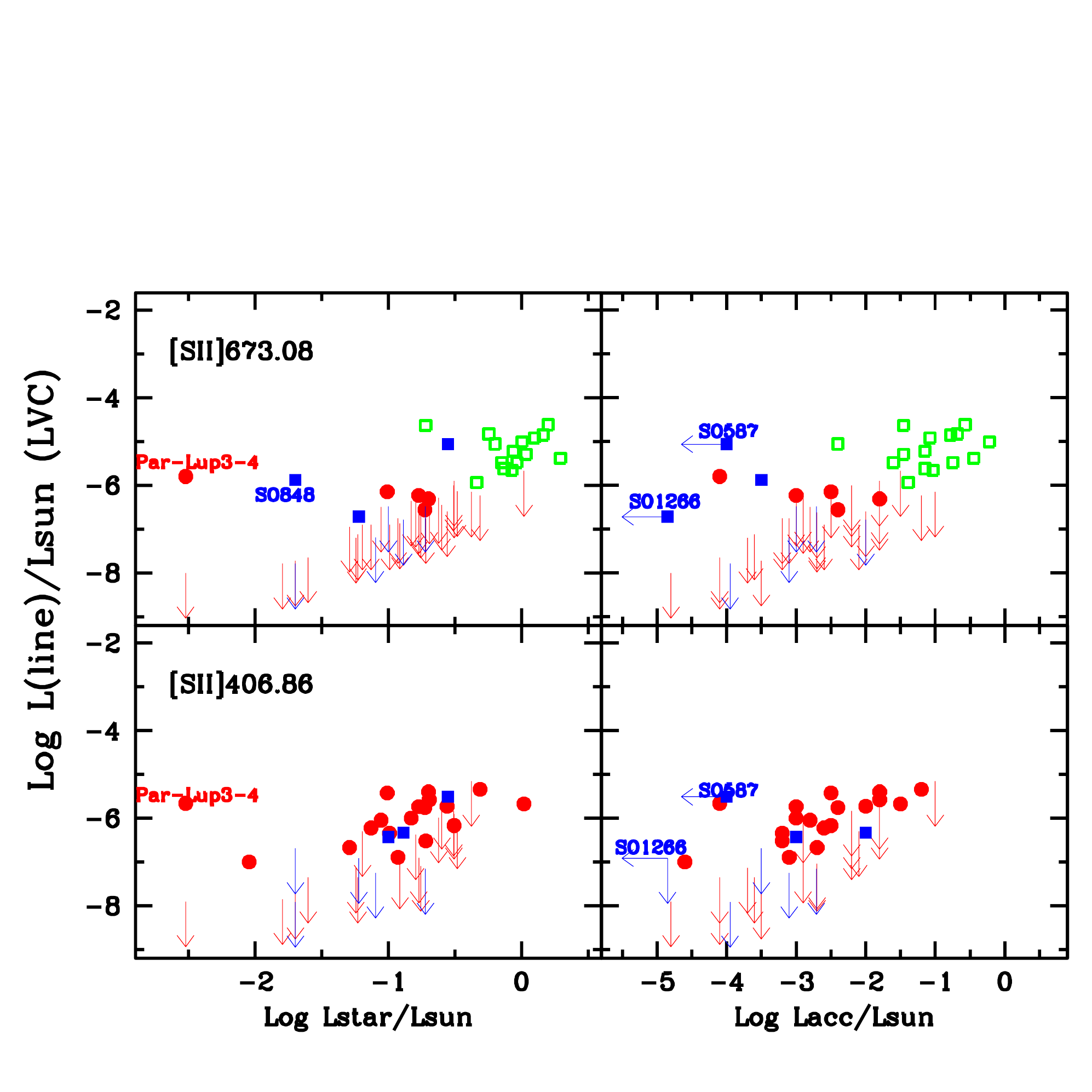}
	\end{center}
	\caption{Line luminosities for the \SIIA\ and \SIIB\ lines are plotted as function of the stellar 
	        luminosity (left Panels) and of the accretion luminosity (right Panels). Red dots are the 
		Lupus objects, blue filled squares are objects in \sig; arrows show 3$\sigma$ upper limits. 
		Green open squares show the position of Taurus TTS from \citet[][]{gullbring98} and \citet[][]{hartigan95}, 
		as described in the text.}
\label{fig_llines_SII}
\end{figure*}

%It is significantly flatter than many other
%results in  
%the literature, where a linear relation is claimed (ref to Fang et al. (??); Herczeg \& Hillenbrand (2008); 
%others?). We note that we are considering here only the LVC of the lines, not the total luminosity; however, 
%we think that  the discrepancy is more likely due to differences in the analysis and in the way the stellar 
%properties are determined.

As mentioned in \S~3, in the GTO objects there is a rather tight correlation between \Lstar\ and \Lacc.
%\, with slope $1.4\pm 0.1$  and standard deviation 0.46 (Alcal\a' et al. 2013; see Figure~\ref{fig_lacc_lstar}), so that we cannot know 
This makes it very difficult to identify which of the system properties 
controls the emission of the forbidden lines,
either the stellar   or the accretion ones.
It is interesting that the relation to \Lacc\ is almost linear,
as for all the permitted lines that are tracers of the 
accretion process \citep[][]{alcala14}. 

Figure~\ref{fig_llines_SII} plots the luminosity of the \SIIA\ and \SIIB\ lines 
as function of \Lstar\ and \Lacc\, respectively.
As for the \ion{O}{i} lines, there is a good correlation with both, with similar slopes,
but with a larger number of upper limits, especially in the \SIIB\ lines.

\subsection {Individual Objects}

There are few objects that deviate significantly from the general trends, 
and we have identified them in the figures.
In addition to Par-Lup3-4, that we have already discussed, there are 2 stars
in \sori\ with very strong forbidden lines.
One is  SO587 which has an upper limit to \Lacc\ $\sim 10^{-4}$ \Lsun,
and a LVC spectrum with especially strong  high-excitation lines, i.e., 
\SIIB/\OIA $\sim 1.5$, \NII/\OIA $\sim 1.7$ and \NII/\OIA $\sim 3$. 
Also the \SIIA\ line is very strong for the observed \Lacc\ 
 (but not for the \Lstar\ of the object).
No \OIB\ is detected. This object has been studied by \citet[][]{rigliaco09}, 
who interpreted it as a photoevaporative wind heated by the nearby O9.5 star
\sori. The physical conditions in the outer wind should therefore be different 
from those in similar objects where no hot nearby star is present, explaining
the strong emission in the high-excitation lines.

Another object in \sori\ has a rich spectrum of forbidden lines, namely SO848. 
In this case both a LVC and a HVC-blue are detected and the LVC luminosities are 
as expected, given the  \Lstar\ and \Lacc\ of the object. The only exception 
is the \SIIB\ line, which is unusually strong.
It is possible that the presence of the hot star \sori\ affects at some level also
the emission of SO848, which has a projected distance from \sori\ of 0.45 pc (to be
compared with 0.35 for SO587). A deeper analysis 
of this aspect, however, requires line profiles of higher spectral resolution, 
that will allow a more accurate component separation and also the identification 
of the spectral profiles typical of photoevaporated winds heated by an external 
source \citep[see, e.g.,][]{rigliaco09}.

SO1266 is a low luminosity object in \sori, with only an upper limit to \Lacc\ from  
continuum excess measurement \citep[][]{rigliaco12}; it has relatively strong emission 
lines of H and \ion{Ca}{ii}, so that the mass accretion rate derived from the relation 
between line luminosity and \Lacc\ would be about ten times higher than the continuum 
upper limit. \citet[][]{rigliaco12} argue that chromospheric emission contributes about
80\% of  the  line emission \citep[see also][]{manara13}.
We detected LVC of \OIA, \OIB\ and \SIIB\  which are stronger than in other objects with 
similarly low \Lacc, but not when compared to other objects with similar \Lstar. This is 
an interesting object, that needs follow-up studies.

\subsection {Physical conditions in the LVC emitting region}

\begin{figure}
	\begin{center}
		\includegraphics[width=9cm]{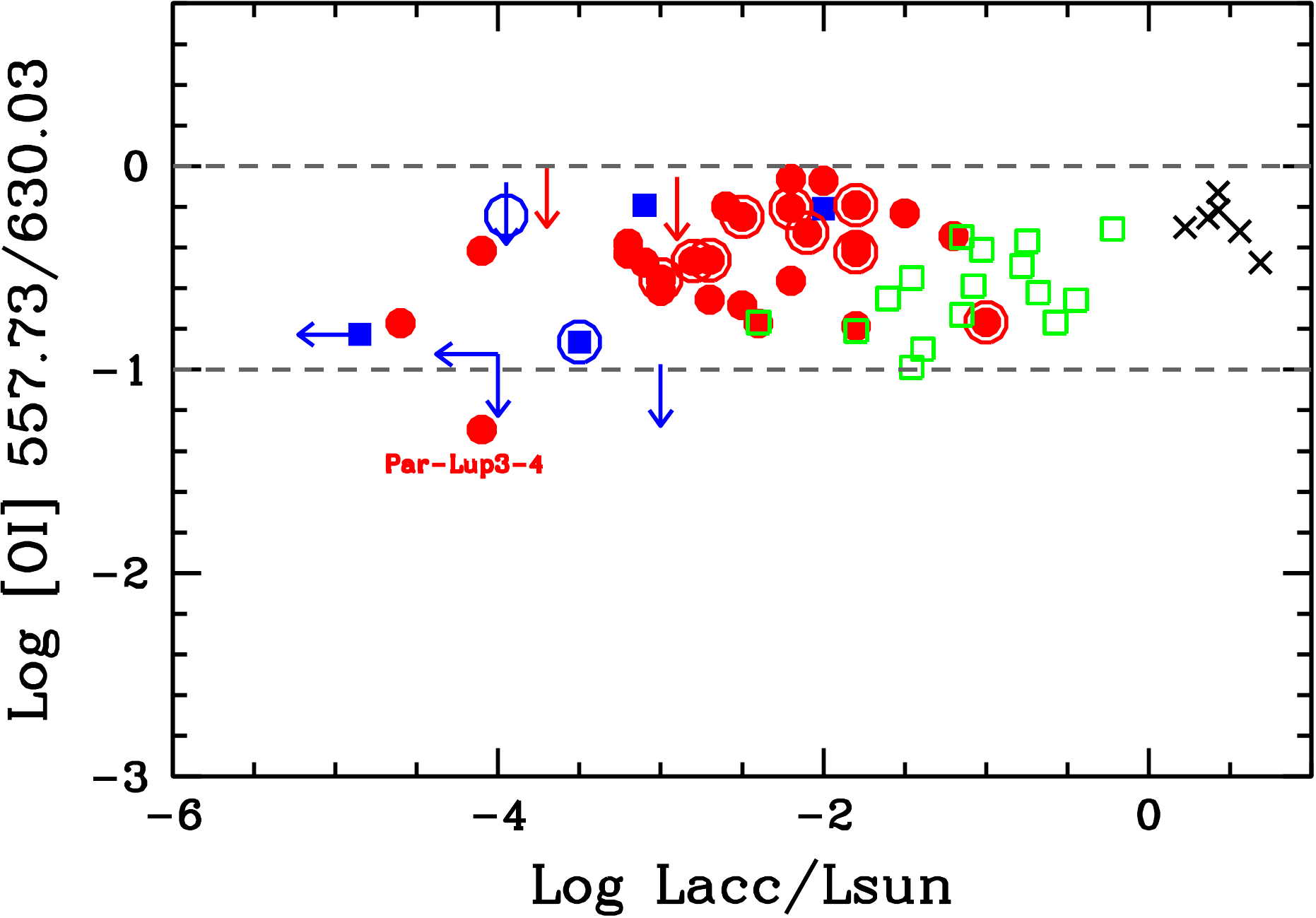}
		\includegraphics[width=9cm]{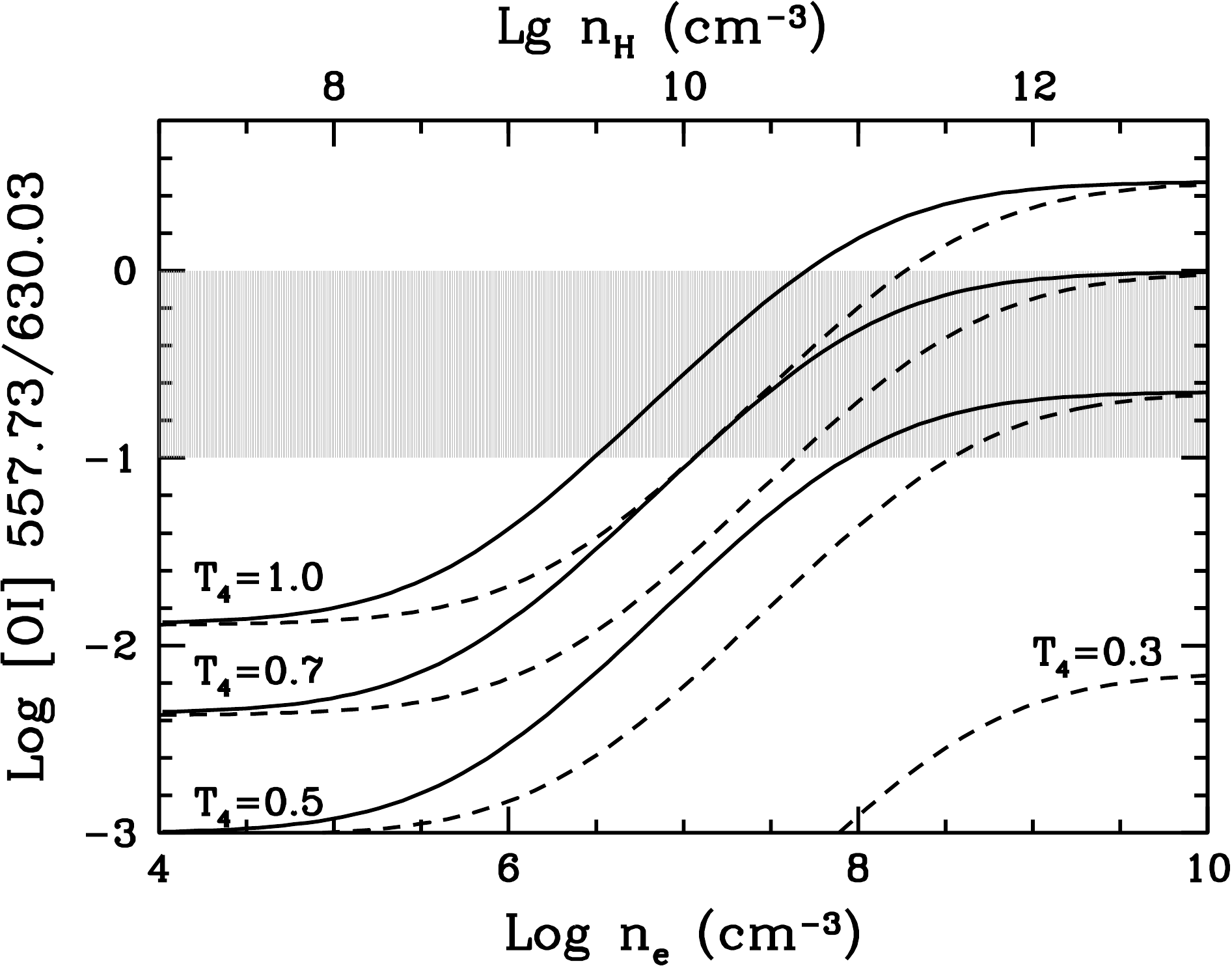}
	\end{center}
	\caption{Top Panel: Ratio \OIB/\OIA\ as function of \Lacc\ for stars in 
	        Lupus (red dots) and \sori (filled blue squares). Arrows indicate 3$\sigma$ upper limits. 
		Objects where both a LVC and a HVC are detected are marked by large open circle. 
		Black crosses show the values for the set of DR Tau spectra obtained at different 
		epochs, green open squares are stars in Taurus (see text). 
		Bottom Panel: Model predictions for collisions with electron (solid curves) and 
		with hydrogen atoms (dashed curves). Different curves refer to different temperatures, 
		as labelled. The $n_e$ scale is shown in the bottom axis, that for $n_H$ on the top 
		axis. The grey area shows the location of the observed values,
		marked by the two dashed lines in the top panel.}
\label{fig_OIOI}
\end{figure}

Line ratios provide very interesting information  on the physical conditions 
of the emitting region. Figure~\ref{fig_OIOI} top Panel plots the values of the 
ratio of the two [\ion{O}{i}] lines at 557.73 and 630.03  as function of \Lacc\
for the GTO sample and for the
more luminous TTS from the literature described in the previous section.
The ratio is impressively constant over the whole sample, with values ranging from
$\sim 0.1$ to $\sim 1$. There is  no correlation with either the \Lstar\
or the intensity of the \OIA\ line. This is very similar to what has
been found by \citet[][]{hartigan95} for higher luminosity TTS, also shown
in the figure.

The observed values can be compared in Figure~\ref{fig_OIOI}, bottom panel,
to the predictions of homogeneous and isothermal models where the 
excitation is due to electron collisions (solid curves, bottom axis) or
to collisions with neutral hydrogen (dashed curves, top axis).
Different curves correspond to different temperatures, as labelled.
Note that the model predictions for the intensity of the \OIB\ when collisional excitation by H dominates  are 
uncertain, as the de-excitation cross section   
of the level $^1S_0$ is not known.
The calculations in Figure~\ref{fig_OIOI} have been performed assuming the
same rate of the $1D_2$ level \citep[see also the discussion in][]{gorti11}. 
The grey area shows the location of the observed ratios; one can see that,
unless the temperature is significantly higher than $10^4$,  they
are consistent with a rather dense gas, with $n_e \gg  n_{cr}$(\OIA) $\simgreat 10^7$ \cmc, 
or, alternatively, $n_H \gg 10^{10}$ \cmc, if collisions with H dominate the
level excitation. 
We have limited our models to $T\le 10^4$ K, but we note that
at the high
temperature
$T=30000$ K, $n_e=10^5$ \cmc, the ratio \OIA/\OIB\  is only  $\sim 0.1$ 
barely consistent with the low end of the observed interval, and would decrease
further for lower values of $n_e$.
 Temperatures below $\sim 5000$K cannot reproduce the observed ratios, no matter the value of the density. In the following, we will assume that the region
emitting the [OI] lines  have density 
$n_e \ge 10^7$ \cmc\ (or $n_H \ge 10^{10}$ \cmc) and temperature $5000 \le T \le 10000$K.

%\begin{figure}
%	\begin{center}
%		\includegraphics[width=9cm]{OIOI_density}
%	\end{center}
%	\caption{ Ratio \OIB/\OIA\  for collisions with electron (solid curves) and with hydrogen atoms (dashed curves). Different curves refer to different temperatures, as labelled. The $n_e$ scale is shown in the bottom axis, that for
%$n_H$ on the top axis.
%The grey area shows the location of the observed values.
% }
%\label{fig_OIOI_dens}
%\end{figure}

\begin{figure}
	\begin{center}
		\includegraphics[width=9cm]{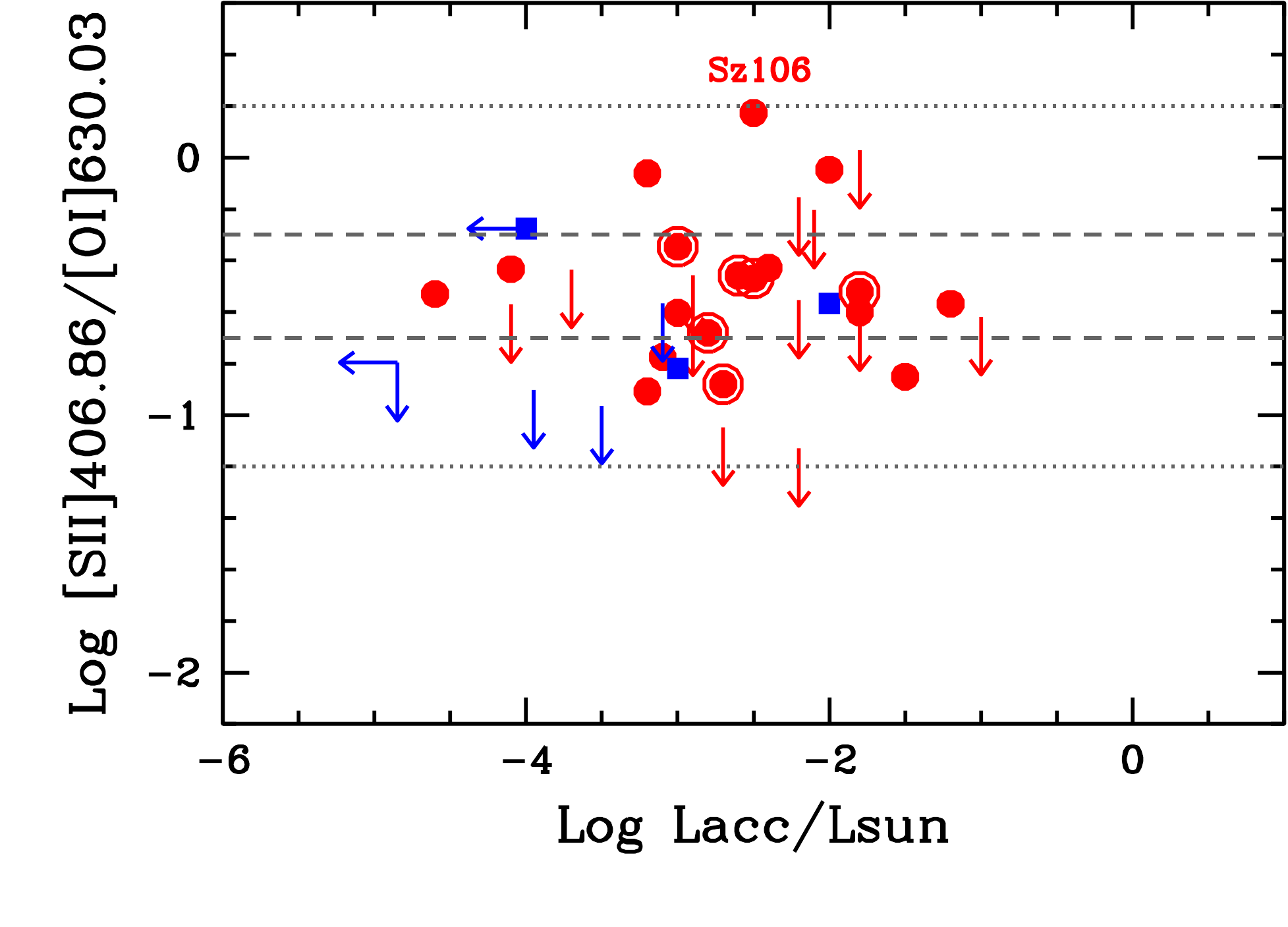}
		\includegraphics[width=9cm]{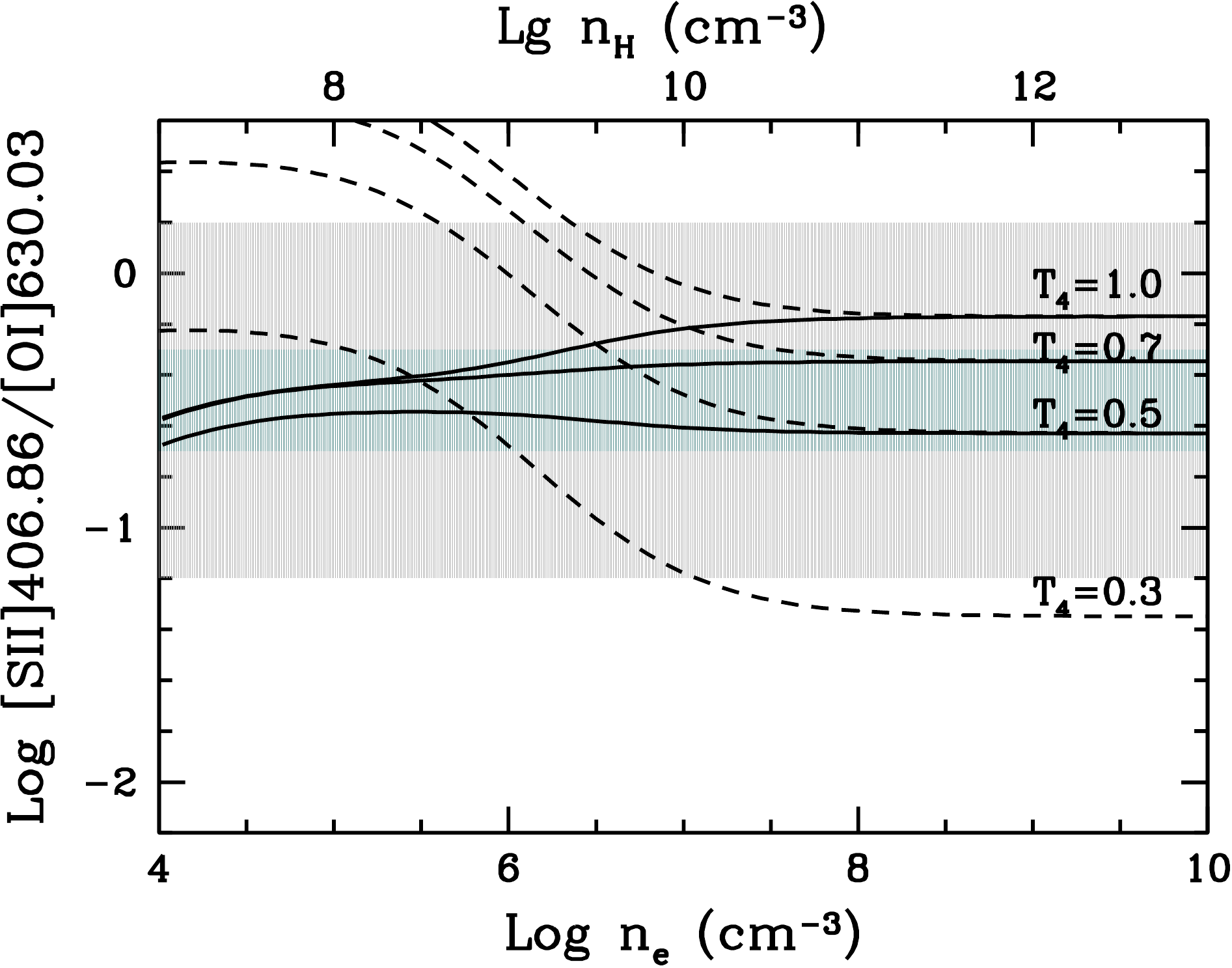}
	\end{center}
	\caption{Same as Figure~\ref{fig_OIOI}  for the ratio  [\ion{S}{ii}]\,406.86/[\ion{O}{i}]\,630.03. In the bottom Panel, 
	        the dark-grey area shows the location of 65\% of the observed values (dashed lines in the top panel). Three points 
		lie in the upper ligh-grey area, the rest in the lower light-grey one (dotted lines in the top panel).}
\label{fig_SIIOI}
\end{figure}

The third most frequently observed line is \SIIA.
Figure~\ref{fig_SIIOI}, top panel plots the observed ratio of the intensity of the 
 \SIIA\ to  \OIA\ line as function of \Lacc. 
We have 
marked  the  objects where there is unambiguous evidence that
the two lines form in the LVC, i.e., both LVC and HVC are detected
in the \OIA\ line, as detailed in \S~5.1. 
Figure~\ref{fig_SIIOI}, bottom panel
plots the predictions for electronic collisions (solid curves) and
atomic hydrogen collisions (dashed curves; same as Figure~\ref{fig_OIOI}). 
The \SIIA\ line has a critical density for electronic
collisions very similar to that of the \OIA\ line
and a difference in the energy of the upper level of $\sim 10000$ K only, so that
their ratio is expected to vary little with the density and temperature
of the emitting region. 
We have assumed that
all oxygen is neutral, all solphur is \ion{S}{ii} and the
\citet[][]{asplund05}  solar elemental abundances ( $\alpha$(O)$=4.5\times 10^{-4}$, $\alpha$(S)$=1.4\times 10^{-5}$).
The ratio S/O is very similar if we take, instead, the proto-solar abundances of \citet[][]{lodders03}.
The results in the case of collisions with neutral hydrogen are very uncertain, as no de-excitation 
rates for collisions with neutral hydrogen exist. The curves in the bottom panel of Figure~\ref{fig_SIIOI}  
have been computed in the orbiting collision approximation
(see eq. 2.34 of \citet[][]{draine11}.

For the density and temperature values that account for the observed
\OIA/\OIB\ ratios, the models predict ratios   \SIIA/\OIA $\sim$0.2-0.5.
This interval is shown by
the dashed horizontal lines in the top panel and the
dark-shaded region in the bottom panel.
About 65\%\ of the measured values (and 5/6 of those with both components)
lie in this interval, only 3 objects have ratios higher than 0.5, while
5 measurements plus 5  upper limits are lower than 0.2.
%The range 0.2--0.5, which comprises 65\% of the observed values, is shown by the
%dark grey box, 
%Ratios that fall within the dark grey box
%agree very well with the model predictions, for the same range of 
%parameters that reproduce the \OIB/\OIA\ ratios and
%the assumptions that all oxygen is neutral. 
In the objects within the dark grey region, the strength of the \SIIA\ line is therefore
as  expected if the three lines \OIA, \OIB\ and \SIIA\
are emitted by the same region and are the result of thermal
processes, namely collisional excitation.
We note also that this emitting region cannot be much ionized.
As the oxygen ionization is coupled
to that of hydrogen by charge exchange, an ionization fraction $n_e/n_H \sim 0.5$, for example, would
shift all model predicted ratios upward by 0.3 dex, i.e., above the observed range.
Moreover, if  \ion{O}{ii}/\ion{O}{i} $\sim 0.5$, the \OIIB\ line should have an intensity comparable to that of the \OIB, while it is detected in very few objects (Table~\ref{table_lines}).
Neither of these is a strong quantitative argument, but, all together,  it seems likely that in these objects the LVC emitting region is mostly neutral.

Very few objects have ratios \SIIA/\OIA\ larger than
$\sim 0.5$; in these cases, it is possible  that 
a significant fraction of oxygen is ionized. 
The largest ratio is
observed in Sz106, where we detect  also  the \OII\   and \NII\ lines at roughly
zero velocity, and which is likely an object where the
forbidden line emission is dominated by a jet aligned
with the plane of the sky, where the ionization fraction (and therefore the
ratio \ion{O}{ii}/\ion{O}) could be of 20-30\% \citep[][]{bacciotti99}.
Note, however, that the large values of the observed \SIIA/\OIA\ ratio
($\sim 1.6$) requires an ionization fraction $n_e/n_H\simgreat 0.5$.

Observed \SIIA/\OIA\  below $\sim 0.2$ are difficult to explain if the
line emission is thermal,
even when the rather large errors of the observed points are taken into account.
One would need
to assume that a significant fraction of sulphur is either neutral or doubly
ionized, which seems unlikely, or that the temperature is
significantly lower than 5000 K, which is inconsistent with the constraints
obtained from the \OIB/\OIA\ ratio. 
In these objects, it is possible that a dominant part of the \ion{O}{i} emission
is due to non-thermal processes, namely photodissociation of OH, as
advocated for TW Hya by \citet[][]{gorti11}; we will go back to this point
in \S~\ref{Sec:WindModels}.

\begin{figure}
	\begin{center}
	\end{center}
		\includegraphics[width=9cm]{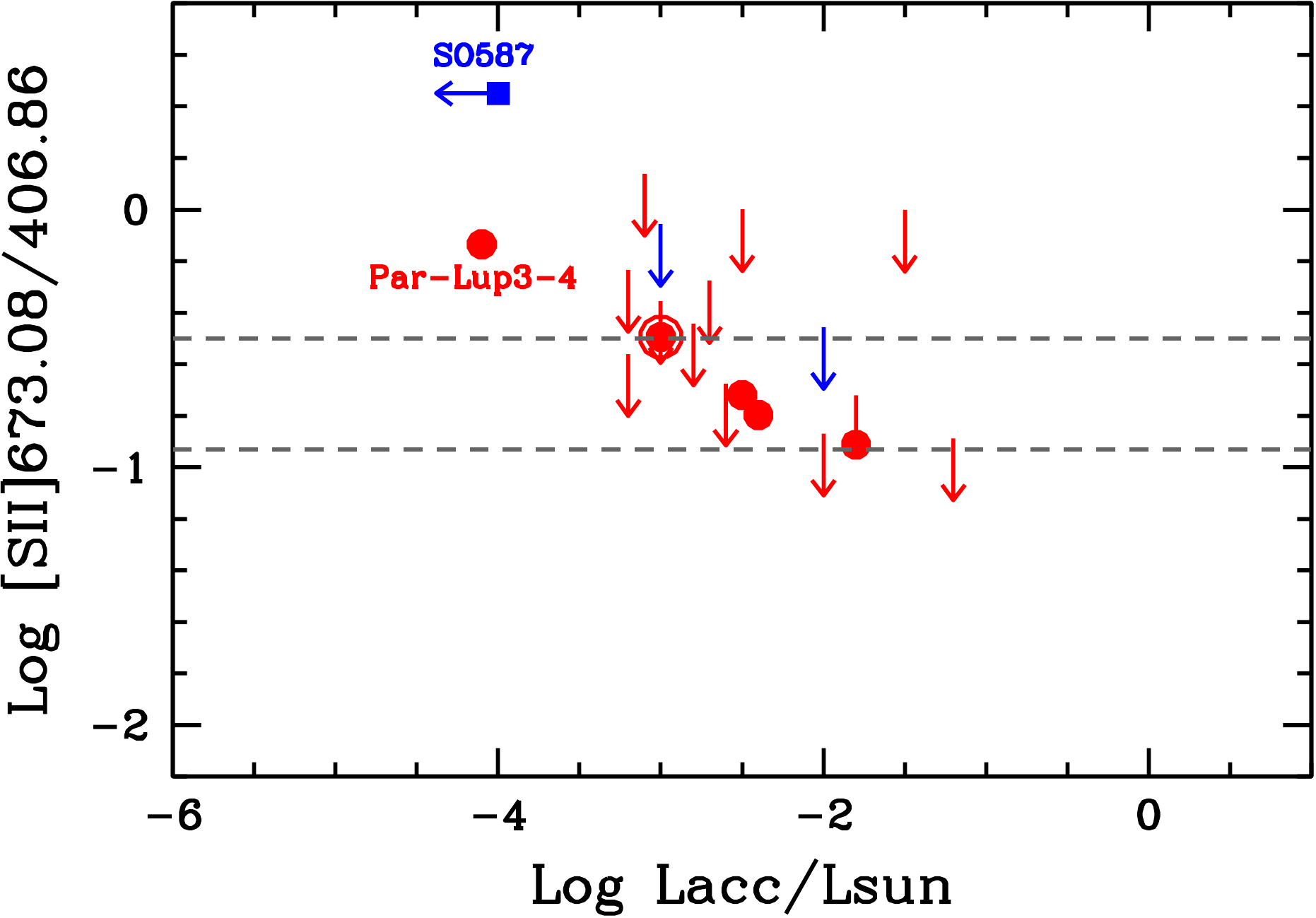}
		\includegraphics[width=9cm]{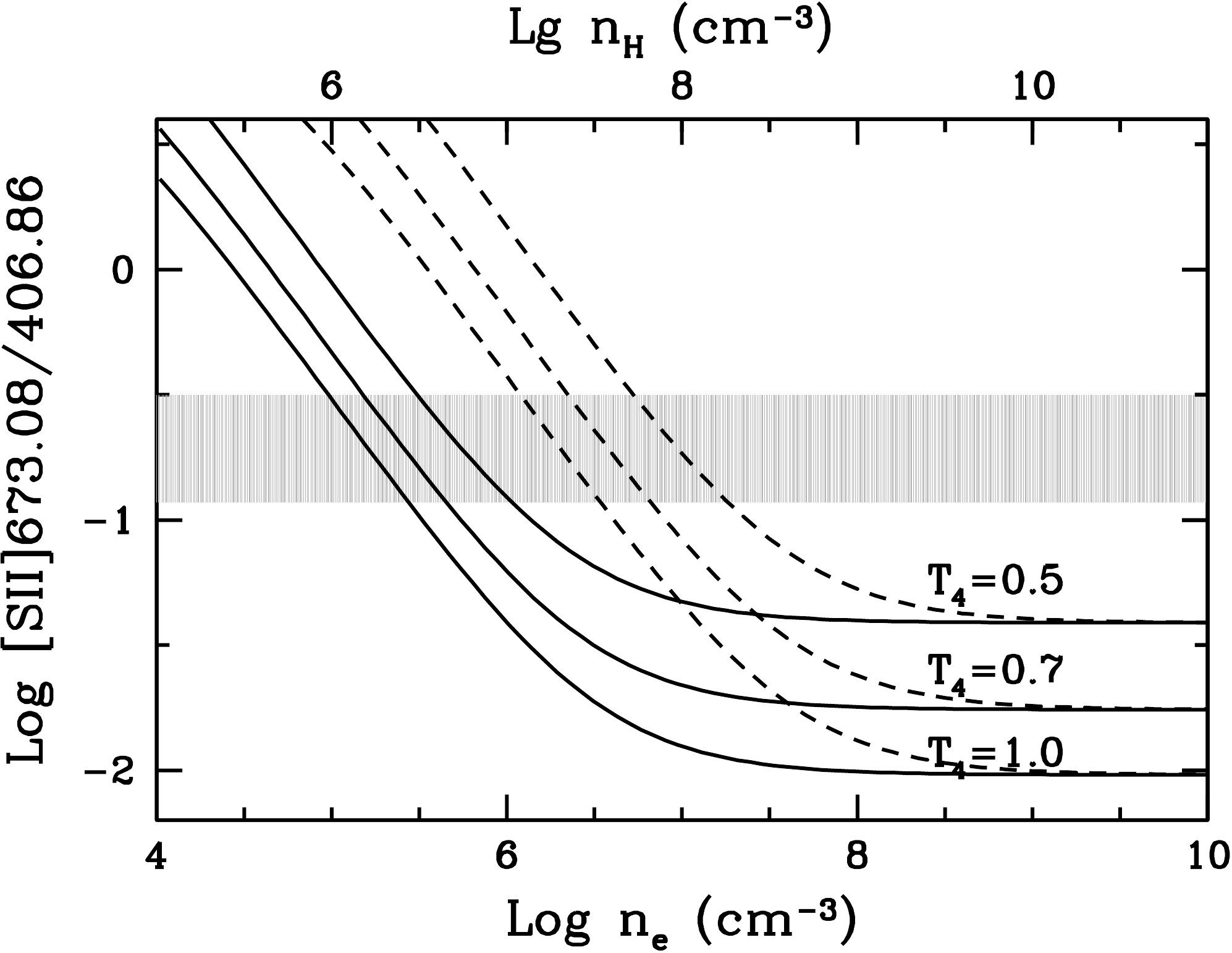}
	\caption{Same as Figure~\ref{fig_OIOI} for the ratio of the LVC [\ion{S}{ii}]\,673.08/[\ion{S}{ii}]\,406.86. 
	        The shaded region in the bottom panel shows the locus of the four detections 
		in the upper panel (excluding the special cases of Par-Lup3-4 and SO587, see text), between the two dashed lines in the top panel.}
\label{fig_SIISII}
\end{figure}

A LVC  \SIIB\ is detected in few GTO objects.
This line has a critical density much lower than either the \SIIA\ or the \OIA\ lines 
and is relatively stronger in lower density regions. Figure~\ref{fig_SIISII}, top panel 
plots the   \SIIB/\SIIA\ ratio as function of \Lacc, and Figure~\ref{fig_SIISII}, bottom 
panel the predictions of the models. The ratios tend to be higher than expected for 
the physical conditions that account for the \OIA/\OIB\ and \SIIB/\OIA\ intensity ratios,
suggesting that a region of lower electron density (but still higher than $\sim 10^5$~\cmc)
is emitting most of the \SIIB\ line. The two sources with the highest ratios \SIIB/\SIIA\ 
are Par-Lup3-4 and SO587, that we have already discussed.

%Fig.~\ref{fig_SIIred_blue_components} shows the ratio of the two \ion{S}{ii} lines in the LVC,
%blue-shifted HVC and red-shifted HVC. 
%The HVC ratios tend to be higher than the LVC ones, as expected in a lower density region, but 
%there is
%only one object (Sz100) where both \ion{S}{ii} lines  have  a LVC and a HVC;
%in this case, indeed, the ratio \SIIB/\SIIA\ is $\sim$0.3 in the LVC and $\sim$1 in the HVC.
%{\bf Non so bene cosa voglio far vedere in questa figura.}

%\begin{figure}
%	\begin{center}
%		\includegraphics[width=9cm]{SIIOI_density}
%	\end{center}
%	\caption{Same as Fig.~\ref{OIOI_dens}  for the ratio  [\ion{S}{ii}]406.86/[\ion{O}{i}]630.03.}
%\label{fig_SIIOI_dens}
%\end{figure}

%\begin{figure}
%	\begin{center}
%	\end{center}
%		\includegraphics[width=9cm]{SIISII_density}
%	\caption{As Fig.~\ref{fig_OIOI_dens} for the ratio of the LVC [\ion{S}{ii}]673.08/[\ion{S}{ii}]406.86.
%}
%\label{fig_SIISII_dens}
%\end{figure}

%\begin{figure}
	%\begin{center}
		%\includegraphics[width=9cm]{SIIred_blue_components}
	%\end{center}
	%\caption{Ratio of the  [\ion{S}{ii}]406.86/[\ion{O}{i}]630.03 in the LVC (dots),
%blue-shifted HVC (squares) and red-shifted HVC (pentagons) for objects where both lines are detected. The Sz100 values are joined by a vertical line. Red and blue symbols refer to objects in Lupus and \sori,
%respectively.}
%\label{fig_SIIred_blue_components}
%\end{figure}

\subsection {Mass, Momentum and Volume}

The analysis of the \ion{O}{i} and \ion{S}{ii} lines in the previous section indicates that
in 65\% of the objects we have evidence that the emission is coming from a 
warm ($T \sim 5000-10000$ K), dense ($n_e \simgreat 10^6$ \cmc), mostly
neutral gas. 
As  the density is larger than the critical density, the mass of gas in the emitting region can be 
derived from the luminosity of any forbidden line and is function of the temperature only. If we 
consider the \OIA\ line, which is detected in the largest number of stars, it is:

\begin {equation}
M_{gas} = \mu m_H {{L({\rm [\ion{O}{i}]\,630.03})}\over{j({\rm [\ion{O}{i}]\,630.03}) \,\alpha(\ion{O}{i})}} 
\end{equation}
where $\mu m_H = 2\times 10^{-24}$ is the mean molecular weight of neutral gas, L(\OIA) is the line 
luminosity, $j({\rm [\ion{O}{i}]\,630.03})$ is the line emissivity and $\alpha(\ion{O}{i})$ is the 
fractional abundance of neutral oxygen.  For $T\sim 5000-10000$ K,
in the high density limit
$j({\rm [\ion{O}{i}]\,630.03})$ ranges from $1.2\times 10^{-16}$ 
to $1.1\times 10^{-15}$ erg s$^{-1}$ per \ion{O}{i} atom. Assuming an average value $\sim 6\times 10^{-16}$ 
and that most oxygen is \ion{O}{i}, it is:

\begin{equation}
M_{gas} \sim 1.4 \times 10^{-11} M_\odot \big({{L({\rm [\ion{O}{i}]\,630.03})}\over{10^{-6}\,\rm L_\odot}}\big)
\end{equation}

In the GTO sample, $M_{gas}$ increases from $\sim 1.4\times 10^{-12}$ for the
lowest luminosity objects (L(\OIA)/\Lsun=$10^{-7}$) to $\sim 1.4\times 10^{-9}$ \Msun\ 
for the objects with the strongest \OIA.
The corresponding momentum, assuming a velocity $V_{peak} \sim 10$ km/s,
goes from $\sim 1.4\times 10^{-11}$ to $\sim 1.4\times 10^{-8}$ \Msun\,km/s. 

The volume occupied by the emitting gas can also be derived from the line luminosity 
if the gas density is known: 

\begin {equation}
Vol_{gas} =  \big({{L({\rm [\ion{O}{i}]\,630.03})}\over{n_H\, j({\rm[\ion{O}{i}]\,630.03})\, \alpha(\ion{O}{i})}}\big) 
\end{equation}

If electron collisions dominate the level excitation, we can derive a very conservative 
upper limit of $n_H \sim$$10^8$~\cmc, assuming $n_e \sim$$10^7$~\cmc\ and an ionization 
fraction $\sim$$0.1$. The minimum $n_H$ will be at least hundred times higher if atomic 
hydrogen collisions dominate. Therefore, $Vol_{gas}$ ranges from $\simgreat 1.4 \times 10^{37}$~cm$^3$ 
for objects with L(\OIA)$\sim$$10^{-7}$~\Lsun\ to $\simgreat 1.4\times 10^{40}$~cm$^3$ when
L(\OIA)$\sim$$10^{-4}$~\Lsun. This would translate in a linear dimension of 0.16\,AU to 1.6\,AU, 
assuming  spherical geometry.

\begin{figure}
	\begin{center}
		\includegraphics[width=9cm]{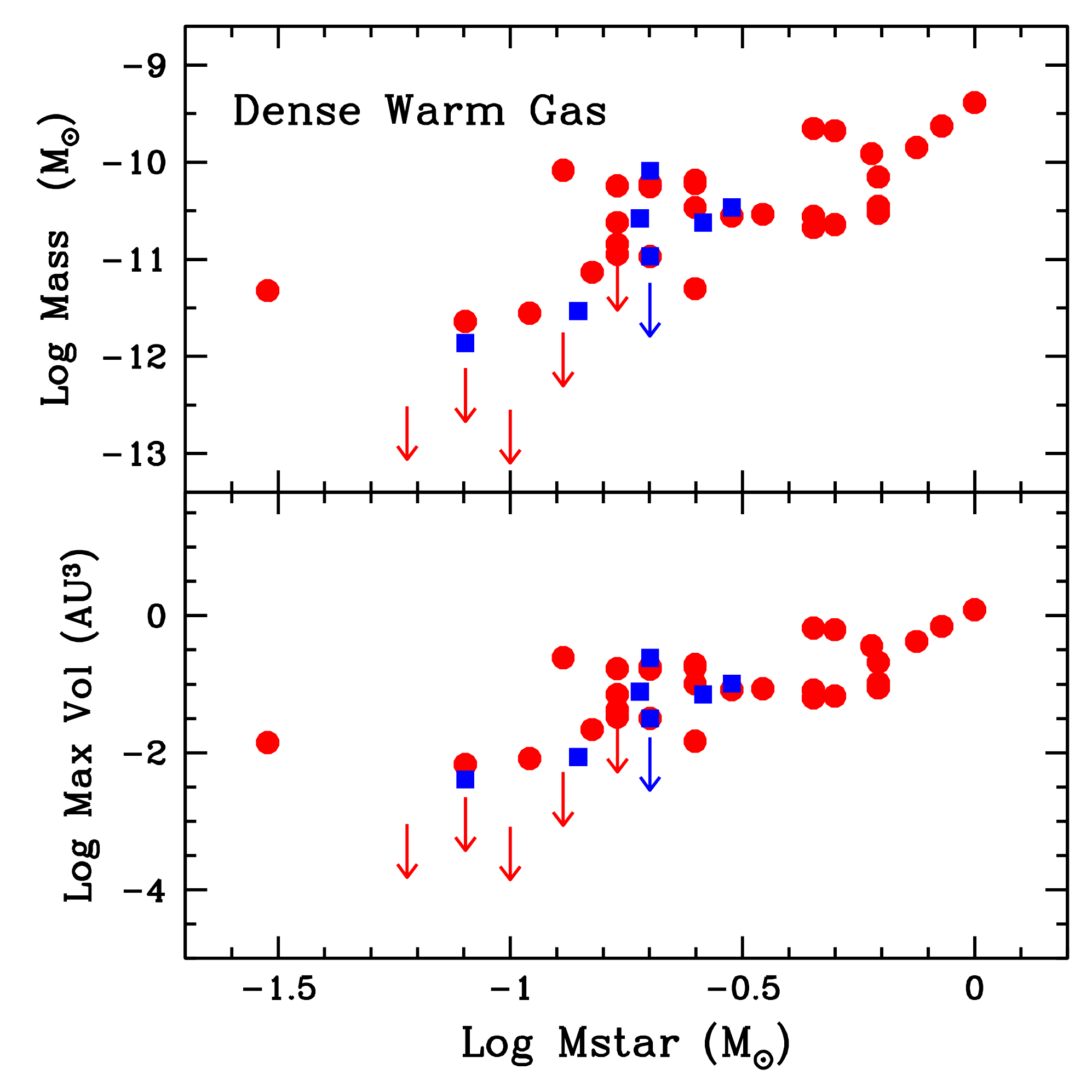}
	\end{center}
	\caption{Properties of the dense, warm gas as function of the stellar mass for the GTO sample. Red dots are objects in Lupus, blue filled squares in \sori; 
arrows are objects with upper limits of the \OIA\ luminosity, from which we derive the properties of the emitting region. 
The top panel plots the mass of the gas in the region, the bottom panel the upper limit to the volume, computed for an electron density $n_e=10^7$ \cmc, ionization fraction 0.1 ($n_H=10^8$ \cmc). 
In both panels we have assumed a line emissivity which is the average between the values for T=10000 K ($1.1 \times 10^{-15}$ erg $s^{-1}$ per \ion{O}{i}) and T=5000 K ($1.2 \times 10^{-16}$ erg $s^{-1}$ per \ion{O}{i}) once $n_e \gg n_{cr}$. 
}
\label{fig_mass}
\end{figure}

The values of the mass of the dense, warm gas and the maximum values it occupies 
are plotted in Figure~\ref{fig_mass} as function of the mass of the central object.

%To compute the mass-loss rate  one needs an estimate of the
%flow velocity, that we can take to be of about 10 km/s, and of
%a typical length, which, in the case of the LVC, is probably
%determined by a change in the required physical conditions, and
%is   not known (see Hartigan et al. 1995).
%Assuming, for simplicity, a cilindrical geometry for the flow, it is:
%
%\begin{equation}
%\dot M_{loss} = (\mu m_H)\> n_H \> V_{peak} \> A 
%\end {equation}
%where $V_{peak}\sim 10$~km/s is the flow velocity and $A$ is the cilinder section.
%We can explore the constraints that our results set on the values of
%the unknown quantity  $A$.

%Fig.~\ref{mloss-macc} shows the values of \Mloss\ as function of the mass accrettion rate of the GTO object assuming 
%The results are shown in Fig.~\ref{fig_mloss} as function of the
%mass accretion rate of the object for 
%a cilinder radius $\sim 1$ AU. Note that if $n_H$ increases by a factor 10, so will \Mloss. Similarly, \Mloss\ is proportional to the area, i.e., to the
%extension of the disk region from which the wind is emitted.
%Taken at face value, the results in Fig.~\ref{fig_mloss} indicate \Mloss/\Macc\ 
%ratios in the range 0.01--1. Note that the trend of increasing ratios for
%increasing \Macc\ may be an artifact of the assumption of constant density
%and area.
%

%\section {Odd Objects}

%Should we discuss here (briefly) objects as SO586, SO848, Par-Lup-3-4? Or others that stick out in the various diagrams?

\section{Discussion}

The small, blue-shifted velocity of the LVC line peaks
is the main indication of their origin in a wind, ejected from the disk surface
with low velocity \citep[see][]{hartigan95}. The quality of our spectra does
not allow us to constrain the peak velocity to an accuracy better than $\sim 10$ km/s, 
but the blue-shifts are clearly detected in all lines, and we will therefore discuss 
their properties in the context of the  wind models.

\subsection {Mass-loss rate}

The mass-loss rate is the  fundamental global quantity that 
characterizes the mass-loss process and its inpact on the disk evolution. All the existing estimates of \Mloss\ have been derived for the HVC 
(\citet[]{hartigan95};see \citet{cabrit02} for a discussion) and do not necessarily apply to the outflowing gas traced by the LVC emission, which can trace a different mechanism of mass-loss.

In spite of the number of spectroscopic data on the LVC, deriving a meaningful  mass-loss rate is difficult, as different assumptions on the geometry of the ouflowing matter lead to very different estimates.
A crude estimate of \Mloss\ can be obtained by dividing the
mass of the emitting region (eq.(6)) by a timescale $\tau=L/{\rm v}$ where $L$
and ${\rm v}$ are measured along the flow direction. 
Let us assume  ${\rm v} \sim V_{peak}$, and the  maximum value of the volume
occupied by the emitting matter  derived in \S~6.4.
Values of \Mloss\ obtained assuming  an outflow spherical geometry 
are in the range $\sim 2\times 10^{-11} - 10^{-9}$ \Myr, approximately bewteen
1 and 0.1 \Macc.
If the outflow is lounched from a disk area $A$ in  the vertical direction and has a  cylindrical geometry,
then $L=Vol/A$.
We have considered two possibilities. 
The first is that the outflow is emitted by a disk annulus at a distance from the star such that the keplerian velocity $V_{kepler}=V_{peak}$;
assuming an annulus  width $\Delta R/R$=0.1, we find that $L$ is typically 10-30 times smaller than before and  \Mloss$\gg$ \Macc.
The second possibility, suggested by models of photoevaporative winds, is that the emitting region is at a distance from the star of $\sim 1.8$~(\Mstar/\Msun) AU
\citep[][]{alexander14}. In this case $L$ and \Mloss\ are
similar to those obtained for spherical geometry.

In fact, it is likely that $L$ is much larger than our estimates. If the flow velocity is not orthogonal to the disk, but has a large tangential component, as in a conical wind, then $L$ can be much larger and is defined
by a change in the physical conditions, density in particular, that
are traced by the forbidden lines we  observe.
A detail study of high-spectral and spatial resolution  line profiles is needed
to  obtain an estimate of the mass-loss rate of the slow wind
component.

\subsection {Wind models}
\label{Sec:WindModels}

MHD disk winds have been associated to 
the LVC of the TTS forbidden
lines \citep[][]{kwanandtademaru95}, as at their base, when the disk material becomes 
unbound, they have high density and low velocities
(see, e.g., .
\citep[][]{ferreira06, 
suzuki09, bai13, zanniandferreira13, lesur13, fromang13, bai14,
kurosawa12}.
It is, however, very difficult to estimate if any of these models can explain
the observed properties of the LVC emitting region
derived in this paper, or, even more, to use the observations to discriminate between the various mechanisms that can trigger a MHD wind and constrain
the model parameters.
In very few cases the results of the MHD models and simulations have been coupled with 
calculations of the temperature 
\citep[][]{safiera, safierb, shang98, shang02, cabrit99, garcia01, pesenti04, pyo03, 
pyo06, cabrit07, panoglou12}, and have in general focussed on jet-tracing lines for 
luminous, highly accreting TTS. 

Photoevaporative winds are also characterised by a high-density, low velocity
region at their base and,
in recent years, they have been associated
to the LVC of the forbidden lines. 
Models that take into account the effect
of the EUV, X-ray and FUV radiation on the disk have been
computed by various groups, using different simplifications and making 
different choises for the parameters \citep[e.g.,]{font04, 
hollenbachandgorti09, ercolano09, owen10, ercolanoandowen10}.
% (e.g., Font et al., Hollenbach, Ercolano etc.). 
The results from different models  still show large discrepancies, as discussed most 
recently by \citet[][]{alexander14}. In addition, few models predict luminosities 
of the optical forbidden lines, with the exception of the \OIA\ line, and are 
limited to a small range of stellar, disk and radiation field properties.

The largest set of line luminosity calculations is from \citet[][]{ercolanoandowen10}, 
who computed the forbidden line luminosity expected in a photoevaporative wind
around a star of 0.7~\Msun\ due to the combined effect of EUV and X-ray photons
(no FUV photons were included). 
They found that the optical forbidden lines are produced in a warm neutral gas
with temperatures of 3000--5000~K, where collisions with atomic hydrogen control 
the level populations.
For the largest  $L_x \sim 2 \times 10^{30}$~erg/s the line luminosity is 
of $\sim 10^{-5}$~\Lsun\ for the \OIA, \SIIA\ and \SIIB, roughly as observed. 
However, the predicted \OIB\ is about two orders of magnitude  lower than
observed. Ercolano \&\ Owen point 
out that in their models the excitation of the upper level of the \OIB\ transition 
is highly underestimated, since the collision rate with neutral hydrogen are unknown 
and could therefore not be included in their excitation calculations. The calculations
of the \OIA/\OIB\ ratios in \S~6.2, albeith with a very crude approximation for the
neutral hydrogen collisional de-excitation rates of the \ion{O}{i} $^1D_2$ level, suggest that
one needs a gas density $n_H \simgreat 10^{10}-10^{11}$~\cmc\ for
$T\sim 7000-5000$~K, conditions that do not seem to occur in the models.
These models reproduce quite well the small blue-shift of the line peaks,
but tend to underestimate the line width, which, for the \OIA\ line, is
always predicted to be smaller than $\sim 20$~\kms.
Photoevaporative wind models by \citet[][]{hollenbachandgorti09}, which also include
EUV and X-ray but make different assumptions and approximations give somewhat
lower luminosity for the \OIA\ line. However, for a very soft X-ray spectrum,
L(\OIA)$\sim 10^{-5}$~\Lsun\ for log\Lx=30.3 \citep[as in][]{ercolanoandowen10}.
%This is consistent with the measured values for \Lstar$\sim 1$ \Lsun, i.e.
%\Lx/\Lstar$\sim 10^{-3}$ ({\bf ?}). 
\citet[][]{hollenbachandgorti09} do not  compute any other optical forbidden line.

\begin{figure}
	\begin{center}
	 \includegraphics[width=9cm]{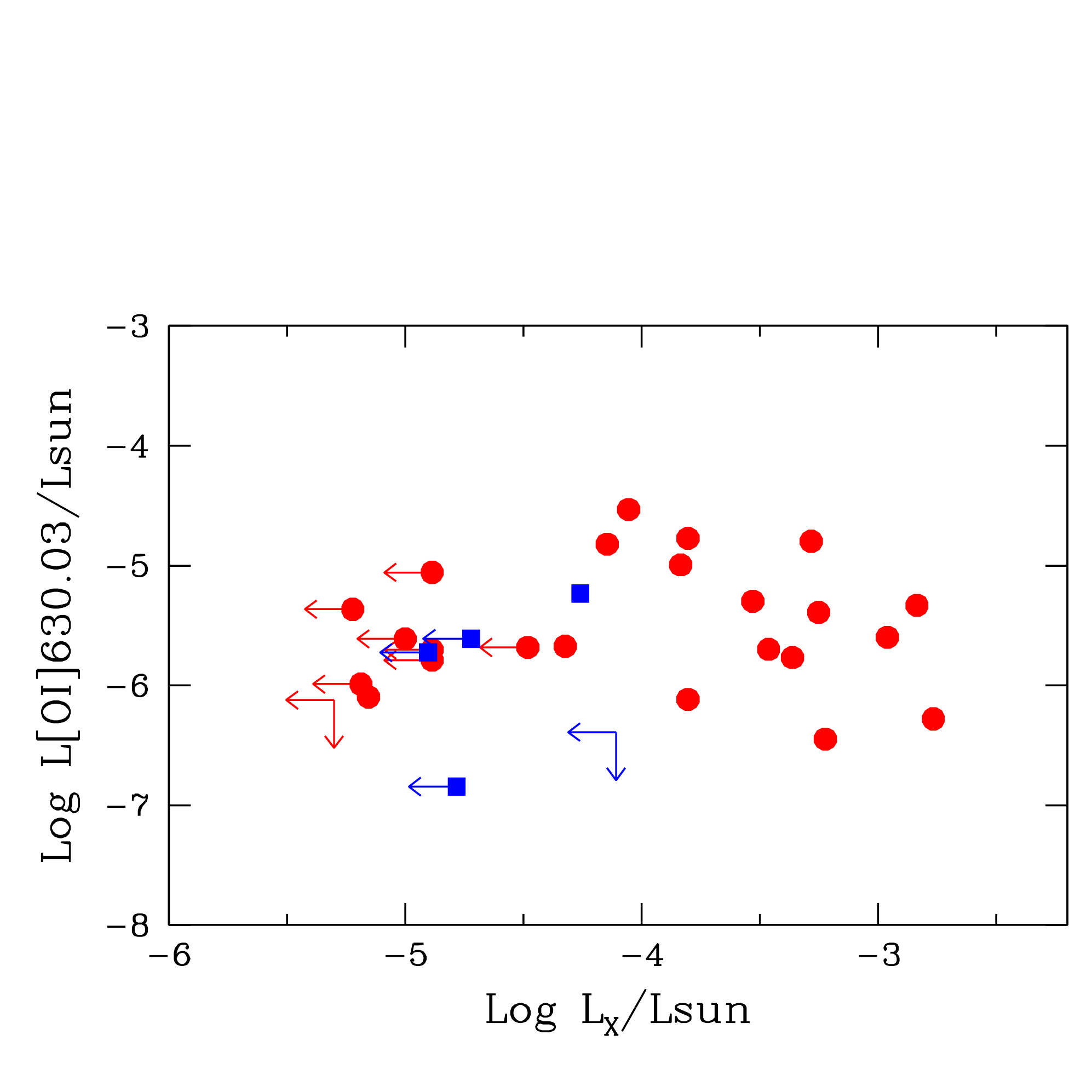}
	\end{center}
        \caption{Luminosity of the \OIA\ line  plotted as a function of X-ray luminosity, 
	when available, for the GTO sample. Red dots are the Lupus objects, blue filled squares are 
	objects in \sori. The X-ray luminosities were collected from the ROSAT observations 
	by  \citet[][]{krautter97} for Lupus and from the XMM-Newton observations by 
	\citet[][]{franciosini06} for \sori.}
\label{fig_xray}
\end{figure}

The effect of the FUV photons (both of accretion and chromospheric origin) has been 
discussed by \citet[][]{gortiandhollenbach09} for a range of stellar masses; they do 
not predict the line spectrum but only the disk-integrated mass-loss rate. More recent 
models by \citet[][]{owen12} compare the results for two stellar masses (0.1 and 0.7~\Msun) 
and include FUV photons as well; also in this case, no line luminosities are computed. 
In these models, as long as the ratio FUV/X is small, the disk integrated mass-loss 
rate is roughly proportional to the X-ray luminosity and has a very weak dependence on 
the stellar mass \citep[]{gortiandhollenbach09}. Assuming that also in these 
more realistic models the line luminosity is roughly proportional to $L_X$, as in 
\citet[][]{ercolanoandowen10}, one would expect a relation between, e.g., L(\OIA) and 
$L_X$, which is not seen \citep[Figure~\ref{fig_xray}; see also][]{rigliaco13}.
It is possible that, in fact, the FUV flux controls the wind temperature, and that
the gas mass in the wind with density and temperature high enough to emit the optical forbidden 
lines is larger for more luminous, more accreting stars. We cannot estimate if this 
is indeed the case from the published results of \citet[][]{owen12}, but we think that 
it is a possibility that should be explored further. 
 
%However, this explanation would also 
% require that the line luminosities are proportional to \Mloss.
%As discussed in Sec.7.4, this relationship holds only if the physical
%conditions and the
%geometry of the wind do not change with stellar and disk properties,
%and this does not seem to be the case in the Owen et al.~(2012) models.
%In particular,
%when scaled to the gravitational radius, which in its turn is  inversely
%proportional to the stellar mass, temperature and mass-loss profile 
%do not depend on \Mstar, while the density structure scale as \Mstar$^{-2}$.
%We expect, then, that winds from lower mass stars will be less dense,
%making the condition that the two \ion{O}{i} lines (\OIA\ and \OIB) have similar
%luminosity even harder to meet than in more massive TTS. 

 The difficulty of wind models to reproduce the LVC spectrum, and in 
particular the strength of the \OIB\ line, pushed some authors to consider a different mechanism to excite the OI levels other than collisions with electrons and/or hydrogen atoms, namely photodissociation of OH in the disk surface layers
\citep[][]{stoerzer00, acke05, rigliaco13}.
The most detailed study of the LVC spectrum in a TTS has been carried on for  
TW Hya (\Mstar$\sim 0.7$~\Msun, \Lstar$\sim 0.53$~\Lsun, \Macc$\sim 10^{-9}$~\Myr) 
by \citet[][]{gorti11} and \citet[][]{pascucci11}.
TW Hya  has a L(\OIA)$\sim 1.1 10^{-5}$~\Lsun, a ratio \OIA/\OIB$\sim 7$, \SIIA/\OIA$\sim 0.13$, 
\SIIB/\SIIA$\simless 0.13$, not very different from the results we find in the GTO stars. 
\citet[][]{gorti11} models that include X-ray and FUV photons fail to reproduce the  
\OIB\ observed luminosity; this
leads the authors to conclude that  the strongest contribution to the
two \ion{O}{i} lines comes from  photodissociation of OH,
taking place on the disk surface, and \ion{O}{i} lines do not trace a wind component.
Only 20\%\ of the \OIA\ line  is due to thermal processes
in the dust-depleted inner disk, where also the \SIIA\ line is emitted; the predicted
global ratio \SIIA/\OIA$\sim 0.5$,  is roughly consistent with the observations.
Among TTS, TW Hya has rather unusual line profiles, with narrow 
lines (FWHM$\sim 10$~km/s) centered at zero velocity \citep[][]{pascucci11}, which 
may support an origin in the disk surface, rather than in an unbound flow  \citep[but see][]{owen10}. 
Also, the presence of the dust-depleted inner ragion of radius $\sim 4$~AU seems
crucial to the \citet[][]{gorti11} model results.

The \SIIA\ line  provides the strongest discriminant between the two modes
of line excitation, as it can only be produced by collisional excitation of
\ion{S}{ii}.
As discussed in \S 6.3, in 65\% of the GTO stars the luminosity of the three
lines \OIA, \OIB\ and \SIIA\ is consistent with the emission expected in a high
density, warm and neutral gas, where the lines are {\it all} thermally excited: 
 a significant
non-thermal contribution to the \ion{O}{i} lines will reduce the \SIIA/\OIA\ 
well below the observed values. We think that in these objects at least
a non-thermal origin of the \ion{O}{i} lines could be ruled out.
There is, however,
 a small but not negligible fraction of objects (10 in total)
with ratios \SIIA/\OIA\ lower than predicted by thermal excitation, where
models like that of TW Hya may apply. On the other hand,
they do not seem to be in any way exceptional; they are distributed over the whole range
of \Lstar, \Lacc\ and \Lx\ and 
the line profiles are similar to those of the other objects, broad with
slightly blue-shifted peaks. Also, 
there is no evidence of  an inner dust-depleted region 
similar to that in TW Hya
but in a couple of cases \citep[][]{rigliaco12}; Alcal\'a et al. in preparation).

An obvious caveat is in order, namely that the GTO spectra have relatively
low spectral resolution; any detailed analysis and firm conclusion require
higher quality data, to validate our assumption that all the lines are emitted by the same kinematical component.
In view of its importance,
it is interesting that in the past
\SIIA\ has not been searched for as extensively as the \OIA\ line, even if one
expects that their intensity should be similar.

%This may be due, in part,
%to the fact that the red part of the visible spectrum is richer of lines,
%diagnostics of accretion and/or winds, and that the blue one is more affected
%by extinction.

%\begin{figure}
%	\begin{center}
%		\includegraphics[width=9cm]{dimensions}
%	\end{center}
%	\caption{ Mass-loss rates as function of the mass-accretion rate.
%}
%\label{fig_dimensions}
%\end{figure}

\subsection {Bound or Unbound gas?}

The small velocity shift of the peak of the optical forbidden lines and their large width is
difficult to understand in any wind model.  \citet[][]{rigliaco13} suggest that in two rather luminous TTS
for which they have very high spectral resolution profiles of the \OIA\ line, the  LVC can be deconvolved into two
different components, one narrower and slightly blueshifted, and one much broader and symmetric. The broad
component, that contributes to $\sim$40\% of the line intensity, does not trace unbound gas, but rather
the emission coming from the inner region of the disk (a few tenth of AU), where the gas is
gravitationally bound and in keplerian rotation around the star. The narrow component, on the other hand,
could trace outflowing gas, ejected further out in the disk at the  low velocity traced by the peak
velocity shift.

This may be a general occurrence in TTS disks; \citet[][]{acke06} propose a disk surface origin 
for the \OIA\ line in HAe stars and in two cases convincingly
derive the keplerian rotation and the disk orientation 
from high resolution $\Delta {\rm v} \sim 4$ km/s) spectra. 
We note that, if the \citet[][]{rigliaco13} hypothesis is true for all objects,
there should be an anticorrelation between the width of the broad, symmetric
LVC component, which tends to zero if the disk is seen face-on, and the velocity of the HVC, which is maximum 
for face-on disk, as the jet is then expected to be
oriented in the direction of the observer.
A detailed study of high resolution profiles 
of the optical forbidden lines detected in the LVC of the GTO spectra, including the two \OIA, \OIB\ 
and the \SIIA\ lines, will be very valuable also from this point of view.

%HeI absorption in Edwards et al. (2006), Kwan et al. (2007), 
%They find that blue-shifted absorption in the HeI10830 line can be due to a disk wind that intercept the emission from the accretion shock in about 30\% of cases. However, it seems that they need a hot gas for the HeI line to be optically thick.
%Kurosawa et al.  (2011) model the HeI10830 line profile with a disk wind of fixed T=9000K, small emitting region etc., but they have to add an additional sourceof ionization/excitation ($L_x$). I wonder ?
%

%For example, it would be extremely useful to compute, for any given model,
%the mass of gas with density larger than a given value as function of the velocity.
%And, of course, models with no prediction of the temperature will always be useless!

 %Aargh!

\section {Summary and conclusions}

We have examined in this paper the forbidden line spectrum of a sample of
44 low-mass TTS in the star forming regions Lupus and \sori. The spectra have been obtained 
with X-Shooter, as part of the Italian GTO, and have been analyzed to derive stellar and 
accretion properties by \citet[][]{alcala14} and \citet[][]{rigliaco12}.
All the GTO objects have a disk, as inferred from  their IR excess emission,
and are all accretors, with the exception of two stars in \sori, where only
upper limits to \Lacc\ have been measured.

We detect forbidden lines of \ion{O}{i}, \ion{O}{ii}, \ion{S}{ii}, \ion{N}{i} and \ion{N}{ii} 
in a number of stars. The line profiles show the presence of two
components, one peaked close to zero velocity (LVC) and one with a large 
peak velocity
shift (>40~km/s) to the blue and/or to the red (HVC).
%This is typical of all  TTS \citep[]{hartigan95}. 
The HVC  has been identified with the emission of a high velocity jet, 
while the origin of the LVC is uncertain \citep []{hartigan95}.

We focus our analysis on the LVC. The most commonly observed line is the
\OIA, which is detected in 38/44 objects, followed by the other neutral oxygen 
line \OIB\ and the \SIIA. In very few objects we have clear evidence of \SIIB\ LVC
emission. Higher excitation lines are detected with small peak velocity shifts
in few objects, and probably trace a jet oriented along the plane of the sky.
The spectral resolution of the GTO X-Shooter spectra ($\sim 30-60$ km/s) does not allow 
us to perform a detailed study of the gas kinematics. However, we can
measure with reasonable accuracy the shift of the line peak, which is of few km/s 
to the blue for all the lines we have studied, and the line width, which is
of $\sim 30-100$ km/s, with very few cases of unresolved lines.
We confirm, for our low-mass sample, the properties of the LVC previously
observed in solar-mass TTS; 
the small blue shift of the line peaks, in particular, seem to indicate their origin 
in outflowing, low velocity matter, which we call {\it slow wind}. We do not find any 
correlation between  line profiles and  stellar properties.

The GTO spectra show that slow winds are ubiquitous in Class II stars, i.e., in stars 
with circumstellar disks, independently of their stellar and/or accretion properties, 
down to the lowest mass stars in our sample (approximately  0.1 \Msun, \Macc $\sim 10 ^{-11}-10^{-12}$ \Myr.
The slow wind, whose properties  we derive from the analysis of the LVC
of forbidden lines of \ion{O}{i} and \ion{S}{ii}, is prevalently neutral,
dense ($n_H \simgreat 10^8$ \cmc), warm (T$\sim 5000-10000$ K). These physical conditions 
do not seem to depend on the stellar and/or accretion properties.  However, the mass of 
such gas varies by several orders of magnitude, from $\sim 10^{-12}$~\Msun\ for the lowest 
mass stars in our sample to $\sim 10^{-9}$~\Msun\ for solar-mass TTS. This may be accounted 
for by an increase in the density and/or the volume of the emitting matter.

We note that in about 65\%\ of the objects the ratio of the
\SIIA\ to the \OIA\ and \OIB\ lines  agrees with the predictions
for a collisionally excited gas with the physical properties given above.
This seems to rule out (at least in these objects) a non-thermal origin
of the \ion{O}{i} lines, as proposed by \citet[][]{hollenbachandgorti09}.
However, in about 30\%\ of cases, where we measure a relatively low \SIIA\ 
intensity, this remains a viable hypothesis.

Line luminosities are strongly correlated with the stellar luminosity
 and with~\Lacc. In the latter case, the relation is not linear, but 
has a slope of $\sim 0.81 \pm 0.09$.
The results that the line luminosities correlate
equally well with \Lstar\ and \Lacc\ is not surprising, as these two quantities
are themselves tightly correlated
(\Lacc\ $\propto$ \Lstar$^{1.53\pm 0.18}$).
As a consequence, we cannot establish which of these two properties drives the 
correlation with the line luminosity.

%We have compared our results to the predictions of
%photoevaporative wind models and MHD winds. This turns out
%to be difficult, as the models are limited in the range of parameters
%they explore and in the information they provide.
%The origin of the
%TTS slow winds remain elusive. 

The comparison with the prediction of outflow models is difficult and, so far,
the origin of the slow TTS winds remains elusive. In this context, the
derivation of  
the mass and minimum volume of the low-velocity, dense and warm outflowing gas
over a large range of stellar masses, luminosities and mass-accretion rates
in \S 6.4
could provide crucial tests for wind models, even when calculations of the
ionization and excitation conditions are not available.

On the observational side, we note that higher quality spectra, with higher spectral resolution 
and better sensitivity are necessary to confirm  the results of our analysis and to derive
the detailed kinematics and the origin of the line broadening \citep[see, e.g.,][]{rigliaco13}.
Variability studies of the line profiles could also provide very valuable information on the ejection mechanisms.

%As we have noted before,
%if we add the more luminous TTS of Hartigan et al. (1995), the mass
%of gas in the slow wind  must increases by {\bf ??} order of magnitude, maintaining
%similar temperatures and ionization conditions, in such a way
%that a tight correlation between the gas mass and \Lstar\ and/or the mass accretion 
%rate is established. This is the most important  aspect of our results, that should 
%be accounted for.

\appendix

\section{Stellar Properties}
The stellar and accretion parameters for the GTO sample are listed in 
Table~\ref{table_stars} \citep[see][]{alcala14, rigliaco12}.
%table
\begin{table*}
   \centering
      \caption[]{Stellar Properties}
         \label{table_stars}
    \begin{tabular}{l c c l c c c c c c c c c c c }
\hline

Name& $\alpha$(J2000)& $\delta$(J2000)&   ST&        T$_\star$& A$_{\rm V}$ &  L$_\star$&     M$_\star$& Log L$_{acc}$& Log M$_{acc}$&   D\\
 &  & &  &  (K) & & (L$_\odot$)& (M$_\odot$)& (L$_\odot$)& (M$_\odot$/y)& (pc) \\
\hline
        Sz66 &  15 39 28.28  & -34 46 18.0 &  M3.0 &  3415 & 1.00  & 0.200 &   0.45  &  -1.8 &   -8.73 &  150 \\
   AKC2006-19&  15 44 57.90  & -34 23 39.5 &  M5.0 &  3125 & 0.00  & 0.016 &   0.10  &  -4.1 &  -10.85 &  150 \\
        Sz69 &  15 45 17.42  & -34 18 28.5 &  M4.5 &  3197 & 0.00  & 0.088 &   0.20  &  -2.8 &   -9.50 &  150 \\
        Sz71 &  15 46 44.73  & -34 30 35.5 &  M1.5 &  3632 & 0.50  & 0.309 &   0.62  &  -2.2 &   -9.23 &  150 \\
        Sz72 &  15 47 50.63  & -35 28 35.4 &  M2.0 &  3560 & 0.75  & 0.252 &   0.45  &  -1.8 &   -8.73 &  150 \\
        Sz73 &  15 47 56.94  & -35 14 34.8 &	K7 &  4060 & 3.50  & 0.419 &   1.00  &  -1.0 &   -8.26 &  150 \\
        Sz74 &  15 48 05.23  & -35 15 52.8 &  M3.5 &  3342 & 1.50  & 1.043 &   0.50  &  -1.5 &   -8.09 &  150 \\
        Sz83 &  15 56 42.31  & -37 49 15.5 &	K7 &  4060 & 0.00  & 1.313 &   1.15  &  -0.3 &   -7.37 &  150 \\
        Sz84 &  15 58 02.53  & -37 36 02.7 &  M5.0 &  3125 & 0.00  & 0.122 &   0.17  &  -2.7 &   -9.24 &  150 \\
       Sz130 &  16 00 31.04  & -41 43 37.2 &  M2.0 &  3560 & 0.00  & 0.160 &   0.45  &  -2.2 &   -9.23 &  150 \\
       Sz88A &  16 07 00.54  & -39 02 19.3 &	M0 &  3850 & 0.25  & 0.488 &   0.85  &  -1.2 &   -8.31 &  200 \\
       Sz88B &  16 07 00.62  & -39 02 18.1 &  M4.5 &  3197 & 0.00  & 0.118 &   0.20  &  -3.1 &   -9.74 &  200 \\
        Sz91 &  16 07 11.61  & -39 03 47.1 &	M1 &  3705 & 1.20  & 0.311 &   0.62  &  -1.8 &   -8.85 &  200 \\
      Lup713 &  16 07 37.72  & -39 21 38.8 &  M5.5 &  3057 & 0.00  & 0.020 &   0.08  &  -3.5 &  -10.08 &  200 \\
      Lup604s&  16 08 00.20  & -39 02 59.7 &  M5.5 &  3057 & 0.00  & 0.057 &   0.11  &  -3.7 &  -10.21 &  200 \\
        Sz97 &  16 08 21.79  & -39 04 21.5 &  M4.0 &  3270 & 0.00  & 0.169 &   0.25  &  -2.9 &   -9.56 &  200 \\
        Sz99 &  16 08 24.04  & -39 05 49.4 &  M4.0 &  3270 & 0.00  & 0.074 &   0.17  &  -2.6 &   -9.27 &  200 \\
       Sz100 &  16 08 25.76  & -39 06 01.1 &  M5.5 &  3057 & 0.00  & 0.169 &   0.17  &  -3.0 &   -9.47 &  200 \\
       Sz103 &  16 08 30.26  & -39 06 11.1 &  M4.0 &  3270 & 0.70  & 0.188 &   0.25  &  -2.4 &   -9.04 &  200 \\
       Sz104 &  16 08 30.81  & -39 05 48.8 &  M5.0 &  3125 & 0.00  & 0.102 &   0.15  &  -3.2 &   -9.72 &  200 \\
      Lup706 &  16 08 37.30  & -39 23 10.8 &  M7.5 &  2795 & 0.00  & 0.003 &   0.06  &  -4.8 &  -11.63 &  200 \\
       Sz106 &  16 08 39.76  & -39 06 25.3 &  M0.5 &  3777 & 1.00  & 0.098 &   0.62  &  -2.5 &   -9.83 &  200 \\
  Par-Lup3-3 &  16 08 49.40  & -39 05 39.3 &  M4.0 &  3270 & 2.20  & 0.239 &   0.25  &  -2.9 &   -9.49 &  200 \\
  Par-Lup3-4 &  16 08 51.43  & -39 05 30.4 &  M4.5 &  3197 & 0.00  & 0.003 &   0.13  &  -4.1 &  -11.37 &  200 \\
       Sz110 &  16 08 51.57  & -39 03 17.7 &  M4.0 &  3270 & 0.00  & 0.276 &   0.35  &  -2.0 &   -8.73 &  200 \\
       Sz111 &  16 08 54.69  & -39 37 43.1 &	M1 &  3705 & 0.00  & 0.330 &   0.75  &  -2.2 &   -9.32 &  200 \\
       Sz112 &  16 08 55.52  & -39 02 33.9 &  M5.0 &  3125 & 0.00  & 0.191 &   0.25  &  -3.2 &   -9.81 &  200 \\
       Sz113 &  16 08 57.80  & -39 02 22.7 &  M4.5 &  3197 & 1.00  & 0.064 &   0.17  &  -2.1 &   -8.80 &  200 \\
     J160859 &  16 08 59.53  & -38 56 27.6 &  M8.5 &  2600 & 0.00  & 0.009 &   0.03  &  -4.6 &  -10.80 &  200 \\
    c2dJ1609 &  16 09 01.40  & -39 25 11.9 &  M4.0 &  3270 & 0.50  & 0.148 &   0.20  &  -3.0 &   -9.59 &  200 \\
       Sz114 &  16 09 01.85  & -39 05 12.4 &  M4.8 &  3175 & 0.30  & 0.312 &   0.30  &  -2.5 &   -9.11 &  200 \\
       Sz115 &  16 09 06.21  & -39 08 51.8 &  M4.5 &  3197 & 0.50  & 0.175 &   0.17  &  -2.7 &   -9.19 &  200 \\
     Lup818s &  16 09 56.29  & -38 59 51.7 &  M6.0 &  2990 & 0.00  & 0.025 &   0.08  &  -4.1 &  -10.63 &  200 \\
      Sz123A &  16 10 51.34  & -38 53 14.6 &	M1 &  3705 & 1.25  & 0.203 &   0.60  &  -1.8 &   -8.93 &  200 \\
      Sz123B &  16 10 51.31  & -38 53 12.8 &  M2.0 &  3560 & 0.00  & 0.051 &   0.50  &  -2.7 &  -10.03 &  200 \\
\medskip 
  SST-Lup3-1 &  16 11 59.81  & -38 23 38.5 &  M5.0 &  3125 & 0.00  & 0.059 &   0.13  &  -3.6 &  -10.17 &  200 \\

       SO397 &  05 38 13.18  & -02 26 08.6 &  M4.5 &  3200 & 0.00  & 0.19  &    0.20  &  -2.71   &  -9.42	 & 360 \\
       SO490 &  05 38 23.58  & -02 20 47.5 &  M5.5 &  3060 & 0.00  & 0.08  &    0.14  &  -3.10   &  -9.97	 & 360 \\
       SO500 &  05 38 25.41  & -02 42 41.2 &    M6 &  2990 & 0.00  & 0.02  &    0.08  &  -3.95   & -10.27	 & 360 \\
       SO587 &  05 38 34.04  & -02 36 37.3 &  M4.5 &  3200 & 0.00  & 0.28  &    0.20  & $<$-4    & $<$-10.41 & 360 \\
       SO646 &  05 38 39.01  & -02 45 32.0 &  M3.5 &  3350 & 0.00  & 0.10  &    0.30  &  -3.00   &  -9.68	 & 360 \\
       SO848 &  05 39 01.94  & -02 35 02.8 &    M4 &  3270 & 0.00  & 0.02  &    0.19  &  -3.50   & -10.39	 & 360 \\
      SO1260 &  05 39 53.63  & -02 33 42.9 &    M4 &  3270 & 0.00  & 0.13  &    0.26  &  -2.00   &  -8.97	 & 360 \\
      SO1266 &  05 39 54.22  & -02 27 32.9 &  M4.5 &  3200 & 0.00  & 0.06  &    0.20  & $<$-4.85 & $<$-11.38 & 360 \\
\hline
        \end{tabular}
\end{table*}

\newpage
%
%\begin{figure}
%	\begin{center}
%		\includegraphics[width=9cm]{Sz100_profili}
%	\end{center}
%	\caption{Same as Fig.~\ref{fig_Sz69} for Sz100. The \OIA\ has an extended blue wing 
%	        that is clearly seen also in the \SIIA\ line and, elbeith noisy, in the \SIIB, 
%		which has also a LVC component. The \NII\ and \OII\ blueshifted components 
%		are also detected. What is the redshifted \OII? }
%\label{fig_Sz100}
%\end{figure}
%
%\begin{figure}
%	\begin{center}
%		\includegraphics[width=9cm]{Sz114_profili}
%	\end{center}
%	\caption{Same as Fig.~\ref{fig_Sz69} for Sz114. Two components, a LVC and a blueshifted 
%	         one are clearly detected in both \OIA\ and \SIIA. No detections of \SIIB, \NII, \OII.} 
%\label{fig_Sz114}
%\end{figure}
%
%\begin{figure}
%	\begin{center}
%		\includegraphics[width=9cm]{Sz123A_profili}
%	\end{center}
%	\caption{Same as Fig.~\ref{fig_Sz69} for Sz123A. The  LVC has an extended blue wing 
%	        in both  \OIA\ and \SIIA. Blue-shifted emission in \NII\ is detected at 
%		3$\sigma$ level; no detections of \SIIB\ and \OII.} 
%\label{fig_Sz123A}
%\end{figure}
%
%\begin{figure}
%	\begin{center}
%		\includegraphics[width=9cm]{Sz123B_profili}
%	\end{center}
%	\caption{Same as Fig.~\ref{fig_Sz69} for Sz123B. The \OIA\ has an extended blue wing 
%	        that is clearly seen as a second component in the \SIIA line. Blue-shifted 
%		emission in \NII\ is detected at 3$\sigma$ level; no detections of \SIIB\ and \OII.} 
%\label{fig_Sz123B}
%\end{figure}
%

\section {Line properties}

Tables \ref{tab_tabellone_1}, \ref{tab_tabellone_2},
\ref{tab_tabellone_3} and \ref{tab_tabellone_4} list the properties of the LVC, HVC-blue shifted and HVC-red shifted components of the lines \OIB, \OIA, \OII,
\OIIB, \SIIA, \SIIB, and \NII. For each component the tables give the intensity,
the uncertainty, the velocity shift of the line peak and the FWHM of the observed component. When not detected, we give $3\sigma$ upper limit for the LVC
only.

\onecolumn

%%%%%%%%%%%%%%%%%%%%%%%%%%%%%%%%%%%%%%%%%%%%%%%%%%%%%%
\setlength{\tabcolsep}{5pt}

\begin{landscape}
\scriptsize
% \begin{longtable}{l|c|c|c|c|c|c|c|c|c|c|c|c}
\begin{longtable}{l ||c c c|| c c c| c c c| c c c}
\caption[ ]{\label{tab_tabellone_1} Fluxes, velocity peak and full-width half maximun of [\ion{O}{i}] lines}\\
\hline
   name &   \multicolumn{3}{c||}{\OIB} &   \multicolumn{9}{c}{\OIA}  \\ \cline{2-13}
        &     \multicolumn{3}{c||}{LVC}                   &        \multicolumn{3}{c|}{LVC}  & \multicolumn{3}{c|}{HVC-Blue}  &  \multicolumn{3}{c}{HVC-Red}  \\ \cline{2-13}
              &   flux                    &V$_{\rm peak}$ & FWHM       &      flux  & V$_{\rm peak}$    & FWHM   &	 flux        & V$_{\rm peak}$  &  FWHM  &  flux  &  V$_{\rm peak}$  &  FWHM  \\
              &(erg\,s$^{-1}$\,cm$^{-2}$) & (km\,s$^{-1}$) & (km\,s$^{-1}$) &	(erg\,s$^{-1}$\,cm$^{-2}$)  &	(km\,s$^{-1}$)  & (km\,s$^{-1}$)  & (erg\,s$^{-1}$\,cm$^{-2}$)  & (km\,s$^{-1}$) & (km\,s$^{-1}$)    &	 (erg\,s$^{-1}$\,cm$^{-2}$) & (km\,s$^{-1}$) & (km\,s$^{-1}$)	  \\
%              &                     &              &	    	&			 &	   &	    &			    &		&	    &			    &	      & 	  \\
\hline
        Sz66  & 3.68($\pm$0.20)e-15 &  -13.2       &  72.4  	&   2.25($\pm$0.06)e-14  &  -19.4  &  48.6  &	   ...  	    &  ...	&  ...      &	...		    &  ...    &   ...	  \\
   AKC2006-1  & $<$1.50e-17	    &  ...         & ...    	&   $<$2.85e-17 	 &   ...   & ...    &	   ...  	    &  ...	&  ...      &	...		    &  ...    &   ...	  \\
        Sz69  & 2.08($\pm$0.16)e-15 &  -10.9       & 59$^\dagger$  	&   6.09($\pm$0.20)e-15  &   -7.7  &  55.5  &  3.44($\pm$0.10)e-15  &	-92.9	&   73.5    &  1.34($\pm$0.50)e-15  &  58.8   &   48.3    \\
        Sz71  & 2.60($\pm$0.50)e-15 &  -13.9       &  68.3  	&   3.00($\pm$0.10)e-15  &   -1    &  97.5  &	   ...  	    &  ...	&  ...      &	...		    &  ...    &   ...	  \\
        Sz72  & 1.80($\pm$0.60)e-15 &  -15.2       &  72.1  	&   2.80($\pm$0.40)e-15  &  -10.8  &  42.2  &  4.57($\pm$0.70)e-15  &  -125.0	&   72.3    &	 ...		    &  ...    &   ...	  \\
        Sz73  & 7.10($\pm$0.25)e-15 &  -27.5       & 59$^\dagger$  	&   4.15($\pm$0.02)e-14  &  -18.9  &  60.8  &  9.32($\pm$0.04)e-14  &	-94.2	&   59.1    &	 ...		    &  ...    &   ...	  \\
        Sz74  & 1.25($\pm$0.30)e-14 &	 7.3       &  76.5  	&   2.13($\pm$0.25)e-14  &   -7.5  &  40.5  &	   ...  	    &  ...	&  ...      &	...		    &  ...    &   ...	  \\
        Sz83  &     ...	            &  ...         & ...    	&     ...		 &   ...   & ...    &	   ...  	    &  ...	&  ...      &	...		    &  ...    &   ...	  \\
        Sz84  & 3.20($\pm$0.80)e-16 &	-8.4       &  76.7  	&   1.45($\pm$0.15)e-15  &   -7.4  &  41.4  &	   ...  	    &  ...	&  ...      &	...		    &  ...    &   ...	  \\
       Sz130  & 1.33($\pm$0.18)e-15 &	-8.1       &  74.9  	&   2.14($\pm$0.30)e-15  &   17.7  &  72.9  &  4.79($\pm$0.60)e-15  &	-49.3	&   57.4    &	 ...		    &  ...    &   ...	  \\
       Sz88A  & 6.07($\pm$0.30)e-15 &	-1.9       &  62.7  	&   1.34($\pm$0.10)e-14  &   -1.7  &  53.5  &	   ...  	    &  ...	&  ...      &	...		    &  ...    &   ...	  \\
       Sz88B  & 2.04($\pm$0.40)e-16 &	-7.1       &  67.9  	&   6.07($\pm$0.80)e-16  &   -3.9  &  55.6  &	   ...  	    &  ...	&  ...      &	...		    &  ...    &   ...	  \\
        Sz91  & 1.65($\pm$0.30)e-15 &	-2.3       &  70.2  	&   4.00($\pm$0.48)e-15  &   -3.4  & 34$^\dagger$&	   ...  	    &  ...	&  ...      &	...		    &  ...    &   ...	  \\
      Lup713  & $<$8.00e-17	    &   ...        & ...    	&   $<$4.30e-17 	 &    4.2  & 34$^\dagger$&	   ...  	    &  ...	&  ...      &	...		    &  ...    &   ...	  \\
     Lup604s  & $<$1.60e-16	    &   ...        & ...    	&   1.58($\pm$0.16)e-16  &   -2.5  &  50.1  &	   ...  	    &  ...	&  ...      &	...		    &  ...    &   ...	  \\
        Sz97  & $<$2.50e-16	    &   ...        & ...    	&   2.83($\pm$0.80)e-16  &  -36.4  &  58.4  &	   ...  	    &  ...	&  ...      &	...		    &  ...    &   ...	  \\
        Sz99  & 8.60($\pm$0.80)e-16 &  -10.0       &  92.0  	&   1.36($\pm$0.09)e-15  &    3.3  &  77.2  &	   ...  	    &  ...	&  ...      & 2.20($\pm$0.60)e-16   &  72.8   &   51.0    \\
       Sz100  & 9.01($\pm$0.90)e-16 &  -16.0       &  73.6  	&   3.24($\pm$0.20)e-15  &   -6.6  &  43.7  &  3.98($\pm$0.20)e-15  &	-54.6	&   84.6    & 4.06($\pm$1.30)e-16   &  56.2   &   50.2    \\
       Sz103  & 6.24($\pm$0.90)e-16 &	-9.0       &  73.3  	&   3.70($\pm$0.30)e-15  &  -32.4  &  54.9  &	   ...  	    &  ...	&  ...      &	...		    &  ...    &   ...	  \\
       Sz104  & 1.74($\pm$0.50)e-16 &	-6.0       & 59$^\dagger$	&   4.18($\pm$0.50)e-16  &   -6.8  &  40.1  &	   ...  	    &  ...	&  ...      &	...		    &  ...    &   ...	  \\
      Lup706  & $<$6.50e-17	    &   ...        & ...    	&   $<$1.74e-17 	 &  ...    & ...    &	   ...  	    &  ...	&  ...      &	...		    &  ...    &   ...	  \\
       Sz106  & 4.14($\pm$0.50)e-16 &  -14.5       &  70.2  	&   2.00($\pm$0.20)e-15  &   -9.6  & 108.1  &	   ...  	    &  ...	&  ...      &	...		    &  ...    &   ...	  \\
  Par-Lup3-3  & $<$3.00e-15	    &   ...        & ...    	&   3.41($\pm$0.50)e-15  &   -1.4  &  64.8  &	   ...  	    &  ...	&  ...      &	...		    &  ...    &   ...	  \\
  Par-Lup3-4  & 2.38($\pm$0.22)e-16 &	-4.6       &  78.2  	&   4.69($\pm$0.50)e-15  &    0.6  &  65.6  &	   ...  	    &  ...	&  ...      &	...		    &  ...    &   ...	  \\
       Sz110  & 1.40($\pm$0.28)e-15 &	 4.9       &  98.3  	&   1.65($\pm$0.25)e-15  &   -7.1  &  66.7  &	   ...  	    &  ...	&  ...      &	...		    &  ...    &   ...	  \\
       Sz111  & 2.20($\pm$0.60)e-15 &	 0.6       &  63.6  	&   8.06($\pm$0.80)e-15  &   -2.9  &  46.8  &	   ...  	    &  ...	&  ...      &	...		    &  ...    &   ...	  \\
       Sz112  & 7.19($\pm$1.00)e-16 &	 0.2       &  68.2  	&   1.94($\pm$0.15)e-15  &   -2.4  &  49.6  &	   ...  	    &  ...	&  ...      &	...		    &  ...    &   ...	  \\
       Sz113  & 3.00($\pm$1.00)e-16 &  -10.0       &  66.0    	&   6.37($\pm$0.60)e-16  &  -12.3  &  34$^\dagger$ &  2.17($\pm$0.15)e-15  &  -122.6	&   54.6    &	 ...		    &  ...    &   ...	  \\
     J160859  & 4.57($\pm$1.00)e-17 &  -40.6       & 59$^\dagger$  	&   2.69($\pm$0.30)e-16  &  -16.8  &  58.3  &	   ...  	    &  ...	&  ...      &	...		    &  ...    &   ...	  \\
    c2dJ1609  & 7.75($\pm$1.00)e-16 &	-9.1       &  81.2   	&   3.18($\pm$0.40)e-15  &   -9.9  &  74.8  &	   ...  	    &  ...	&  ...      &	...		    &  ...    &   ...	  \\
       Sz114  & 8.95($\pm$1.50)e-16 &	 0.5       &  48.2  	&   1.59($\pm$0.20)e-15  &  -10.1  &  43.9  &  1.32($\pm$0.20)e-15  &	-93.3	&   77.3    &	 ...		    &  ...    &   ...	  \\
       Sz115  & $<$7.00e-16	    &   ...        & ...    	&   $<$6.00e-16 	 &   ...   & ...    &	   ...  	    &  ...	&  ...      &	...		    &  ...    &   ...	  \\
     Lup818s  & 5.00($\pm$1.70)e-17 &  -40.0       & 59$^\dagger$  	&   1.30($\pm$0.17)e-16  &   -7.2  &  43.3  &	   ...  	    &  ...	&  ...      &	...		    &  ...    &   ...	  \\
      Sz123A  & 2.63($\pm$0.80)e-15 &  -14.4       &  81.6  	&   6.96($\pm$0.40)e-15  &  -10.1  &  64.8  &  2.39($\pm$0.15)e-15  &	-61.9	&   61.4    &	 ...		    &  ...    &   ...	  \\
      Sz123B  & 4.46($\pm$0.40)e-16 &	-6.8       &  78.2  	&   1.29($\pm$0.08)e-15  &  -10.6  &  66.2  &  4.15($\pm$1.00)e-16  &	-69.6	&   65.8    &	 ...		    &  ...    &   ...	  \\
  SST-Lup3-1  & $<$1.40e-16	    &   ...        & ...    	&   $<$1.00e-16 	 &   ...   & ...    &	   ...  	    &  ...	&  ...      &	...		    &  ...    &   ...	  \\
       SO397  & $<$4.00e-17	    &   ...        & ...    	&   $<$1.00e-16 	 &   ...   & ...    &	   ...  	    &  ...	&  ...      &	...		    &  ...    &   ...	  \\
       SO490  & 3.30($\pm$0.40)e-17 &  -21.0       &  61.9  	&   5.15($\pm$1.00)e-17  &  -22.7  &  62.8  &	   ...  	    &  ...	&  ...      &	...		    &  ...    &   ...	  \\
       SO500  & $<$2.00e-17	    &   ...        & ...    	&   2.40($\pm$0.70)e-17  &  -21.2  &  80.9  &  3.53($\pm$1.00)e-18  &	-70.3	&   16.8    &	 ...		    &  ...    &   ...	  \\
       SO587  & $<$1.70e-16	    &   ...        & ...    	&   1.43($\pm$0.07)e-15  &   -7.5  &  56.0  &	   ...  	    &  ...	&  ...      &	...		    &  ...    &   ...	  \\
       SO646  & $<$6.40e-17	    &   ...        & ...    	&   6.00($\pm$0.60)e-16  &   -9.3  &  71.8  &	   ...  	    &  ...	&  ...      &	...		    &  ...    &   ...	  \\
       SO848  & 6.30($\pm$0.40)e-17 &	-2.5       &  63.9  	&   4.61($\pm$0.15)e-16  &   -8.5  &  42.2  &  3.60($\pm$0.10)e-16  &	-51.8	&   48.6    &	 ...		    &  ...    &   ...	  \\
      SO1260  & 2.60($\pm$0.30)e-16 &	-5.8       & 59$^\dagger$  	&   4.20($\pm$0.35)e-16  &   -6.6  &  41.7  &	   ...  	    &  ...	&  ...      &	...		    &  ...    &   ...	  \\
      SO1266  & 2.80($\pm$1.00)e-17 &	-9.1       & 59$^\dagger$  	&   1.88($\pm$0.20)e-16  &   -7.5  &  49.2  &	   ...  	    &  ...	&  ...      &	...		    &  ...    &   ...	  \\
\hline			    			     	  
\end{longtable}
\footnotesize{
%Notes:
\begin{itemize}
\item $^\dagger$ : Not resolved line; the FWHM is the instrumental resolution (e.g., 59 km/s for the \OIB\ line  and 34 km/s for the \OIA\ line).
\end{itemize}
}
\end{landscape}
%%%%%%%%%%%%%%%%%%%%%%%%%%%%%%%%%%%%%%%%%%%%%%%%%%%%%%%%%%%%%%%%%%%%%%%%%%

%%%%%%%%%%%%%%%%%%%%%%%%%%%%%%%%%%%%%%%%%%%%%%%%%%%%%%
\setlength{\tabcolsep}{3pt}

\begin{landscape}
\scriptsize
\begin{longtable}{l|| c c c | c c c | c c c || c c c | c c c| c c c}
\caption[ ]{\label{tab_tabellone_2} Fluxes, velocity peak and full-width half maximun of [\ion{O}{ii}] lines}\\
\hline
   name &   \multicolumn{9}{c||}{\OII} &   \multicolumn{9}{c}{\OIIB}  \\ \cline{2-19}
         &     \multicolumn{3}{c|}{LVC}                 &    \multicolumn{3}{c|}{HVC-Blue} & \multicolumn{3}{c||}{HVC-Red}  &  \multicolumn{3}{c|}{LVC} & \multicolumn{3}{c|}{HVC-Blue} & \multicolumn{3}{c}{HVC-Red}  \\ \cline{2-19}
              &   flux                 &V$_{\rm peak}$ & FWHM       &   flux  & V$_{\rm peak}$  & FWHM   & flux    & V$_{\rm peak}$  &  FWHM  &  flux  &  V$_{\rm peak}$  &  FWHM & flux    & V$_{\rm peak}$  &  FWHM  &  flux  &  V$_{\rm peak}$  &  FWHM  \\
              &(erg\,s$^{-1}$\,cm$^{-2}$) & (km\,s$^{-1}$) & (km\,s$^{-1}$) &	(erg\,s$^{-1}$\,cm$^{-2}$)  &	(km\,s$^{-1}$)  & (km\,s$^{-1}$)  & (erg\,s$^{-1}$\,cm$^{-2}$)  & (km\,s$^{-1}$) & (km\,s$^{-1}$)    &	 (erg\,s$^{-1}$\,cm$^{-2}$) & (km\,s$^{-1}$) & (km\,s$^{-1}$)	& (erg\,s$^{-1}$\,cm$^{-2}$)  & (km\,s$^{-1}$) & (km\,s$^{-1}$)    &	 (erg\,s$^{-1}$\,cm$^{-2}$) & (km\,s$^{-1}$) & (km\,s$^{-1}$)  \\
\hline
              &                        &           &	      & 			&	  &	      & 		      & 	 &	      & 			&	      & 	    &			 &		 &		 &	   &		 &	       \\ 
        Sz66  &   4.27($\pm$1.00)e-16  &   -36.4   & 	62.8  &       ...		& ...	  &	 ...  &       ...	      &  ...	 &	...   & 	$<$1.91e-15	&    ...      &       ...   &	   ...  	 &    ...	 &     ...	 &  ...     &	   ...    &	    ...  \\   
   AKC2006-1  & 	$<$1.50e-17    &   ...     & 	 ...  &       ...		& ...	  &	 ...  &       ...	      &  ...	 &	...   & 	$<$9.42e-17	&    ...      &       ...   &	   ...  	 &    ...	 &     ...	 &  ...     &	   ...    &	    ...  \\   
        Sz69  & 	$<$2.00e-16    &   ...     & 	 ...  &       ...		& ...	  &	 ...  &  4.30($\pm$1.20)e-16  &  107.5   &     76.6   & 	$<$2.08e-16	&    ...      &       ...   &	   ...  	 &    ...	 &     ...	 &  ...     &	   ...    &	    ...  \\   
        Sz71  & 	$<$1.90e-16    &   ...     & 	 ...  &       ...		& ...	  &	 ...  &       ...	      &  ...	 &	...   & 	$<$1.91e-15	&    ...      &       ...   &	   ...  	 &    ...	 &     ...	 &  ...     &	   ...    &	    ...  \\   
        Sz72  & 	$<$1.50e-15    &   ...     & 	 ...  &       ...		& ...	  &	 ...  &  4.20($\pm$0.50)e-15  &  101.7   &     86.4   & 	$<$1.70e-15	&    ...      &       ...   &	   ...  	 &    ...	 &     ...	 &  ...     &	   ...    &	    ...  \\   
        Sz73  & 	$<$4.00e-15    &   ...     & 	 ...  &   6.00($\pm$1.80)e-15	& -96.1   &	76.9  &       ...	      &  ...	 &	...   &   3.00($\pm$0.50)e-15	&   -10.0     &      45.0   & 7.6($\pm$0.50)e-15 &   -109.0	 &	37.0	 &  ...     &	   ...    &	    ...  \\   
        Sz74  & 	$<$1.60e-15    &   ...     & 	 ...  &       ...		& ...	  &	 ...  &       ...	      &  ...	 &	...   &   $<$6.00e-15	&   ...     &      ...   &	   ...  	 &    ...	 &     ...	 &  ...     &	   ...    &	    ...  \\   
        Sz83  & 		...    &   ...     & 	 ...  &       ...		& ...	  &	 ...  &       ...	      &  ...	 &	...   & 	  ...	       &     ...      &       ...   &	   ...  	 &    ...	 &     ...	 &  ...     &	   ...    &	    ...  \\   
        Sz84  &   7.69($\pm$2.5)e-17   &   -16.8   & 	59$^\dagger$&       ...		& ...	  &	 ...  &       ...	      &  ...	 &	...   & 	$<$7.72e-16	&    ...      &       ...   &	   ...  	 &    ...	 &     ...	 &  ...     &	   ...    &	    ...  \\   
       Sz130  & 	$<$6.00e-16    &   ...     & 	 ...  &       ...		& ...	  &	 ...  &       ...	      &  ...	 &	...   & 	$<$8.33e-16	&    ...      &       ...   &	   ...  	 &    ...	 &     ...	 &  ...     &	   ...    &	    ...  \\   
       Sz88A  & 	$<$3.70e-15    &   ...     & 	 ...  &       ...		& ...	  &	 ...  &       ...	      &  ...	 &	...   &   $<$1.00e-15	&   ...     &      ...   &	   ...  	 &    ...	 &     ...	 &  ...     &	   ...    &	    ...  \\   
       Sz88B  & 	$<$4.00e-17    &   ...     & 	 ...  &       ...		& ...	  &	 ...  &       ...	      &  ...	 &	...   & 	$<$5.08e-16	&    ...      &       ...   &	   ...  	 &    ...	 &     ...	 &  ...     &	   ...    &	    ...  \\   
        Sz91  & 	$<$1.70e-16    &   ...     & 	 ...  &       ...		& ...	  &	 ...  &       ...	      &  ...	 &	...   & 	$<$1.61e-15	&    ...      &       ...   &	   ...  	 &    ...	 &     ...	 &  ...     &	   ...    &	    ...  \\   
      Lup713  & 	$<$1.60e-17    &   ...     & 	 ...  &       ...		& ...	  &	 ...  &       ...	      &  ...	 &	...   & 	$<$4.01e-17	&    ...      &       ...   &	   ...  	 &    ...	 &     ...	 &  ...     &	   ...    &	    ...  \\   
     Lup604s  & 	$<$3.10e-17    &   ...     & 	 ...  &       ...		& ...	  &	 ...  &       ...	      &  ...	 &	...   & 	$<$1.29e-16	&    ...      &       ...   &	   ...  	 &    ...	 &     ...	 &  ...     &	   ...    &	    ...  \\   
        Sz97  & 	$<$1.00e-16    &   ...     & 	 ...  &       ...		& ...	  &	 ...  &       ...	      &  ...	 &	...   & 	$<$4.17e-16	&    ...      &       ...   &	   ...  	 &    ...	 &     ...	 &  ...     &	   ...    &	    ...  \\   
        Sz99  & 		...    &   ...     & 	 ...  &       ...		& ...	  &	 ...  &  3.12($\pm$0.50)e-16  &  83.1	 &     73.5   & 	$<$1.61e-16	&    ...      &       ...   &	   ...  	 &    ...	 &     ...	 &  ...     &	   ...    &	    ...  \\   
       Sz100  & 		...    &   ...     & 	 ...  &   2.99($\pm$0.70)e-16	& -64.0   &	59$^\dagger$ &       ...	      &  ...	 &	...   & 	$<$5.29e-16	&    ...      &       ...   &	   ...  	 &    ...	 &     ...	 &  ...     &	   ...    &	    ...  \\   
       Sz103  & 	$<$1.50e-16    &   ...     & 	 ...  &       ...		& ...	  &	 ...  &       ...	      &  ...	 &	...   & 	$<$6.65e-16	&    ...      &       ...   &	   ...  	 &    ...	 &     ...	 &  ...     &	   ...    &	    ...  \\   
       Sz104  & 	$<$4.80e-17    &   ...     & 	 ...  &       ...		& ...	  &	 ...  &       ...	      &  ...	 &	...   & 	$<$4.01e-16	&    ...      &       ...   &	   ...  	 &    ...	 &     ...	 &  ...     &	   ...    &	    ...  \\   
      Lup706  & 	$<$1.00e-17    &   ...     & 	 ...  &       ...		& ...	  &	 ...  &       ...	      &  ...	 &	...   & 	$<$1.19e-17	&    ...      &       ...   &	   ...  	 &    ...	 &     ...	 &  ...     &	   ...    &	    ...  \\   
       Sz106  &   4.96($\pm$0.30)e-16  &   -0.9    & 	 120.0 &       ...		& ...	  &	 ...  &       ...	      &  ...	 &	...   &   $<$2.28e-16	&  ...      &      ...   &	   ...  	 &    ...	 &     ...	 &  ...     &	   ...    &	    ...  \\   
  Par-Lup3-3  & 	$<$6.00e-16    &   ...     & 	 ...  &       ...		& ...	  &	 ...  &       ...	      &  ...	 &	...   & 	$<$6.25e-16	&    ...      &       ...   &	   ...  	 &    ...	 &     ...	 &  ...     &	   ...    &	    ...  \\   
  Par-Lup3-4  &   3.01($\pm$1.00)e-17  &   28.0    & 	 92.4 &       ...		& ...	  &	 ...  &       ...	      &  ...	 &	...   &   5.00($\pm$1.30)e-17	&    0.0     &      69.0   &	   ...  	 &    ...	 &     ...	 &  ...     &	   ...    &	    ...  \\   
       Sz110  & 	$<$3.50e-16    &   ...     & 	 ...  &       ...		& ...	  &	 ...  &       ...	      &  ...	 &	...   & 	$<$1.09e-15	&    ...      &       ...   &	   ...  	 &    ...	 &     ...	 &  ...     &	   ...    &	    ...  \\   
       Sz111  & 	$<$1.20e-16    &   ...     & 	 ...  &       ...		& ...	  &	 ...  &       ...	      &  ...	 &	...   & 	$<$1.33e-15	&    ...      &       ...   &	   ...  	 &    ...	 &     ...	 &  ...     &	   ...    &	    ...  \\   
       Sz112  & 	$<$5.30e-17    &   ...     & 	 ...  &       ...		& ...	  &	 ...  &       ...	      &  ...	 &	...   & 	$<$1.19e-15	&    ...      &       ...   &	   ...  	 &    ...	 &     ...	 &  ...     &	   ...    &	    ...  \\   
       Sz113  & 	$<$4.00e-15    &   ...     & 	 ...  &       ...		& ...	  &	 ...  &       ...	      &  ...	 &	...   & 	$<$2.26e-16	&    ...      &       ...   &	   ...  	 &    ...	 &     ...	 &  ...     &	   ...    &	    ...  \\   
     J160859  & 	$<$2.00e-17    &   ...     & 	 ...  &       ...		& ...	  &	 ...  &       ...	      &  ...	 &	...   & 	$<$2.45e-17	&    ...      &       ...   &	   ...  	 &    ...	 &     ...	 &  ...     &	   ...    &	    ...  \\   
    c2dJ1609  & 	$<$2.00e-16    &   ...     & 	 ...  &       ...		& ...	  &	 ...  &       ...	      &  ...	 &	...   & 	$<$6.25e-16	&    ...      &       ...   &	   ...  	 &    ...	 &     ...	 &  ...     &	   ...    &	    ...  \\   
       Sz114  & 	$<$1.10e-16    &   ...     & 	 ...  &       ...		& ...	  &	 ...  &       ...	      &  ...	 &	...   & 	$<$1.75e-15	&    ...      &       ...   &	   ...  	 &    ...	 &     ...	 &  ...     &	   ...    &	    ...  \\   
       Sz115  & 	$<$1.00e-16    &   ...     & 	 ...  &       ...		& ...	  &	 ...  &       ...	      &  ...	 &	...   & 	$<$5.94e-16	&    ...      &       ...   &	   ...  	 &    ...	 &     ...	 &  ...     &	   ...    &	    ...  \\   
     Lup818s  & 	$<$3.30e-17    &   ...     & 	 ...  &       ...		& ...	  &	 ...  &       ...	      &  ...	 &	...   & 	$<$7.57e-17	&    ...      &       ...   &	   ...  	 &    ...	 &     ...	 &  ...     &	   ...    &	    ...  \\   
      Sz123A  & 	$<$1.20e-15    &   ...     & 	 ...  &       ...		& ...	  &	 ...  &       ...	      &  ...	 &	...   & 	$<$1.32e-15	&    ...      &       ...   &	   ...  	 &    ...	 &     ...	 &  ...     &	   ...    &	    ...  \\   
      Sz123B  & 	$<$1.20e-16    &   ...     & 	 ...  &       ...		& ...	  &	 ...  &       ...	      &  ...	 &	...   & 	$<$3.36e-16	&    ...      &       ...   &	   ...  	 &    ...	 &     ...	 &  ...     &	   ...    &	    ...  \\   
  SST-Lup3-1  & 	$<$3.00e-17    &   ...     & 	 ...  &       ...		& ...	  &	 ...  &       ...	      &  ...	 &	...   & 	$<$2.62e-16	&    ...      &       ...   &	   ...  	 &    ...	 &     ...	 &  ...     &	   ...    &	    ...  \\   
       SO397  & 	$<$1.60e-17    &   ...     & 	 ...  &       ...		& ...	  &	 ...  &       ...	      &  ...	 &	...   & 	$<$1.90e-16	&    ...      &       ...   &	   ...  	 &    ...	 &     ...	 &  ...     &	   ...    &	    ...  \\   
       SO490  & 	$<$9.10e-18    &   ...     & 	 ...  &       ...		& ...	  &	 ...  &       ...	      &  ...	 &	...   & 	$<$6.35e-17	&    ...      &       ...   &	   ...  	 &    ...	 &     ...	 &  ...     &	   ...    &	    ...  \\   
       SO500  & 	$<$1.00e-17    &   ...     & 	 ...  &       ...		& ...	  &	 ...  &       ...	      &  ...	 &	...   & 	$<$1.04e-17	&    ...      &       ...   &	   ...  	 &    ...	 &     ...	 &  ...     &	   ...    &	    ...  \\   
       SO587  &   2.43($\pm$0.04)e-15  &   -6.5    & 	 63.3 &       ...		& ...	  &	 ...  &       ...	      &  ...	 &	...   & 	$<$1.39e-16	&    ...      &       ...   &	   ...  	 &    ...	 &     ...	 &  ...     &	   ...    &	    ...  \\   
       SO646  & 	$<$2.40e-17    &   ...     & 	 ...  &       ...		& ...	  &	 ...  &       ...	      &  ...	 &	...   & 	$<$1.15e-16	&    ...      &       ...   &	   ...  	 &    ...	 &     ...	 &  ...     &	   ...    &	    ...  \\   
       SO848  & 	$<$1.50e-16    &   ...     & 	 ...  &   9.42($\pm$0.50)e-16	& -40.4   &	79.0  &       ...	      &  ...	 &	...   &   1.60($\pm$0.18)e-16	&   -26.0     &      69.0   &	   ...  	 &    ...	 &     ...	 &  ...     &	   ...    &	    ...  \\   
   %   SO1260  &   4.54($\pm$1.50)e-17  &   36.8    & 	$<$38 &       ...		& ...	  &	 ...  &       ...	      &  ...	 &	...   & 	$<$1.86e-16	&    ...      &       ...   &	   ...  	 &    ...	 &     ...	 &  ...     &	   ...    &	    ...  \\   
      SO1260  &                 ...    &   ...     & 	 ...  &       ...		& ...	  &	 ...  &       ...	      &  ...	 &	...   & 	$<$1.86e-16	&    ...      &       ...   &	   ...  	 &    ...	 &     ...	 &  ...     &	   ...    &	    ...  \\   
      SO1266  & 		...    &   ...     & 	 ...  &   1.04($\pm$0.24)e-16	& -64.0   &	59$^\dagger$ &       ...	      &  ...	 &	...   & 	$<$9.67e-17	&    ...      &       ...   &	   ...  	 &    ...	 &     ...	 &  ...     &	   ...    &	    ...  \\   
\hline			    			     	  
\end{longtable}
\footnotesize{
%Notes:
\begin{itemize}
\item $^\dagger$ : Not resolved line; the FWHM is the instrumental resolution (e.g., 59 km/s for the \OII\ line  and 34 km/s for the \OIIB\ line).
\end{itemize}
}
\end{landscape}
%%%%%%%%%%%%%%%%%%%%%%%%%%%%%%%%%%%%%%%%%%%%%%%%%%%%%%%%%%%%%%%%%%%%%%%%%%

%%%%%%%%%%%%%%%%%%%%%%%%%%%%%%%%%%%%%%%%%%%%%%%%%%%%%%
\setlength{\tabcolsep}{3pt}

\begin{landscape}
\scriptsize
\begin{longtable}{l|| c c c | c c c | c c c || c c c | c c c| c c c}
\caption[ ]{\label{tab_tabellone_3} Fluxes, velocity peak and full-width half maximun of [\ion{S}{ii}] lines}\\
\hline
   name &   \multicolumn{9}{c||}{\SIIA} &   \multicolumn{9}{c}{\SIIB}  \\ \cline{2-19}
              &     \multicolumn{3}{c|}{LVC}                 &    \multicolumn{3}{c|}{HVC-Blue} & \multicolumn{3}{c||}{HVC-Red}  &  \multicolumn{3}{c|}{LVC} & \multicolumn{3}{c|}{HVC-Blue} & \multicolumn{3}{c}{HVC-Red}  \\ \cline{2-19}
              &   flux                 &V$_{\rm peak}$ & FWHM       &   flux  & V$_{\rm peak}$ & FWHM   & flux    & V$_{\rm peak}$  &  FWHM  &  flux  &  V$_{\rm peak}$  &  FWHM & flux    & V$_{\rm peak}$  &  FWHM  &  flux  &  V$_{\rm peak}$  &  FWHM  \\
              &(erg\,s$^{-1}$\,cm$^{-2}$) & (km\,s$^{-1}$) & (km\,s$^{-1}$) &	(erg\,s$^{-1}$\,cm$^{-2}$)  &	(km\,s$^{-1}$)  & (km\,s$^{-1}$)  & (erg\,s$^{-1}$\,cm$^{-2}$)  & (km\,s$^{-1}$) & (km\,s$^{-1}$)    &	 (erg\,s$^{-1}$\,cm$^{-2}$) & (km\,s$^{-1}$) & (km\,s$^{-1}$)	& (erg\,s$^{-1}$\,cm$^{-2}$)  & (km\,s$^{-1}$) & (km\,s$^{-1}$)    &	 (erg\,s$^{-1}$\,cm$^{-2}$) & (km\,s$^{-1}$) & (km\,s$^{-1}$)  \\
\hline
              &                        &           &             &                         &         &           &                       &          &            &                         &             &             &                    &               &               &         & 	    &		  \\ 
        Sz66  & 5.63($\pm$0.40)e-15	& -22.7    &  60.0  &	       ...	     &   ...	 &    ...   &	   ...         &   ...    &	...  &   6.89($\pm$2.00)e-16   & -19.5   &    38.2  &	    ... 	    &  ...    &     ...  &	 ...		& ...	 &     ...    \\
   AKC2006-1  &       $<$2.00e-17	& ...	   &   ...  &	       ...	     &	 ...	 &    ...   &	   ...         &   ...    &	...  &	    $<$2.30e-17        &  ...	 &     ...  &	    ... 	    &  ...    &     ...  &	 ...		& ...	 &     ...    \\
        Sz69  & 1.27($\pm$0.15)e-15	&  -8.4    &  81.0  &	6.80($\pm$0.80)e-16  &  -99.6	 &    67.5  &  5.50(2.00)e-17  &  80.1    &    22.0  &	    $<$4.60e-16        &  ...	 &     ...  &	2.69($\pm$0.50)e-16 & -103.2  &    51.6  &  3.88($\pm$0.70)e-16 & 62.2   &    37.5    \\
        Sz71  &       $<$2.10e-15	& ...	   &   ...  &	       ...	     &	 ...	 &    ...   &	   ...         &   ...    &	...  &	    $<$1.40e-15        &  ...	 &     ...  &	    ... 	    &  ...    &     ...  &	 ...		& ...	 &     ...    \\
        Sz72  &       $<$3.00e-15	& ...	   &   ...  &	3.20($\pm$1.00)e-15  & -133.2	 &   117.7  &	   ...         &   ...    &	...  &	    $<$5.00e-16        &  ...	 &     ...  &	1.50($\pm$0.30)e-15 & -132.2  &    40.8  &	 ...		& ...	 &     ...    \\
        Sz73  &       $<$1.00e-14	& ...	   &   ...  &	2.60($\pm$0.40)e-14  &  -96.3	 &    79.8  &	   ...         &   ...    &	...  &	    $<$1.00e-15        &  ...	 &     ...  &	8.40($\pm$2.00)e-15 &  -98.3  &    64.9  &	 ...		& ...	 &     ...    \\
        Sz74  & 3.00($\pm$0.60)e-15	& -15.8    &  59$^\dagger$ &	       ...	     &	 ...	 &    ...   &	   ...         &   ...    &	...  &	    $<$3.00e-15        &  ...	 &     ...  &	    ... 	    &  ...    &     ...  &	 ...		& ...	 &     ...    \\
        Sz83  &      ...		& ...	   &   ...  &	       ...	     &	 ...	 &    ...   &	   ...         &   ...    &	...  &	    ... 	       &  ...	 &     ...  &	    ... 	    &  ...    &     ...  &	 ...		& ...	 &     ...    \\
        Sz84  &       $<$1.30e-16	& ...	   &   ...  &	       ...	     &	 ...	 &    ...   &	   ...         &   ...    &	...  &	    $<$1.90e-16        &  ...	 &     ...  &	    ... 	    &  ...    &     ...  &	 ...		& ...	 &     ...    \\
       Sz130  &       $<$6.00e-16	& ...	   &   ...  &	6.40($\pm$2.00)e-16  &  -49.9	 &    73.8  &	   ...         &   ...    &	...  &	    $<$5.90e-16        &  ...	 &     ...  &	    ... 	    &  ...    &     ...  &	 ...		& ...	 &     ...    \\
       Sz88A  & 3.63($\pm$0.30)e-15	&   3.8    &  65.6  &	       ...	     &	 ...	 &    ...   &	   ...         &   ...    &	...  &	    $<$4.70e-16        &  ...	 &     ...  &	    ... 	    &  ...    &     ...  &	 ...		& ...	 &     ...    \\
       Sz88B  & 1.02($\pm$0.20)e-16	&  -4.2    &  65.5  &	       ...	     &	 ...	 &    ...   &	   ...         &   ...    &	...  &	    $<$1.40e-16        &  ...	 &     ...  &	    ... 	    &  ...    &     ...  &	 ...		& ...	 &     ...    \\
        Sz91  &       $<$1.00e-15	& ...	   &   ...  &	       ...	     &	 ...	 &    ...   &	   ...         &   ...    &	...  &	    $<$1.00e-15        &  ...	 &     ...  &	    ... 	    &  ...    &     ...  &	 ...		& ...	 &     ...    \\
      Lup713  &       $<$1.50e-17	& ...	   &   ...  &	       ...	     &	 ...	 &    ...   &	   ...         &   ...    &	...  &	    $<$1.50e-17        &  ...	 &     ...  &	    ... 	    &  ...    &     ...  &	 ...		& ...	 &     ...    \\
     Lup604s  &       $<$5.80e-17	& ...	   &   ...  &	       ...	     &	 ...	 &    ...   &	   ...         &   ...    &	...  &	    $<$5.00e-17        &  ...	 &     ...  &	    ... 	    &  ...    &     ...  &	 ...		& ...	 &     ...    \\
        Sz97  &       $<$9.90e-17	& ...	   &   ...  &	       ...	     &	 ...	 &    ...   &	   ...         &   ...    &	...  &	    $<$2.40e-16        &  ...	 &     ...  &	    ... 	    &  ...    &     ...  &	 ...		& ...	 &     ...    \\
        Sz99  & 4.74($\pm$1.00)e-16	&  -4.4    &  84.2  &	       ...	     &	 ...	 &    ...   & 9.23(3.00)e-17   &   87.0   &    48.7  &	    $<$1.00e-16        &  ...	 &     ...  &	    ... 	    &  ...    &     ...  &  2.17($\pm$0.70)e-16 & 61.9   &    80.7    \\
       Sz100  & 1.46($\pm$0.15)e-15	& -23.4    &  68.0  &	4.30($\pm$1.00)e-16  &  -90.7	 &    69.1  &	   ...         &   ...    &	...  &	 4.68($\pm$1.20)e-16   & -12.3   &    52.3  &  4.66($\pm$1.20)e-16  & -71.2   &    61.8  &	 ...		& ...	 &     ...    \\
       Sz103  & 1.38($\pm$1.80)e-15	& -29.8    &  63.9  &	       ...	     &	 ...	 &    ...   &	   ...         &   ...    &	...  &	 2.20($\pm$0.70)e-16   & -37.5   &   34$^\dagger$  &	    ... 	    &  ...    &     ...  &	 ...		& ...	 &     ...    \\
       Sz104  & 3.63($\pm$0.80)e-16	&  -1.6    &  58 &	       ...	     &	 ...	 &    ...   &	   ...         &   ...    &	...  &	    $<$1.00e-16        &  ...	 &     ...  &	    ... 	    &  ...    &     ...  &	 ...		& ...	 &     ...    \\
      Lup706  &       $<$1.00e-17	& ...	   &   ...  &	       ...	     &	 ...	 &    ...   &	   ...         &   ...    &	...  &	    $<$7.90e-18        &  ...	 &     ...  &	    ... 	    &  ...    &     ...  &	 ...		& ...	 &     ...    \\
       Sz106  & 2.98($\pm$0.18)e-15	& -34.4    & 136.4  &	       ...	     &	 ...	 &    ...   &	   ...         &   ...    &	...  &	 5.68($\pm$1.00)e-16   &   0.5   &    80.1  &	    ... 	    &  ...    &     ...  &	 ...		& ...	 &     ...    \\
  Par-Lup3-3  &       $<$8.16e-16	& ...	   &   ...  &	       ...	     &	 ...	 &    ...   &	   ...         &   ...    &	...  &	    $<$4.10e-16        &  ...	 &     ...  &	    ... 	    &  ...    &     ...  &	 ...		& ...	 &     ...    \\
  Par-Lup3-4  & 1.73($\pm$0.04)e-15	&   3.9    &  76.4  &	       ...	     &	 ...	 &    ...   &	   ...         &   ...    &	...  &	 1.27($\pm$0.02)e-15   &  10.3   &    70.1  &	    ... 	    &  ...    &     ...  &	 ...		& ...	 &     ...    \\
       Sz110  & 1.48($\pm$0.30)e-15	& -25.7    & 118.4  &	       ...	     &	 ...	 &    ...   &	   ...         &   ...    &	...  &	    $<$2.00e-16        &  ...	 &     ...  &	    ... 	    &  ...    &     ...  &	 ...		& ...	 &     ...    \\
       Sz111  &       $<$6.00e-16	& ...	   &   ...  &	       ...	     &	 ...	 &    ...   &	   ...         &   ...    &	...  &	    $<$5.80e-16        &  ...	 &     ...  &	    ... 	    &  ...    &     ...  &	 ...		& ...	 &     ...    \\
       Sz112  & 2.40($\pm$0.60)e-16	&  -4.0    &  58 &	       ...	     &	 ...	 &    ...   &	   ...         &   ...    &	...  &	    $<$1.40e-16        &  ...	 &     ...  &	    ... 	    &  ...    &     ...  &	 ...		& ...	 &     ...    \\
       Sz113  &       $<$4.00e-16	& ...	   &   ...  &	2.33($\pm$0.50)e-15  & -124.6	 &    73.71 &	   ...         &   ...    &	...  &	    $<$1.00e-16        &  ...	 &     ...  &  7.60($\pm$0.50)e-16  & -120.9  &    65.5  &	 ...		& ...	 &     ...    \\
     J160859  & 7.98($\pm$1.70)e-17	& -38.4    &  69.0  &	       ...	     &	 ...	 &    ...   &	   ...         &   ...    &	...  &	    ... 	       &  ...	 &     ...  &	    ... 	    &  ...    &     ...  &	 ...		& ...	 &     ...    \\
    c2dJ1609  & 7.94($\pm$1.50)e-16	& -12.5    &  95.1  &	       ...	     &	 ...	 &    ...   &	   ...         &   ...    &	...  &	    $<$3.50e-16        &  ...	 &     ...  &	    ... 	    &  ...    &     ...  &	 ...		& ...	 &     ...    \\
       Sz114  & 5.39($\pm$1.30)e-16	&  -1.6    &  59.9  &	       ...	     &	 ...	 &    ...   &	   ...         &   ...    &	...  &	    $<$5.40e-16        &  ...	 &     ...  &	    ... 	    &  ...    &     ...  &	 ...		& ...	 &     ...    \\
       Sz115  &       $<$6.40e-17	& ...	   &   ...  &	       ...	     &	 ...	 &    ...   &	   ...         &   ...    &	...  &	    $<$1.90e-16        &  ...	 &     ...  &	    ... 	    &  ...    &     ...  &	 ...		& ...	 &     ...    \\
     Lup818s  &       $<$3.50e-17	& ...	   &   ...  &	       ...	     &	 ...	 &    ...   &	   ...         &   ...    &	...  &	    $<$1.80e-17        &  ...	 &     ...  &	    ... 	    &  ...    &     ...  &	 ...		& ...	 &     ...    \\
      Sz123A  & 2.10($\pm$0.40)e-15	& -21.7    &  89.7  &	5.80($\pm$1.00)e-16  & -103.3	 &    65.6  &	   ...         &   ...    &	...  &	    $<$4.00e-16        &  ...	 &     ...  &	    ... 	    &  ...    &     ...  &	 ...		& ...	 &     ...    \\
      Sz123B  & 1.70($\pm$0.25)e-16	&   7.4    &  76.3  &	1.00($\pm$0.50)e-16  &  -65.5	 &    58.0  &	   ...         &   ...    &	...  &	    $<$9.00e-17        &  ...	 &     ...  &	    ... 	    &  ...    &     ...  &	 ...		& ...	 &     ...    \\
  SST-Lup3-1  &       $<$3.50e-17	& ...	   &   ...  &	       ...	     &	 ...	 &    ...   &	   ...         &   ...    &	...  &	    $<$6.00e-17        &  ...	 &     ...  &	    ... 	    &  ...    &     ...  &	 ...		& ...	 &     ...    \\
       SO397  &       $<$1.70e-17	& ...	   &   ...  &	       ...	     &	 ...	 &    ...   &	   ...         &   ...    &	...  &	    $<$8.00e-17        &  ...	 &     ...  &	    ... 	    &  ...    &     ...  &	 ...		& ...	 &     ...    \\
       SO490  &       $<$1.40e-17	& ...	   &   ...  &	       ...	     &	 ...	 &    ...   &	   ...         &   ...    &	...  &	    $<$1.60e-17        &  ...	 &     ...  &	    ... 	    &  ...    &     ...  &	 ...		& ...	 &     ...    \\
       SO500  &       $<$3.00e-18	& ...	   &   ...  &	       ...	     &	 ...	 &    ...   &	   ...         &   ...    &	...  &	    $<$4.00e-18        &  ...	 &     ...  &	    ... 	    &  ...    &     ...  &	 ...		& ...	 &     ...    \\
       SO587  & 7.58($\pm$0.70)e-16	&  -8.0    &  74.6  &	       ...	     &	 ...	 &    ...   &	   ...         &   ...    &	...  &	 2.14($\pm$0.02)e-15   &  -9.1   &    48.1  &	    ... 	    &  ...    &     ...  &	 ...		& ...	 &     ...    \\
       SO646  & 9.11($\pm$2.00)e-17	& -26.7    &  90.3  &	       ...	     &	 ...	 &    ...   &	   ...         &   ...    &	...  &	    $<$8.00e-17        &  ...	 &     ...  &	    ... 	    &  ...    &     ...  &	 ...		& ...	 &     ...    \\
       SO848  &       $<$5.00e-17	& ...	   &   ...  &	 3.47($\pm$0.17)e-16 &	-46.9    &    75.6  &	   ...         &   ...    &	...  &	 3.26($\pm$0.25)e-16   &  -3.0   &    50.0  &  4.48($\pm$0.25)e-16  & -47.3   &    49.6  &	 ...		& ...	 &     ...    \\
      SO1260  & 1.14($\pm$0.20)e-16	& -27.2    &  61.0  &	       ...	     &	 ...	 &    ...   &	   ...         &   ...    &	...  &	    $<$4.00e-17        &  ...	 &     ...  &	    ... 	    &  ...    &     ...  &	 ...		& ...	 &     ...    \\
      SO1266  &       $<$3.00e-17	& ...	   &   ...  &	 2.80($\pm$1.00)e-17 &	-68.4    &   59$^\dagger$  &	   ...         &   ...    &	...  &	 4.71($\pm$1.50)e-17   &  -4.1   &   34$^\dagger$  &	    ... 	    &  ...    &     ...  &	 ...		& ...	 &     ...    \\
\hline			    			     	  
\end{longtable}
\footnotesize{
%Notes:
\begin{itemize}
\item $^\dagger$ : Not resolved line; the FWHM is the instrumental resolution (e.g., 59 km/s for the \SIIA\ line  and 34 km/s for the \SIIB\ line).
\end{itemize}
}
\end{landscape}
%%%%%%%%%%%%%%%%%%%%%%%%%%%%%%%%%%%%%%%%%%%%%%%%%%%%%%%%%%%%%%%%%%%%%%%%%%

%%%%%%%%%%%%%%%%%%%%%%%%%%%%%%%%%%%%%%%%%%%%%%%%%%%%%%
\setlength{\tabcolsep}{3pt}

\begin{landscape}
\scriptsize
\begin{longtable}{l|| c c c | c c c | c c c }
\caption[ ]{\label{tab_tabellone_4} Fluxes, velocity peak and full-width half maximun of [\ion{N}{ii}] lines}\\
\hline
   name &   \multicolumn{9}{c}{\NII}  \\ \cline{2-10} 
              &     \multicolumn{3}{c|}{LVC}      &    \multicolumn{3}{c|}{HVC-Blue}             & \multicolumn{3}{c}{HVC-Red}    \\ \cline{2-10} 
% \hline
              &   flux                 &V$_{\rm peak}$ & FWHM       &   flux  & V$_{\rm peak}$ & FWHM   & flux    & V$_{\rm peak}$  &  FWHM    \\
              &(erg\,s$^{-1}$\,cm$^{-2}$) & (km\,s$^{-1}$) & (km\,s$^{-1}$) &	(erg\,s$^{-1}$\,cm$^{-2}$)  &	(km\,s$^{-1}$)  & (km\,s$^{-1}$)  & (erg\,s$^{-1}$\,cm$^{-2}$)  & (km\,s$^{-1}$) & (km\,s$^{-1}$)      \\
\hline
        Sz66  &     1.12($\pm$0.30)e-15	  &    -36.7	&    37.9  &	  ...		 &  ...    &	 ...  &       ...	    & ...   &    ...  \\
   AKC2006-1  &  	$<$5.0e-17	  &	...	&    ...   &	  ...		 &  ...    &	 ...  &       ...	    & ...   &    ...  \\
        Sz69  &         ...     	  &	...	&    ...   & 3.12($\pm$1.00)e-16 & -90.3   &	62.7  & 1.72($\pm$0.50)e-16 & 57.0  &   44.3  \\
        Sz71  &  	$<$8.00e-16	  &	...	&    ...   &	  ...		 &  ...    &	 ...  &       ...	    & ...   &    ...  \\
        Sz72  &  	$<$8.00e-16	  &	...	&    ...   &	  ...		 &  ...    &	 ...  &       ...	    & ...   &    ...  \\
        Sz73  &  	$<$8.00e-15	  &	...	&    ...   & 1.60($\pm$0.30)e-14 & -94.5   &   103.0  &       ...	    & ...   &    ...  \\
        Sz74  &  	$<$3.00e-15	  &	...	&    ...   &	  ...		 &  ...    &	 ...  &       ...	    & ...   &    ...  \\
        Sz83  &  	$<$3.60e-15	  &	...	&    ...   &	  ...		 &  ...    &	 ...  &       ...	    & ...   &    ...  \\
        Sz84  &         ...     	  &	...	&    ...   &	  ...		 &  ...    &	 ...  &       ...	    & ...   &    ...  \\
       Sz130  &  	$<$7.50e-16	  &	...	&    ...   & 1.00($\pm$0.26)e-15 & -56.5   &	40.8  &       ...	    & ...   &    ...  \\
       Sz88A  &  	$<$5.00e-16	  &	...	&    ...   &	  ...		 &  ...    &	 ...  &       ...	    & ...   &    ...  \\
       Sz88B  &  	$<$2.30e-16	  &	...	&    ...   &	  ...		 &  ...    &	 ...  &       ...	    & ...   &    ...  \\
        Sz91  &  	$<$6.40e-16	  &	...	&    ...   &	  ...		 &  ...    &	 ...  &       ...	    & ...   &    ...  \\
      Lup713  &  	$<$1.50e-17	  &	...	&    ...   &	  ...		 &  ...    &	 ...  &       ...	    & ...   &    ...  \\
     Lup604s  &  	$<$9.00e-17	  &	...	&    ...   &	  ...		 &  ...    &	 ...  &       ...	    & ...   &    ...  \\
        Sz97  &  	$<$3.60e-16	  &	...	&    ...   &	  ...		 &  ...    &	 ...  &       ...	    & ...   &    ...  \\
        Sz99  &  	$<$6.70e-17	  &	...	&    ...   &	  ...		 &  ...    &	 ...  &       ...	    & ...   &    ...  \\
       Sz100  &         ...     	  &	...	&    ...   & 4.82($\pm$1.00)e-16 & -70.8   &	34.9  &       ...	    & ...   &    ...  \\
       Sz103  &         ...     	  &	...	&    ...   & 5.37($\pm$1.00)e-16 & -40.2   &   34$^\dagger$  &       ...	    & ...   &    ...  \\
       Sz104  &  	$<$2.00e-16	  &	...	&    ...   &	  ...		 &  ...    &	 ...  &       ...	    & ...   &    ...  \\
      Lup706  &  	$<$7.60e-18	  &	...	&    ...   &	  ...		 &  ...    &	 ...  &       ...	    & ...   &    ...  \\
       Sz106  &     4.50($\pm$1.00)e-16	  &    -1.0	&   85.0   &	  ...		 &  ...    &	 ...  &       ...	    & ...   &    ...  \\
  Par-Lup3-3  &  	$<$3.00e-16	  &	...	&    ...   &	  ...		 &  ...    &	 ...  &       ...	    & ...   &    ...  \\
  Par-Lup3-4  &     2.22($\pm$0.25)e-16   &    10.2	&   70.0   &	  ...		 &  ...    &	 ...  &       ...	    & ...   &    ...  \\
       Sz110  &  	$<$4.00e-16	  &	...	&    ...   &	  ...		 &  ...    &	 ...  &       ...	    & ...   &    ...  \\
       Sz111  &  	$<$6.50e-16	  &	...	&    ...   &	  ...		 &  ...    &	 ...  &       ...	    & ...   &    ...  \\
       Sz112  &  	$<$3.20e-16	  &	...	&    ...   &	  ...		 &  ...    &	 ...  &       ...	    & ...   &    ...  \\
       Sz113  &         ...     	  &	...	&    ...   &  1.29($\pm$4.00)e-16& -125.1  &	40.6  &       ...	    & ...   &    ...  \\
     J160859  &  	$<$2.00e-17	  &	...	&    ...   &	  ...		 &  ...    &	 ...  &       ...	    & ...   &    ...  \\
    c2dJ1609  &     6.89($\pm$1.00)e-16	  &   -33.4	&   46.2   &	  ...		 &  ...    &	 ...  &       ...	    & ...   &    ...  \\
       Sz114  &  	$<$6.70e-16	  &	...	&    ...   &	  ...		 &  ...    &	 ...  &       ...	    & ...   &    ...  \\
       Sz115  &  	$<$3.30e-16	  &	...	&    ...   &	  ...		 &  ...    &	 ...  &       ...	    & ...   &    ...  \\
     Lup818s  &  	$<$4.50e-17	  &	...	&    ...   &	  ...		 &  ...    &	 ...  &       ...	    & ...   &    ...  \\
      Sz123A  &         ...     	  &	...	&    ...   & 9.27($\pm$3.00)e-16 & -55.8   &	38.9  &       ...	    & ...   &    ...  \\
      Sz123B  &         ...     	  &	...	&    ...   & 2.00($\pm$0.40)e-16 & -56.6   &	44.4  &       ...	    & ...   &    ...  \\
  SST-Lup3-1  &  	$<$1.00e-16	  &	...	&    ...   &	  ...		 &  ...    &	 ...  &       ...	    & ...   &    ...  \\
       SO397  &  	$<$1.30e-16	  &	...	&    ...   &	  ...		 &  ...    &	 ...  &       ...	    & ...   &    ...  \\
       SO490  &  	$<$2.80e-17	  &	...	&    ...   &	  ...		 &  ...    &	 ...  &       ...	    & ...   &    ...  \\
       SO500  &  	$<$7.30e-18	  &	...	&    ...   &	  ...		 &  ...    &	 ...  &       ...	    & ...   &    ...  \\
       SO587  &     4.35($\pm$0.25)e-15	  &    -8.9	&   45.2   &	  ...		 &  ...    &	 ...  &       ...	    & ...   &    ...  \\
       SO646  &  	$<$9.40e-17	  &	...	&    ...   &	  ...		 &  ...    &	 ...  &       ...	    & ...   &    ...  \\
       SO848  &     1.67($\pm$0.05)e-15	  &    -0.5	&   37.8   & 6.01($\pm$0.50)e-16 & -47.9   &	50.7  &       ...	    & ...   &    ...  \\
      SO1260  &  	$<$9.20e-17	  &	...	&    ...   &	  ...		 &  ...    &	 ...  &       ...	    & ...   &    ...  \\
      SO1266  &  	$<$6.3	          &	...	&    ...   &	  ...		 &  ...    &	 ...  &       ...	    & ...   &    ...  \\
\hline			    			     	  
\end{longtable}
\footnotesize{Notes:
\begin{itemize}
\item $^\dagger$ : Not resolved line; the FWHM is the instrumental resolution (e.g., 34 km/s for the \NII\ line).
\end{itemize}
}
\end{landscape}
%%%%%%%%%%%%%%%%%%%%%%%%%%%%%%%%%%%%%%%%%%%%%%%%%%%%%%%%%%%%%%%%%%%%%%%%%%

\twocolumn
\section{Examples of line profiles in selected objects}
\label{app_profiles}

In this Appendix we show  examples of the line profiles discussed in this paper.

\begin{figure*}
	\begin{center}
		\includegraphics[width=18cm]{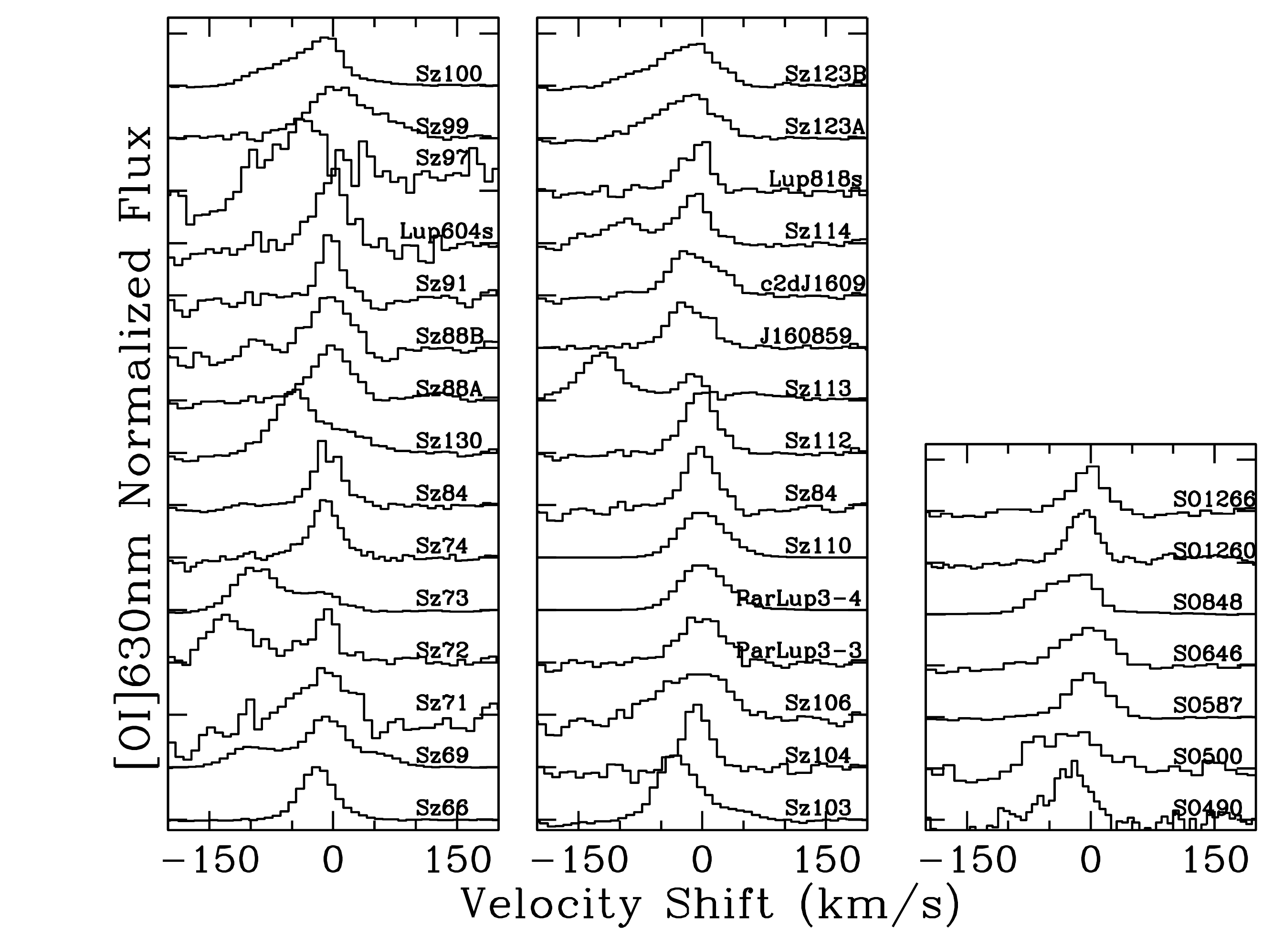}
	\end{center}
	\caption{Normalized, continuum subtracted line profile  of the \OIA\ line. The profiles are shifted vertically for an easier display.}
\label{fig_OI_all}
\end{figure*}

\begin{figure}
	\begin{center}
		\includegraphics[width=9cm]{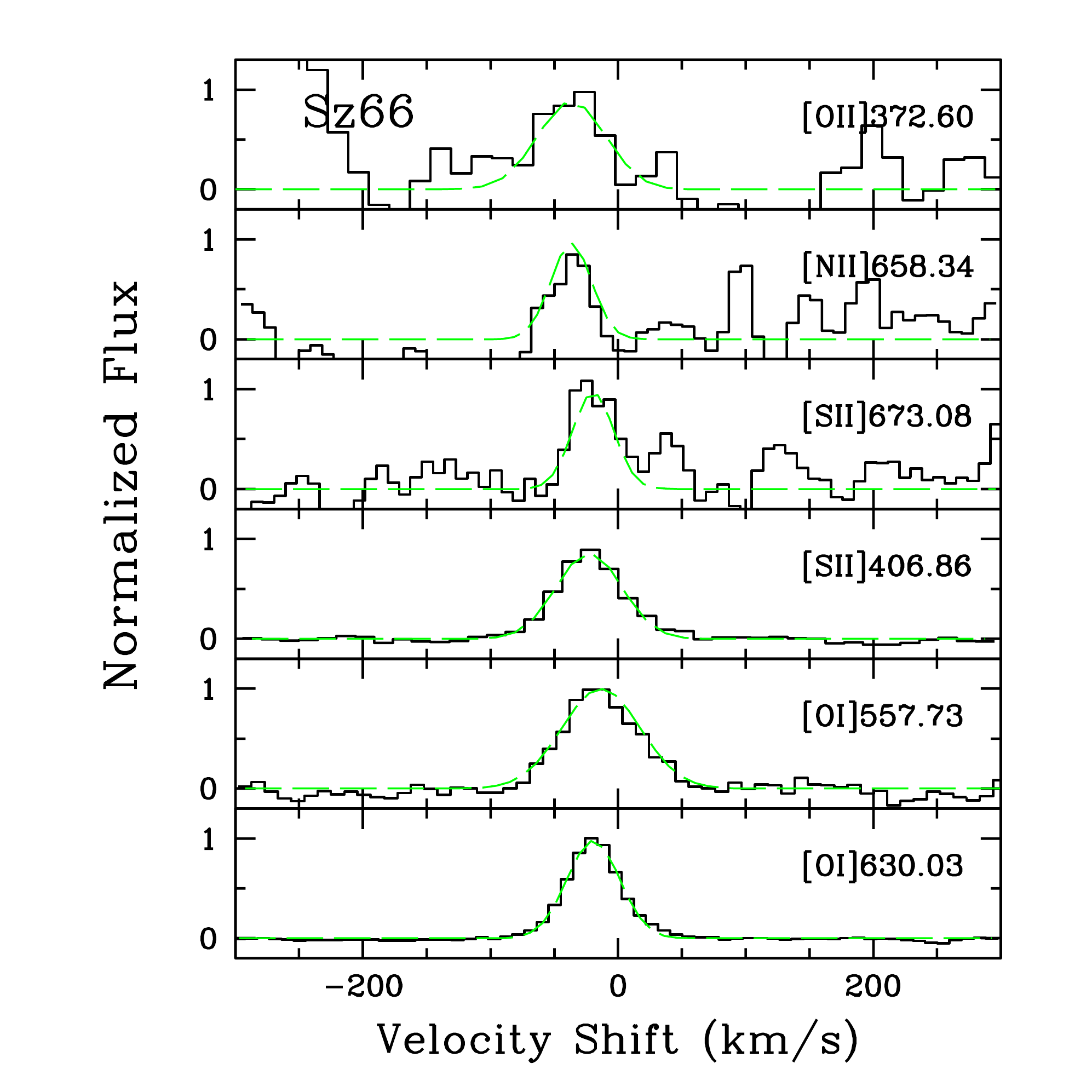}
	\end{center}
	\caption{Normalized, continuum subtracted line profiles (as labelled) for the object Sz66. 
	In this and the following figures 
the dashed colored curves show the LVC (green), HVC-blue shifted (blue) and HVC red-shifted (red);
        the total is given by the dotted (black) line. In this object, only the LVC is detected.
	        }
\label{fig_Sz66}
\end{figure}

\begin{figure}
	\begin{center}
		\includegraphics[width=9cm]{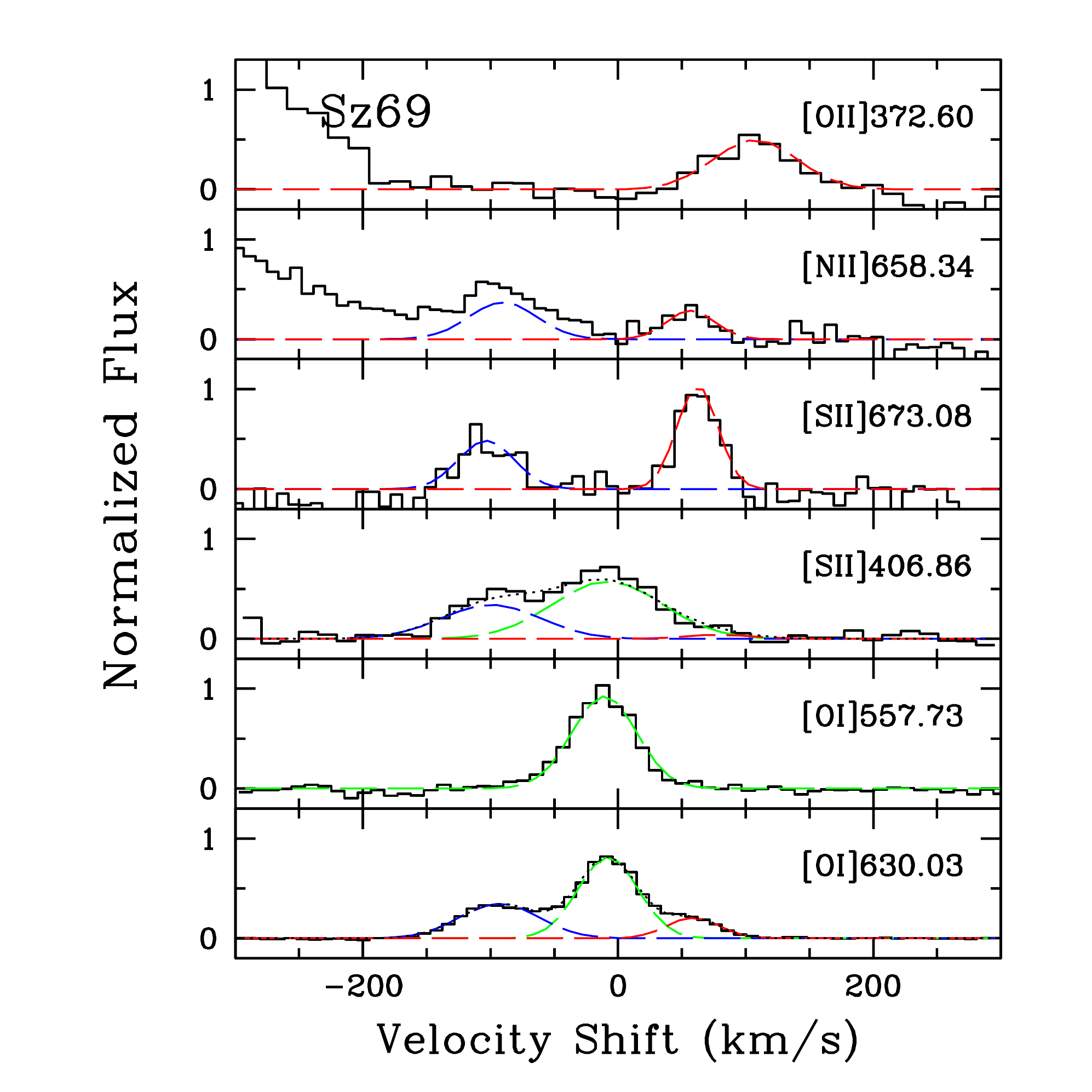}
	\end{center}
	\caption{Same as Fig.~\ref{fig_Sz66} for  Sz69.  }
\label{fig_Sz69}
\end{figure}
\begin{figure}
	\begin{center}
		\includegraphics[width=9cm]{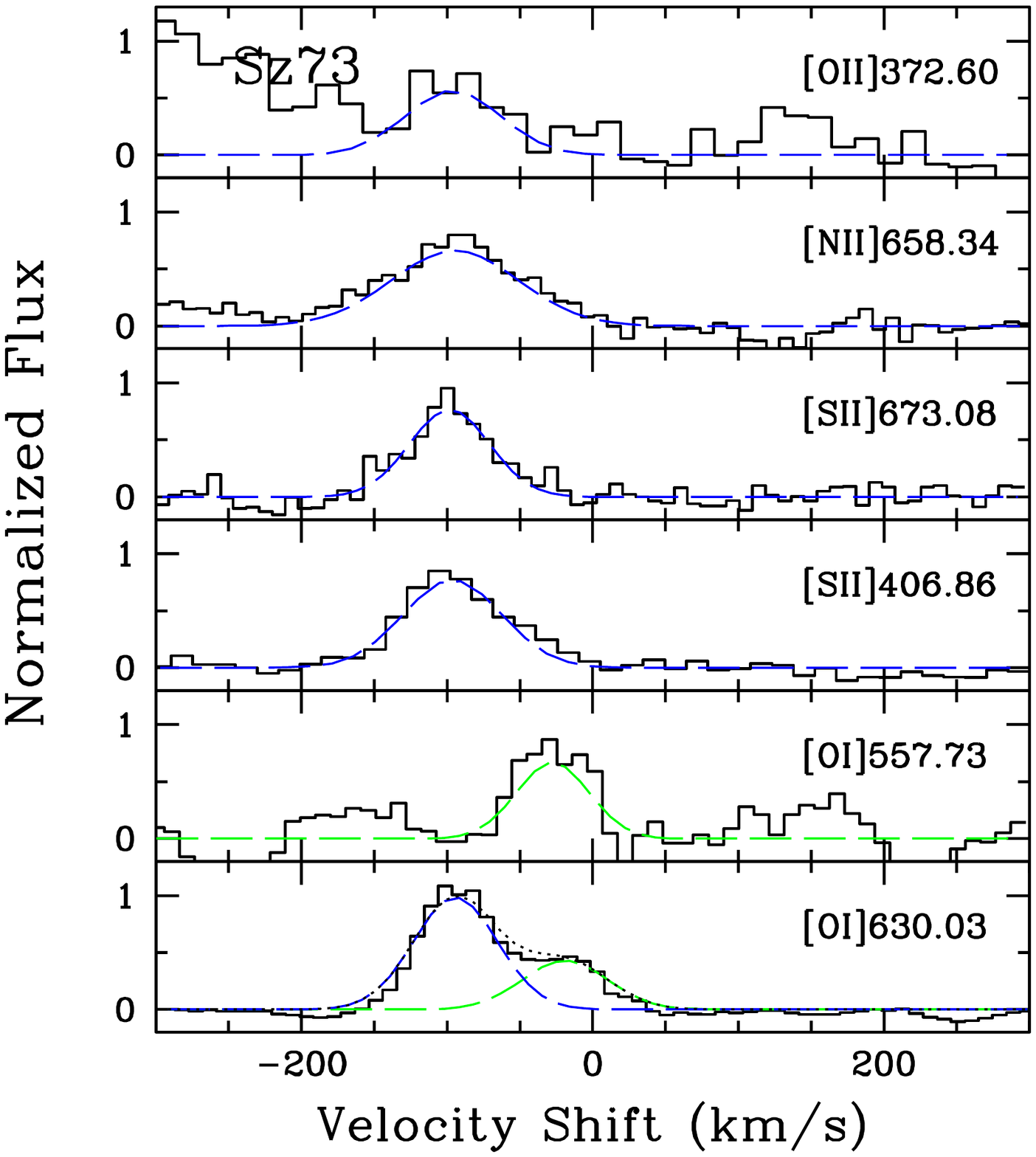}
	\end{center}
	\caption{Same as Fig.~\ref{fig_Sz66} for  Sz73.  }
\label{fig_Sz73}
\end{figure}
\begin{figure}
	\begin{center}
		\includegraphics[width=9cm]{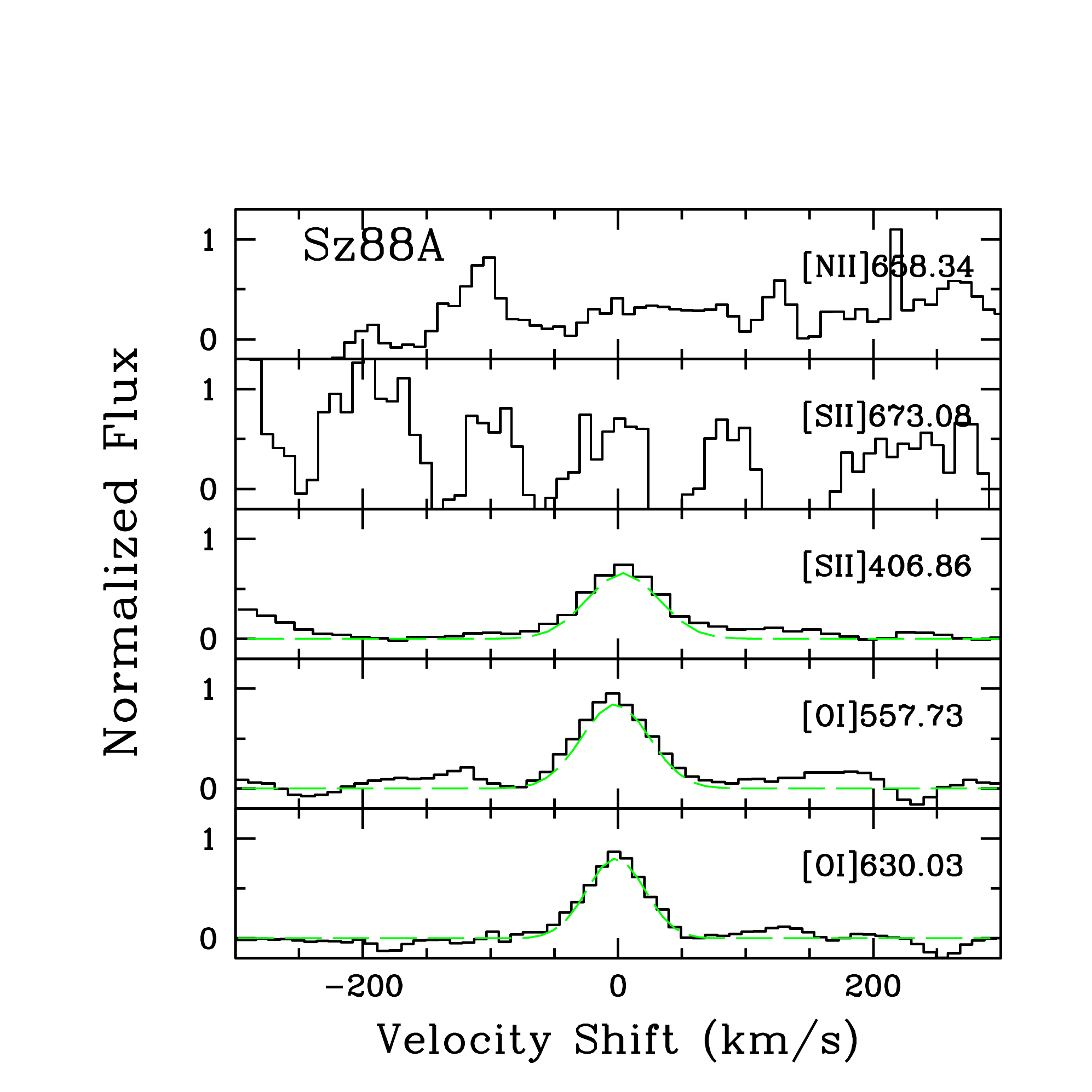}
	\end{center}
	\caption{Same as Fig.~\ref{fig_Sz66} for  Sz88A. The two panels for
the lines \NII\ and \SIIB\ illustrate typical non-detections. 
}
\label{fig_Sz88A}
\end{figure}
\begin{figure}
	\begin{center}
		\includegraphics[width=9cm]{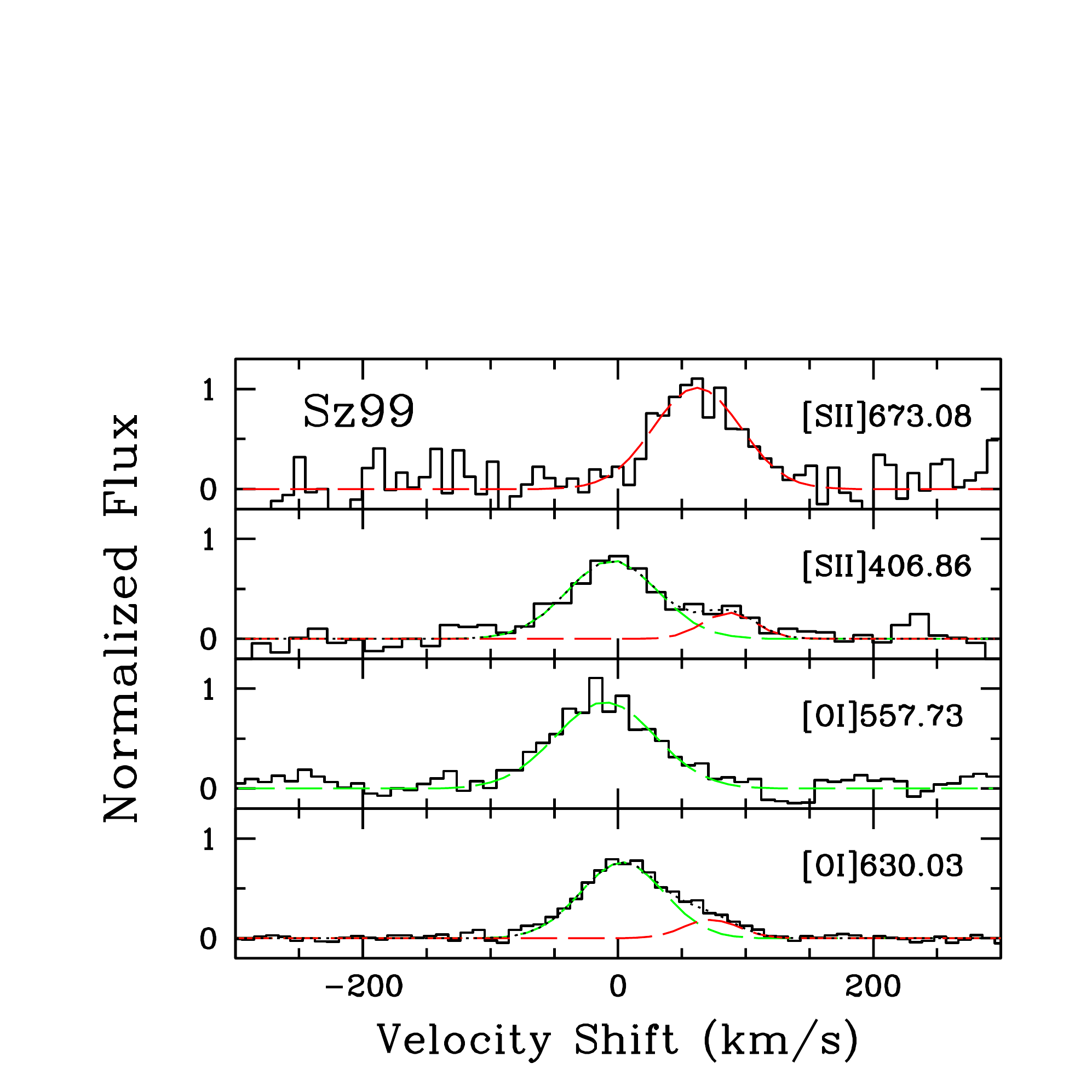}
	\end{center}
	\caption{Same as Fig.~\ref{fig_Sz66} for  Sz99.  }
\label{fig_Sz99}
\end{figure}
\begin{figure}
	\begin{center}
		\includegraphics[width=9cm]{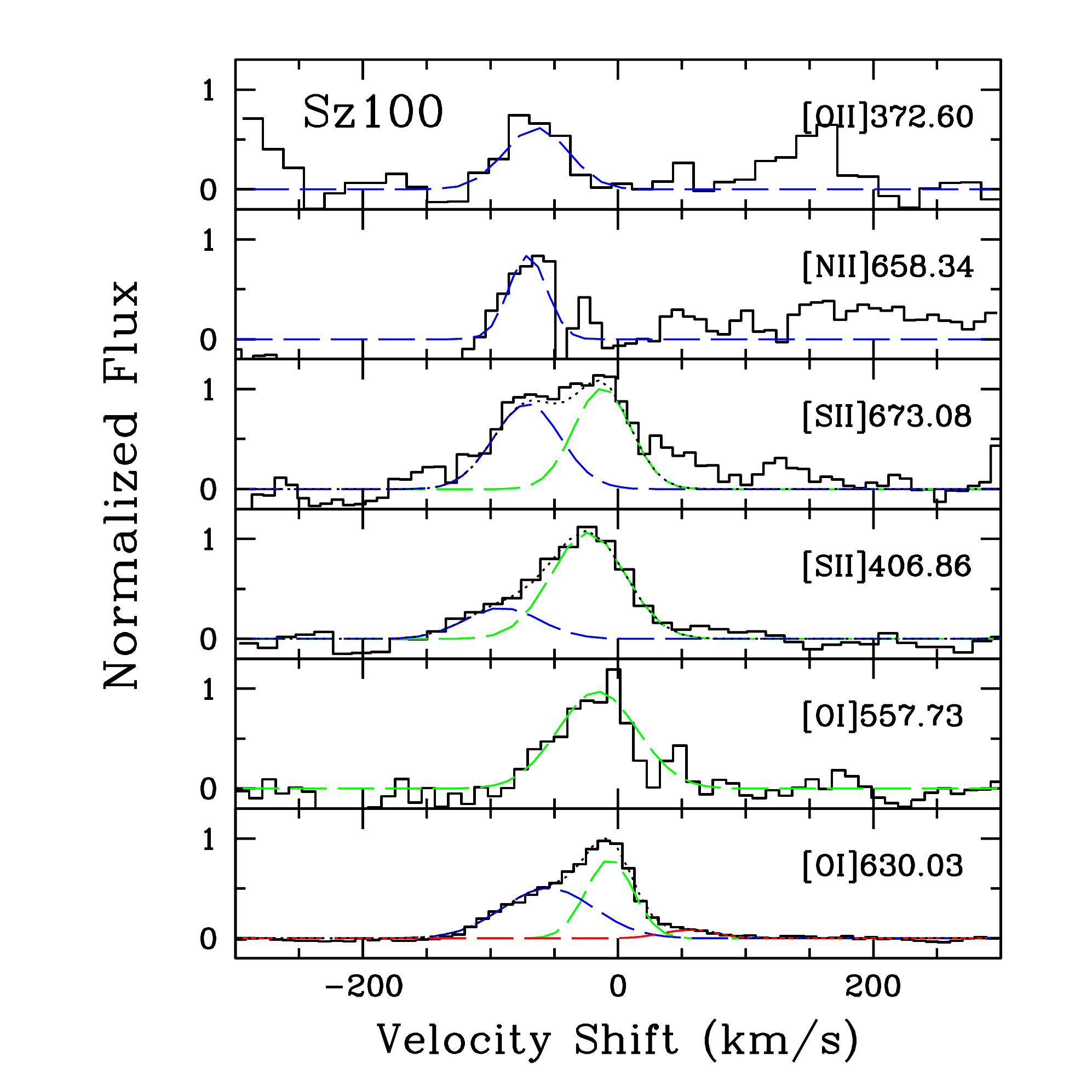}
	\end{center}
	\caption{Same as Fig.~\ref{fig_Sz66} for  Sz100.  }
\label{fig_Sz100}
\end{figure}
\begin{figure}
	\begin{center}
		\includegraphics[width=9cm]{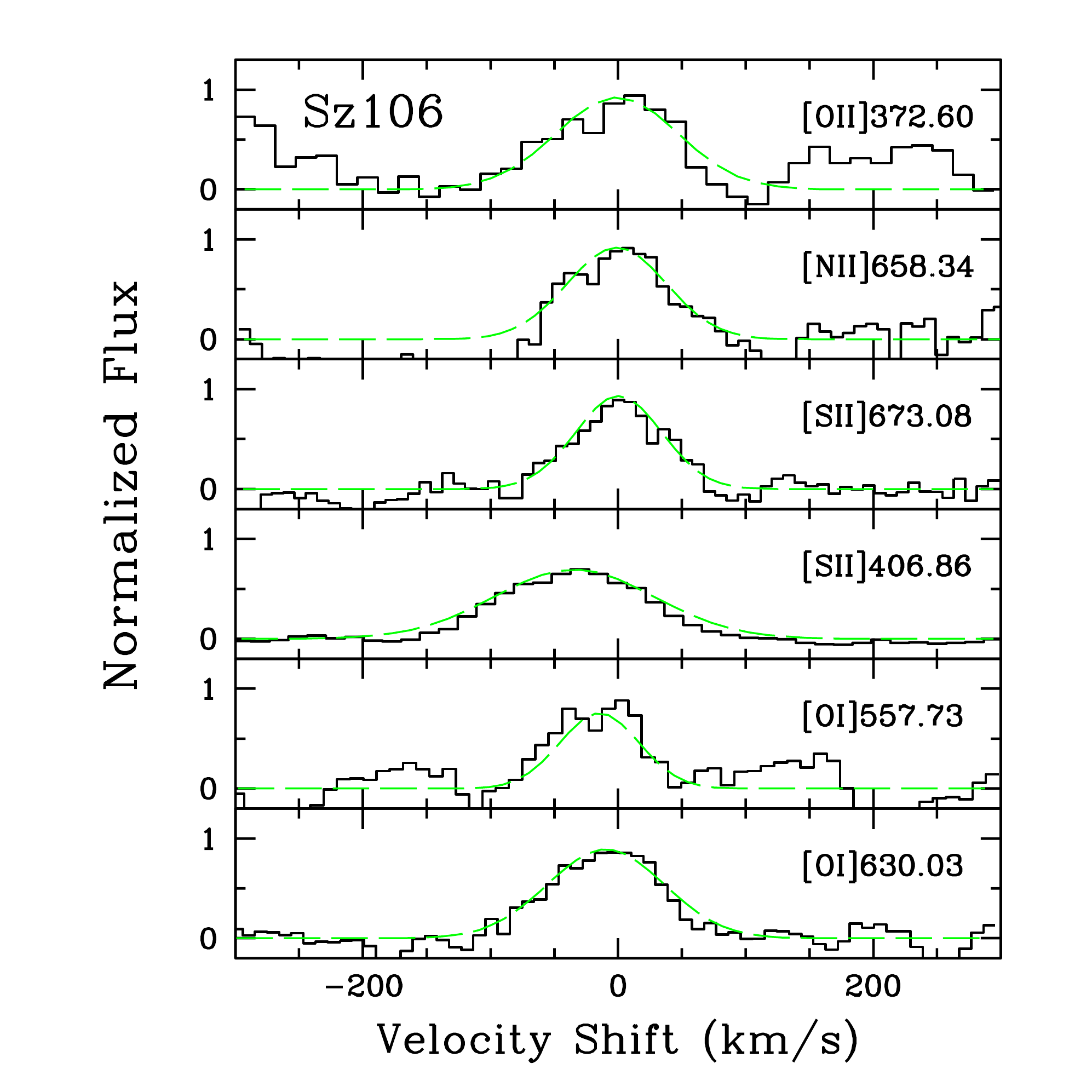}
	\end{center}
	\caption{Same as Fig.~\ref{fig_Sz66} for  Sz106.  }
\label{fig_Sz106}
\end{figure}
\begin{figure}
	\begin{center}
		\includegraphics[width=9cm]{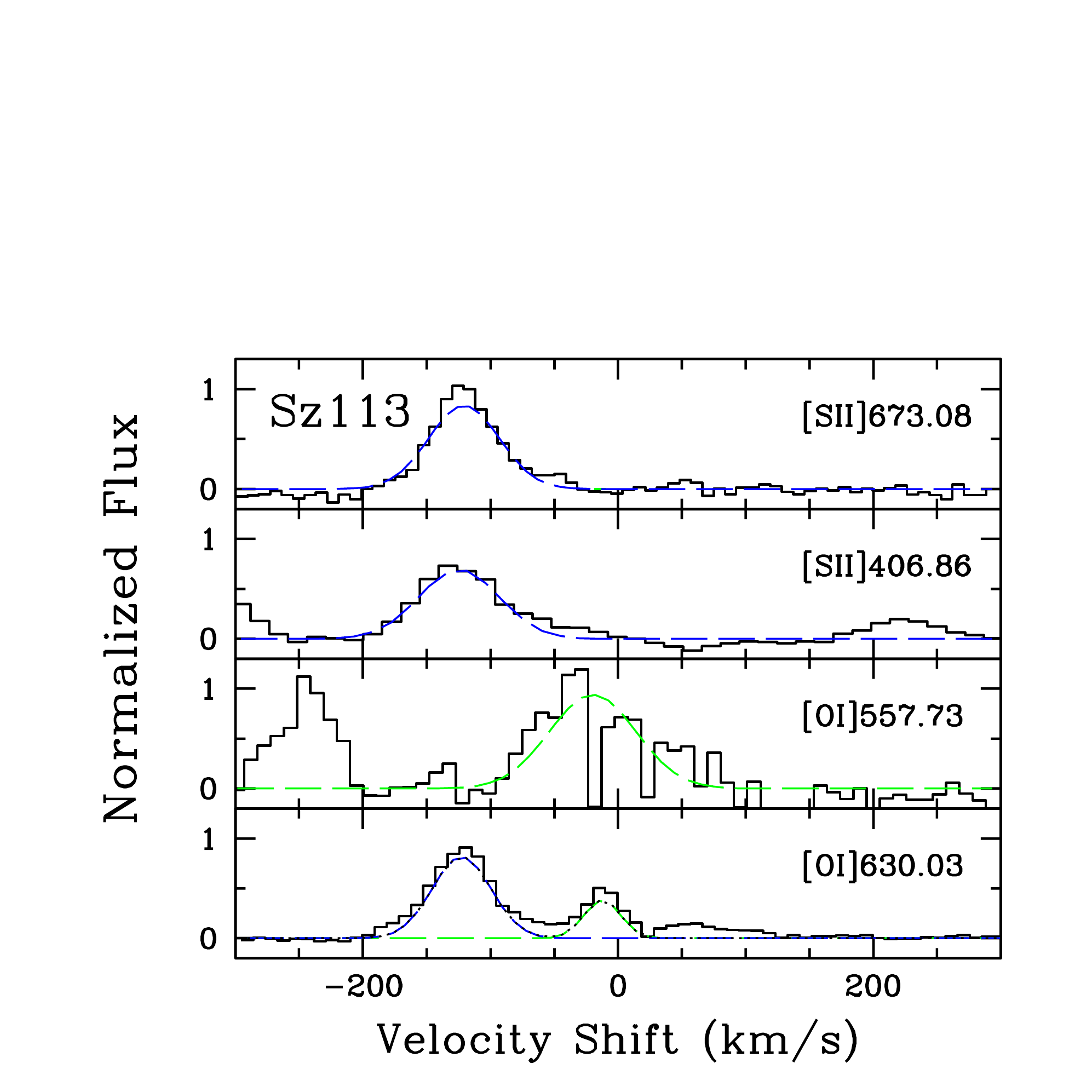}
	\end{center}
	\caption{Same as Fig.~\ref{fig_Sz66} for  Sz113. Note that the \OIB\ is a 3$\sigma$ detection.  }
\label{fig_Sz113}
\end{figure}
\begin{figure}
	\begin{center}
		\includegraphics[width=9cm]{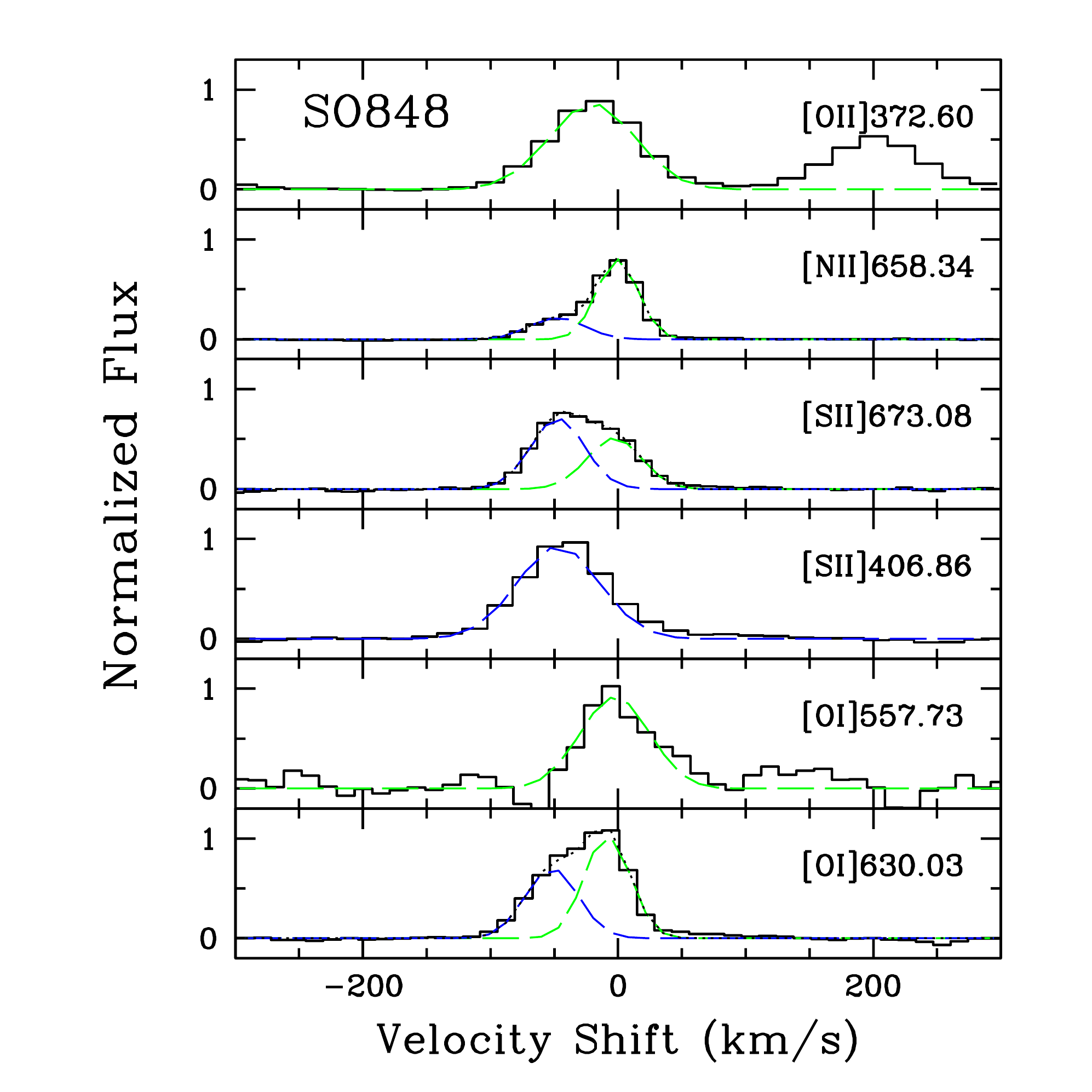}
	\end{center}
	\caption{Same as Fig.~\ref{fig_Sz66} for  SO848.  }
\label{fig_SO848}
\end{figure}

\begin{acknowledgements}
We are indebted to  Bruce Draine for computing the line emissivities used in this work 
and to Francesca Bacciotti, Barbara Ercolano, Ilaria Pascucci and Malcolm Walmsley for 
useful discussions. We thank V. D'Odorico, P. Goldoni and A. Modigliani for their help 
with the X-Shooter pipeline, and F. Getman and G. Capasso for the installation of the
different pipeline versions at Capodimonte. Financial support from INAF, under 
PRIN2013 programe "Disks, jets and the dawn of planets"  is also 
acknowledged.

\end{acknowledgements}

\bibliographystyle{aa}
%\bibliography{winds}

\end{document}